%% file: graphene_damping.tex
\documentclass[aps,prb,
 reprint, 
 tightenlines,
 citeautoscript,%
 showpacs,floatfix,
 superscriptaddress]{revtex4-1}

\usepackage{bm}
\usepackage{amssymb}
\usepackage{graphicx}
\usepackage{amsmath}
\usepackage{textcomp}
\usepackage{multirow}

\begin{document}
\title{Third order nonlinearity of graphene: effects of phenomenological
  relaxation and finite temperature}
\author{J. L. Cheng}
\affiliation{Brussels Photonics Team (B-PHOT), Department of Applied
  Physics and Photonics (IR-TONA), Vrije Universiteit Brussel,
  Pleinlaan 2, 1050 Brussel, Belgium}
\affiliation{Department of Physics and Institute for Optical Sciences,
  University of Toronto, 60 St. George Street, Toronto, Ontario,
  Canada M5S 1A7}
\author{N. Vermeulen}
\affiliation{Brussels Photonics Team (B-PHOT), Department of Applied
  Physics and Photonics (IR-TONA), Vrije Universiteit Brussel,
  Pleinlaan 2, 1050 Brussel, Belgium}
\author{J. E. Sipe}%
\affiliation{Department of Physics and Institute for Optical Sciences,
University of Toronto, 60 St. George Street, Toronto, Ontario, Canada
M5S 1A7}
\date{\today}
\begin{abstract}
We investigate the effect of phenomenological
relaxation parameters on the third order optical nonlinearity of doped
graphene by  perturbatively solving the semiconductor Bloch
equation around the Dirac points. An analytic expression for the
nonlinear conductivity at zero temperature is obtained under the linear dispersion
approximation. With this analytic formula as starting point, we construct the
conductivity at finite temperature and study the optical
response to a laser pulse of finite duration. We illustrate the
dependence of several nonlinear optical effects, such as third harmonic
generation, Kerr effects and two photon absorption, parametric
frequency conversion, and two color coherent current injection,  on
the relaxation parameters, temperature, and pulse duration. In the
special case where one of the electric fields is taken as a dc field, we
investigate the dc-current and dc-field induced second order
nonlinearities, including dc-current induced second harmonic
generation and difference frequency generation. 
\end{abstract}
\maketitle

\section{Introduction}
The optical nonlinearities of monolayer graphene have recently attracted wide
attention  \cite{Rev.Mod.Phys._81_109_2009_CastroNeto,Nat.Photon._4_611_2010_Bonaccorso,Nat.Photon._6_554_2012_Gu},
both experimentally and theoretically. The nonlinear
susceptibility of graphene  \cite{Phys.Rep._535_101_2014_Glazov} is both
strong -- per atom it is orders of magnitude higher than that of common gapped semiconductors
and metals -- and controllable by the chemical
potential  \cite{NewJ.Phys._16_53014_2014_Cheng,Opt.Express_22_15868_2014_Cheng},
which can be tuned by an external gate voltage   \cite{Science_306_666_2004_Novoselov,Science_320_206_2008_Wang} or
chemical doping   \cite{J.Mater.Chem._21_3335_2011_Liu}. With the
possibilities it offers for integration in silicon-based optical integrated
circuits, graphene is an exciting new candidate 
for enhancing nonlinear optical functionalities in silicon-based on-chip optical
devices, such as on-chip broadband light sources, electro-optic
modulators  \cite{Opt.Lett._35_2753_2010_Wuelbern,Opt.Express_22_5252_2014_Matheisen}, 
optical switches  \cite%
{IEEEJ.QuantumElectron._29_2650_1993_Ironside,Appl.Phys.Lett._104_111114_2014_Liu,Opt.Lett._36_3825_2011_Gandomkar}, 
and optical transistors  \cite{Opt.Lett._19_1305_1994_Hagan,ChinesePhys.Lett._30_97301_2013_Ren}. In realizing
 some of these devices  \cite{Opt.Lett._36_3825_2011_Gandomkar}, the presence of second order optical nonlinearities, especially
 second harmonic generation (SHG), is a key requirement. 

The third order optical nonlinearity is described by the susceptibility tensor
$\chi^{(3)}(\omega_1,\omega_2,\omega_3)$ or equivalently the conductivity tensor
$\sigma^{(3)}(\omega_1,\omega_2,\omega_3)$, which has a complex
frequency dependence. It describes different physical effects, such as third harmonic
generation (THG), which is determined by $\chi^{(3)}(\omega,\omega,\omega)$; Kerr effects and two
photon absorption, which are determined by $\chi^{(3)}(-\omega,\omega,\omega)$; two-color
coherent current injection, which is determined by $\chi^{(3)}(-\omega,-\omega,2\omega)$; and parametric frequency
conversion (four wave mixing), which is determined by $\chi^{(3)}(-\omega_s,\omega_p,\omega_p)$. 
Due to the inversion symmetry of its crystal structure, pristine graphene has no
second order optical nonlinearities arising from electric dipole transitions. However, in graphene-based photonic
devices an effective second order susceptibility can arise from the
breaking of inversion symmetry in a number of ways: (1)
the presence of an asymmetric interface between graphene and the substrate   \cite%
{Appl.Phys.Lett._95_261910_2009_Dean,Phys.Rev.B_82_125411_2010_Dean,Phys.Rev.B_85_121413_2012_Bykov,NanoLett._13_2104_2013_An,Phys.Rev.B_89_115310_2014_An,Phys.Rev.B_35_1129_1987_Sipe},
not relevant for normally incident light; (2) the contribution of forbidden transitions
involving the finite wave vector of light
  \cite{Europhys.Lett._79_27002_2007_Mikhailov,J.Phys.Condens.Matter_20_384204_Mikhailov,JETPLett._93_366_2011_Glazov,Phys.Rev.B_84_45432_2011_Mikhailov,Phys.Rep._535_101_2014_Glazov};
(3) the presence of natural curvature fluctuations of suspended
graphene   \cite{Appl.Phys.Lett._105_151605_2014_Lin}; 
(4) the application of a dc electric field to generate an asymmetric steady state
\cite{NanoLett._12_2032_2012_Wu,Phys.Rev.B_85_121413_2012_Bykov,J.Nanophoton._6_61702_2012_Avetissian,NanoLett._13_2104_2013_An,Phys.Rev.B_89_115310_2014_An,Opt.Express_22_15868_2014_Cheng}. 
The last is associated with the third order optical
nonlinearity $\chi^{(3)}(\omega_1,\omega_2,0)$, with one of the
electric fields independent of time. It includes current induced second harmonic
generation  \cite{Appl.Phys.Lett._67_1113_1995_Khurgin} (CSHG) or
electric field induced second harmonic generation (EFISH).

Experimental studies of many of the optical nonlinear effects mentioned above have already demonstrated in graphene.
Typically the experimental data are
analyzed by extracting an effective optical nonlinear susceptibility,
with the graphene monolayer treated as a thin film with a thickness of
$3.3$~\AA  \cite{Phys.Rev.Lett._105_097401_2010_Hendry,ACSNano_7_8441_2013_Saeynaetjoki,Phys.Rev.B_87_121406_2013_Kumar}. In
this way, most of the experimental techniques used to determine the
nonlinear optical response of bulk materials or thin films can be
directly applied to the study of graphene. 
In a gapped semiconductor, third order susceptibilities do not change drastically in the
nonresonant regime, where all photon energies are much lower 
than the energy gap  \cite{boyd_nonlinearoptics}. Yet they show a strong and
complicated photon energy dependence in pristine graphene because resonant transitions
always exist for any photon energy, due to the vanishing gap and the
presence of free carriers, leading to some similarities with a metal film   \cite{Phys.Rev.B_87_121406_2013_Kumar}. These
complexities have been observed in experimental studies of parametric frequency conversion
  \cite{Phys.Rev.Lett._105_097401_2010_Hendry}, 
THG   \cite{ACSNano_7_8441_2013_Saeynaetjoki,Phys.Rev.B_87_121406_2013_Kumar,Phys.Rev.X_3_021014_2013_Hong}, 
 Kerr effects and two photon absorption   \cite{NanoLett._11_2622_2011_Yang,Opt.Lett._37_1856_2012_Zhang,Nat.Photon._6_554_2012_Gu,NanoLett._11_5159_2011_Wu}, 
two color coherent control   \cite{NanoLett._10_1293_2010_Sun,NewJ.Phys._14_105012_2012_Sun,Phys.Rev.B_85_165427_2012_Sun}, 
and SHG
  \cite{Appl.Phys.Lett._95_261910_2009_Dean,Phys.Rev.B_82_125411_2010_Dean,Phys.Rev.B_85_121413_2012_Bykov,NanoLett._13_2104_2013_An,Phys.Rev.B_89_115310_2014_An,Appl.Phys.Lett._105_151605_2014_Lin}
in graphene.

Theoretically, the optical nonlinearities of graphene have
been investigated by perturbative treatments based on Fermi's Golden
Rule, and by density matrix calculations, both of which are standard
methods in studying the optical response of gapped semiconductors. In
an earlier communication we sketched some of the relevant work done
before early 2014   \footnote{Note in particular the footnote on the second page of Cheng {\it et al.} \cite{NewJ.Phys._16_53014_2014_Cheng}, which points out a source of confusion in comparing some of the experimental work with the theoretical study of Hendry {\it et al.}\cite{Phys.Rev.Lett._105_097401_2010_Hendry}{}}; recent contributions include a
calculation by Mikhailov  \cite{Phys.Rev.B_90_241301_2014_Mikhailov} of
THG \footnote{Despite the
  claim\cite{Phys.Rev.B_90_241301_2014_Mikhailov} that the scalar
  potential treatment of THG leads to disagreement with our earlier
  work\cite{NewJ.Phys._16_53014_2014_Cheng}, we find
  \cite{commentcomparison} agreement between the two approaches},\nocite{commentcomparison} and
numerically solution of the equation of motion under strong laser
fields by Avetissian {\it et al.}
\cite{Phys.Rev.B_85_115443_2012_Avetissian,
  J.Nanophoton._6_61702_2012_Avetissian,
  Phys.Rev.B_88_165411_2013_Avetissian,
  Phys.Rev.B_88_245411_2013_Avetissian}. All of
these studies focused on one or a few nonlinear effects. In our
earlier work  \cite{NewJ.Phys._16_53014_2014_Cheng} we performed a perturbative calculation based on a
 density matrix formalism; ignoring all scattering effects, we
 obtained an analytic expression for the general optical sheet conductivity 
$\sigma^{(3)}(\omega_1,\omega_2,\omega_3)$, which can be related to the
effective susceptibility $\chi^{(3)}(\omega_1,\omega_2,\omega_3)$, in
doped graphene at zero temperature. We found that the optical
conductivities depend strongly on the chemical potential and photon
energies, and exhibit many divergences associated with resonant transitions, which occur when photon energies or their
combinations match the chemical potential gap. Taking $\omega_3=0$ and including phenomenological
relaxation times for the generation of both dc and optically induced 
current, we calculated the current induced second order nonlinearities at
zero temperature and obtained an analytic
expression  \cite{Opt.Express_22_15868_2014_Cheng} for
CSHG. The effective susceptibility shows two peaks corresponding to
two resonant transitions induced by the fundamental and the  
second harmonic light, with the
peak values strongly dependent on the relaxation time. Adopting the parameters
used in calculations of bilayer
graphene  \cite{NanoLett._12_2032_2012_Wu}, we obtained a prediction of
a peak susceptibility in monolayer graphene that was similar to that
predicted for the bilayer; the EFISH
contribution was ignored in that calculation.

The importance of the relaxation time demonstrated in that study, and
the desire for more realistic calculations to compare with experiment,
motivates the present work. Here we consider the inclusion of
scattering effects in the semiconductor Bloch equations (SBE) within a
relaxation time approximation, allow for finite temperature to the
extent that it affects the initial state, and explicitly consider the
nonlinear response to pulses of light. 
We obtain an analytic expression for the full nonlinear optical
conductivity $\sigma^{(3)}(\omega_1,\omega_2,\omega_3)$ for optical
transitions around the Dirac points. We discuss
the predictions that follow from this expression for different optical effects, and
we compare with experiment where possible.

Our focus in this work is on doped graphene, where the chemical
potential $\mu\neq 0$. However, the chemical potential dependence of our general
expression for $\sigma^{(3)}(\omega_1,\omega_2,\omega_3)$ allows us to
study the special case of $\mu\to0$. At the very least we might expect
that, for electrons close to the Dirac points, the distinction between ``interband'' and ``intraband'' motion could
be lost. Although different terms that are nominally associated with
interband and intraband motion arise naturally in the development of
the perturbation series, the distinction between those two ``kinds'' of motion
is at best approximate  \cite{Phys.Rev.B_52_14636_1995_Aversa}, and we
indeed find that the way those different formal terms contribute to the final
result for small $\mu$ is nontrivial. More 
importantly, we generally associate the validity of a perturbative
expansion of the optical response with the assumption that the energy
induced by the presence of the optical field is much less than the
energy difference between the bands. In graphene this is always violated for some
states around the Dirac points, regardless of the strength of the optical
field. If these states are occupied by electrons, as they are in undoped
graphene, the reasonableness of a perturbative expansion is in
doubt. Indeed, even a semiclassical treatment of the response to an
applied electric field of electrons near the Dirac points exhibits a
breakdown of the perturbative analysis
  \cite{Europhys.Lett._79_27002_2007_Mikhailov} as $\mu\to0$. We find
evidence of the same kind of behavior 
in the quantum treatment presented here. This has consequences even
for doped graphene if finite temperature is considered, for thermal
fluctuations always place some electrons near the Dirac points.

We organize our paper as follows: In Section~\ref{sec:model} we introduce
the SBE and our approximations for including scattering effects; the
details of the derivation of the nonlinear optical conductivity is
given in Appendix~\ref{app:formula}. The last two subsections of
Section~\ref{sec:model} address the extension of the calculation to
finite temperature, and the treatment of the response to a pulse with finite duration. In Section~\ref{sec:tn} we discuss the
third order nonlinear effects, including THG, Kerr effects and two
photon absorption, two-color coherent current injection, and
parametric frequency conversion; in Section~\ref{sec:sn} we discuss the
current-induced second order nonlinearities, including CSHG, EFISH, and the
nonlinear optical conductivity $\sigma^{(3)}(-\omega_s,\omega_p,0)$
that describes current-induced difference frequency
generation. Throughout the sections we compare with experimental
results when appropriate. We conclude in
Section~\ref{sec:con}. 

\section{Model\label{sec:model}}
We take the Hamiltonian of graphene to be 
\begin{equation}
  H = H_0 + H_{eR} + H_{ep} + H_{ei} + H_{ee}\,,
\end{equation}
Here $ H_0$ is the unperturbed electron Hamiltonian,
\begin{equation}
  H_0 = \sum_{s}\int d\bm k\varepsilon_{s\bm k} a_{s\bm k}^{\dag}a_{s\bm k}\,,
\end{equation}
where the $a_{s\bm k}$ are annihilation operators of Bloch states $|s\bm
k\rangle$ for band $s$ and wave vector $\bm k$, with eigen energy
$\varepsilon_{s\bm k}$. Here $H_{eR}$ describes the interaction with
radiation and in the dipole limit, where the electric field $\bm E(t)$
is approximated as uniform, we have 
\begin{equation}
  H_{eR} = -e\bm E(t)\cdot\sum_{s_1s_2}\int d\bm k a_{s_1\bm
    k}^{\dag} \big(\bm \xi_{s_1s_2\bm k}
  + i\delta_{s_1s_2}\bm\nabla_{\bm k}\big)a_{s_1\bm k}\,,\label{eq:her}
\end{equation}
where $e=-|e|$ and 
\begin{equation}
\bm
\xi_{s_1s_2\bm k} = i\int_{\text{cell}}\frac{d\bm r}{{\cal A}_{\text{cell}}} u_{s_1\bm k}^{\ast}(\bm r) \bm\nabla_{\bm k}
u_{s_2\bm k}(\bm r)
\end{equation}
is the Berry connection between states $|s_1\bm k\rangle$
and $|s_2\bm k\rangle$,  with ${\cal A}_{\text{cell}}$ the unit cell
area and $u_{s\bm k}(\bm r)$ the periodic
part of the Bloch function, $\langle \bm r|s\bm k\rangle=(2\pi)^{-1}e^{i\bm
  k\cdot\bm r}u_{s\bm k}(\bm r, z)$, where $\bm k=(k_x,k_y)$ and $\bm
r=(x,y)$; the graphene is assumed to lie in the $x-y$ plane. We
neglect any response of the system to electric field components in the
$z$ direction. The scattering
terms are given by $H_{ei}$ for the electron-impurity scattering, $H_{ep}$
for the electron-phonon interaction, and $H_{ee}$ for the
carrier-carrier scattering. 

The system is described by a density matrix that is initially
diagonal both in band index and (continuous) wave vector, $\left\langle
a_{s_1{\bm k}_1}^{\dag}a_{s_2{\bm k}_2}\right\rangle
_{t=-\infty }=n_{s_1{\bm k}_1}\delta _{s_{1}s_{2}}\delta ({\bm
  k}_{1}-{\bm k}_{2})$,  where $0\leq n_{s_1{\bm k}_{1}}\leq 1$
describes the initial occupation of the state. In the presence of an applied
uniform electric field it remains diagonal in ${\bm k}_{1}$ and ${\bm
  k}_{2}$ but can acquire off-diagonal elements in $s_{1}$ and
$s_{2}$, describing the correlation between state amplitudes for $\left\vert s_{1}
{\bm k}_{1}\right\rangle $ and $\left\vert s_{2}{\bm k}
_{2}\right\rangle $, $\left\langle a_{s_{1}{\bm k}_{1}}^{\dagger }a_{s_{2}
{\bm k}_{2}}\right\rangle _{t}=\rho _{s_{1}s_{2}{\bm k}_{1}}(t)\delta (
{\bm k}_{1}-{\bm k}_{2})$. We can think of $\rho _{s_{1}s_{2}
{\bm k}}(t)$ as the elements of a $2\times 2$ matrix $\rho _{{\bm k}
}(t)$, and their dynamics are determined by the SBE
\begin{eqnarray}
  \hbar\frac{\partial \rho _{s_1s_2\bm k}}{\partial
    t}&=&-i(\varepsilon_{s_1\bm k}-\varepsilon_{s_2\bm k})
  \rho_{s_1s_2\bm k}\notag\\
  & +& i e E^a(t)\sum_{s}\left(\xi_{s_1s\bm
  k}^a \rho_{ss_2\bm k}-\rho_{s_1s\bm k}\xi_{ss_2\bm
  k}^a\right)  \notag\\
&-& eE^a(t)\frac{\partial \rho_{s_1s_2\bm k}}{\partial
k_a}+
\hbar \left.\frac{\partial \rho_{s_1s_2\bm k}}{\partial
    t}\right|_{\text{scat}}\,.
\label{eq:kbe0}
\end{eqnarray}
Here  $\left.\frac{\partial \rho_{s_1s_2\bm
      k}}{\partial t}\right|_{\text{scat}}$ includes the scattering
terms induced by $H_{ei}+H_{ep}+H_{ee}$, which could in principle be obtained from
well-established treatments of many-particle systems, such as the many-particle density-matrix framework
  \cite{book_HaugKoch,Phys.Rev.B_84_205406_2011_Malic} or the Keldysh Green
function method
  \cite{QuantumKineticsinTransportandOpticsofSemiconductors}. In an
ordinary semiconductor with parabolic band structure, the
current relaxation is mostly caused by carrier-phonon and
carrier-impurity scattering, while carrier-carrier interactions are
less significant due to the approximate equivalence of momentum
conservation and velocity conservation. However, the novel linear
band structure of graphene breaks this equivalence, and the
carrier-carrier interactions play an important role in current
relaxation   \cite{Phys.Rev.B_84_205406_2011_Malic,Phys.Rev.B_87_085319_2013_Zhang};
 thus the full expression for the scattering terms is complicated and
 even hard to solve numerically
   \cite{QuantumKineticsinTransportandOpticsofSemiconductors}. 

\begin{figure}[t]
\centering
\includegraphics[width=3cm]{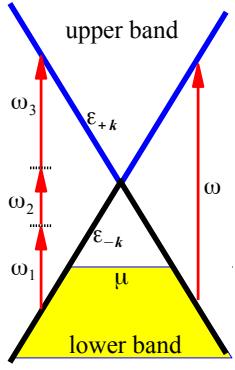}
\caption{(color online) Illustration of the linear dispersion
  approximation of the graphene band structure around the Dirac point. The
  arrows show optical transitions induced by one
  photon with energy $\hbar\omega$ (right) or three photons with
  energy $\hbar\omega_i$ (left).}
\label{fig:illusband}
\end{figure}

We proceed in the standard way by assuming the validity of a
perturbation expansion
\begin{equation}
  \rho_{s_1s_2\bm k}(t) = \sum_{n=0}^{\infty} \rho_{s_1s_2\bm k}^{(n)}(t)\,,
\end{equation}
with $\rho_{s_1s_2\bm k}^{(n)}(t)\propto E^{n}$. Here $\rho_{s_1s_2\bm
  k}^{(0)}(t)=\rho_{s_1s_2\bm k}^0 = \delta_{s_1s_2}n_{s_1\bm k}$ is
the density operator characterizing the equilibrium occupation of
single-particle states at
finite temperature $T$ and chemical potential $\mu$, $n_{s\bm
  k}=[1+e^{(\varepsilon_{s\bm k}-\mu)/(k_BT)}]^{-1}$ is the Fermi-Dirac
distribution with $\beta=1/(k_BT)$  where $k_B$ is Boltzmann's constant. From Eq.~(\ref{eq:kbe0}),
$\rho_{s_1s_2\bm k}^{(n)}(t)$ satisfies
\begin{eqnarray}
  \hbar\frac{\partial \rho _{s_1s_2\bm k}^{(n)}}{\partial
    t}&=&-i(\varepsilon_{s_1\bm k}-\varepsilon_{s_2\bm k})
  \rho_{s_1s_2\bm k}^{(n)}\notag\\
& +& i e E^a(t)\sum_{s}\left(\xi_{s_1s\bm
  k}^a \rho_{ss_2\bm k}^{(n-1)}-\rho_{s_1s\bm k}^{(n-1)}\xi_{ss_2\bm
  k}^a\right)  \notag\\
&-& eE^a(t)\frac{\partial \rho_{s_1s_2\bm k}^{(n-1)}}{\partial
k_a}+
\hbar \left.\frac{\partial \rho_{s_1s_2\bm k}^{(n)}}{\partial
    t}\right|_{\text{scat}}\,,
\label{eq:kbe1}
\end{eqnarray}
where $\rho^{(n)}_{\bm k}\equiv0$ for $n<0$. As a very rough approximation, a relaxation time approximation   \cite{Phys.Rev.B_87_085319_2013_Zhang} can be adopted to give
\begin{eqnarray}
\hbar\left.\frac{\partial \rho_{s_1s_2\bm
      k}^{(n)}}{\partial t}\right|_{\text{scat}} = - \Gamma_{s_1s_2\bm
  k}^{(n)} \rho_{s_1s_2\bm
  k}^{(n)}\,,\text{ for } n\ge 1.\label{eq:dampingrho}
\end{eqnarray}
Here $\Gamma_{s_1s_2\bm k}^{(n)}$ is a relaxation parameter introduced to
describe the dynamics of $\rho_{s_1s_2\bm k}^{(n)}(t)$, and
$\hbar/\Gamma_{s_1s_2\bm k}^{(n)}$ corresponds to a phenomenological relaxation time. In a real
system, $\Gamma_{s_1s_2\bm k}^{(n)}$ can be expected to depend on the temperature,
chemical potential, and external
field  \cite{J.Phys.Soc.Jpn._71_1318_2002_Ando}. Yet because the relaxation plays an important role 
 in optical nonlinearities around resonant transitions, the extremely
phenomenological treatment 
  \cite{Phys.Rev.B_1_2362_1970_Mermin} in
Eq.~(\ref{eq:dampingrho}) can still reveal part of the physics, and in a very
simple way. Even with the use of the six phenomenological
constants $\Gamma_{s\bar{s}\bm k}^{(n)}=\Gamma_{e}^{(n)}$ for
interband transitions and $\Gamma_{ss\bm
  k}^{(n)}=\Gamma_{i}^{(n)}$ for intraband 
transitions, we are still able to obtain an
analytic result for the perturbation calculation within the linear
dispersion approximations around the Dirac points at zero
temperature. From $\rho_{\bm k}^{(n)}(t)$, the (areal) current density,
which in our model has only $x$ and $y$ components, is calculated as $J^d(t) = \sum_{n=1}^{\infty}
J^{(n);d}(t)$ with 
\begin{equation}
J^{(n);d}(t) = e\sum_{s_1s_2}\int\frac{d\bm k}{4\pi^2}
v_{s_2s_1\bm k}^d \rho_{s_1s_2\bm k}^{(n)}(t)\,.
\end{equation}
We give the derivation in Appendix~\ref{app:formula}, where the spin
degeneracy is included. We extract the linear optical conductivity $\sigma^{(1);da}(\omega)$ from 
\begin{equation}
  J^{(1);d}(t) = \int \frac{d\omega}{2\pi} \sigma^{(1);da}(\omega)
  E^a(\omega) e^{-i\omega t}\,,
\end{equation}
where $E^a(\omega) = \int dt E^a(t) e^{i\omega t}$.  In graphene, the hexagonal lattice has $D_{6h}$ (6/mmm)
symmetry   \cite{HandbookofNonlinearOptics}, and there is only one independent nonzero component
$\sigma^{(1);xx} = \sigma^{(1);yy}$. We first consider the zero
temperature results. In this paper, we restrict ourselves to the
neighborhood of the Dirac points (see Fig.~\ref{fig:illusband}), assuming a linear
dispersion relation with two relevant bands that we label $s=+$ (upper)
and $-$ (lower). We recover
the usual result  \cite{J.Phys.Soc.Jpn._71_1318_2002_Ando,Phys.Rev.Lett._99_16803_2007_Mikhailov}
\begin{equation}
  \sigma^{(1);xx}(\omega) =\frac{i\sigma _{0}}{\pi }\left[ \frac{4|\mu |}{%
      \hbar \omega + i\Gamma_{i}^{(1)}}-{\cal G}_\mu\left(\hbar\omega+i\Gamma_{e}^{(1)}\right)\right]\,.\label{eq:sigmaxx}
\end{equation}
Here $\sigma_0={e^2}/{(4\hbar)}$ is the universal conductivity, and
${\cal G}_\mu(\vartheta)$ with $\vartheta=\vartheta_r+i\vartheta_i$ is 
\begin{eqnarray}
  {\cal G}_\mu(\vartheta)
  &=&\ln\left|\frac{2|\mu|+\vartheta}{2|\mu|-\vartheta}\right|
+  i\left(\pi  +
 \arctan\frac{\vartheta_r-2|\mu|}{\vartheta_i} \notag\right.\\
& &\left.- \arctan\frac{\vartheta_r+2|\mu|}{\vartheta_i} \right)\,.
\end{eqnarray}

The third order current is given as 
\begin{eqnarray}
  J^{(3);d}(t) &=& \int \frac{d\omega_1 d\omega_2 d\omega_3}{(2\pi)^3}
  \sigma^{(3);dabc}(\omega_1,\omega_2,\omega_3)\notag\\
&\times &
  E^a(\omega_1)E^b(\omega_2)E^c(\omega_3)
  e^{-i(\omega_1+\omega_2+\omega_3)t}\,,\label{eq:j3rd}
\end{eqnarray}
Here the symmetrized third order optical conductivity  $\sigma^{(3);dabc}$ is 
\begin{eqnarray}
&&  \sigma^{(3);dabc}(\omega_1,\omega_2,\omega_3) \notag\\
&=&
  \frac{1}{6}\bigg[\widetilde{\sigma}^{(3);dabc}(\omega_1,\omega_2,\omega_3)  
+
\widetilde{\sigma}^{(3);dbca}(\omega_2,\omega_3,\omega_1)  \notag\\
&&+
\widetilde{\sigma}^{(3);dcab}(\omega_3,\omega_1,\omega_2) 
+    \widetilde{\sigma}^{(3);dacb}(\omega_1,\omega_3,\omega_2)
\notag\\
&&+
    \widetilde{\sigma}^{(3);dcba}(\omega_3,\omega_2,\omega_1)  +
    \widetilde{\sigma}^{(3);dbac}(\omega_2,\omega_1,\omega_3)
    \bigg]\,,
  \end{eqnarray}
where the unsymmetrized third order optical conductivity is given as
\begin{eqnarray}
  && \widetilde{\sigma}^{(3);dabc}(\omega_1,\omega_2,\omega_3)\notag\\
  &=&i\sigma_3
  \Bigg[\frac{{\cal S}_1^{dabc}}{\nu\nu_0\nu_3}
 + \frac{{\cal S}_2^{dabc}(\vartheta_3)}{\nu\nu_0} + \frac{{\cal S}_3^{dabc}(\vartheta_0)}{\nu\nu_3}  \notag\\
 &&\quad+
 \frac{{\cal S}_4^{dabc}(\vartheta_0,\vartheta_3)}{\nu} + \frac{{\cal S}_5^{dabc}(\vartheta)}{\nu_0\nu_3} + 
 \frac{{\cal S}_6^{dabc}(\vartheta,\vartheta_3)}{\nu_0} \notag\\
&&\quad+ \frac{{\cal S}_7^{dabc}(\vartheta,\vartheta_0)}{\nu_3} + {\cal S}_8^{dabc}(\vartheta,\vartheta_0,\vartheta_3)\Bigg]\,,\label{eq:sigmaall}
\end{eqnarray}
with $\sigma_3\equiv{\sigma_0(\hbar
  v_Fe)^2}/{\pi}$, $\nu_3 \equiv \hbar\omega_3+i\Gamma_{i}^{(1)}$,
$\vartheta_3 \equiv \hbar\omega_3+i\Gamma_e^{(1)}$,
$\nu_0\equiv\hbar\omega_0+i\Gamma_i^{(2)}$,
$\vartheta_0\equiv\hbar\omega_0+i\Gamma_e^{(2)}$, $\nu \equiv \hbar\omega +
i\Gamma_i^{(3)}$, $\vartheta \equiv \hbar\omega+i\Gamma_e^{(3)}$,
$\omega_0\equiv\omega_2+\omega_3$, and
$\omega\equiv\omega_1+\omega_0$. We have followed the standard
convention of nonlinear optics  \cite{boyd_nonlinearoptics} in symmetrizing the terms
$\widetilde{\sigma}^{(3);ijkl}(\omega _{j},\omega _{k},\omega _{l})$ by permuting the indices $%
(jkl)$ to arrive at the nonlinear conductivity
$\sigma^{(3);dabc}(\omega_1,\omega_2,\omega_3)$. The light-matter
interaction in Eq.~(\ref{eq:her}) can be formally separated into an interband
contribution ($s_{1}\neq s_{2}$) and an intraband contribution ($s_{1}=s_{2}$
), and the terms proportional to the different $\mathcal{S}_{i}$ in
$\widetilde{\sigma}^{(3);dabc}$ can be classified according to how many
times each contribution appears   \cite{Phys.Rev.B_52_14636_1995_Aversa}.
The term proportional to ${\cal S}_1$ arises
from only the intraband contributions, and the term proportional to
${\cal S}_6$ arises from only the interband contributions; all others
involve mixtures of both. The quantities ${\cal S}_i^{dabc}$,
$\widetilde{\sigma}^{(3);dabc}$, and $\sigma^{(3);dabc}$ are all fourth order
tensors. Neglecting the optical response in the $z$
direction, there are in all 8 nonzero components for the $D_{6h}$
symmetry, among which
three are independent; they are
\begin{eqnarray}
  {\sigma}^{(3);xxyy} &=&{\sigma}^{(3);yyxx}\,,\notag\\
{\sigma}%
^{(3);xyxy}&=&{\sigma}^{(3);yxyx}\,,\notag\\
{\sigma}^{(3);xyyx}&=&
  {\sigma}^{(3);yxxy}\,,
\end{eqnarray}
and
\begin{eqnarray}
{\sigma}^{(3);xxxx}&=&{\sigma}^{(3);yyyy}\notag\\
&=&{\sigma}%
  ^{(3);xxyy}+{\sigma}^{(3);xyxy}+{\sigma}^{(3);xyyx}\,.
\end{eqnarray}
In the following, we write the independent nonzero components of
fourth rank tensors as column vectors, ordering the independent
components of a fourth rank tensor $T^{dabc}$ as $T=\begin{bmatrix}  T^{(3);xxyy}\\ T^{(3);xyxy} \\ T^{(3);xyyx}
\end{bmatrix}$. By employing the constant vectors 
\begin{equation}
  A_1= \begin{bmatrix}-3\\1\\1
  \end{bmatrix}\,, 
  A_2 = \begin{bmatrix}1\\-3\\1
  \end{bmatrix}\,, 
  A_3 = \begin{bmatrix}1\\1\\-3
  \end{bmatrix}\,,
  A_0 = \begin{bmatrix}1\\1\\1
  \end{bmatrix}\,,
\end{equation}
where note $A_0=-(A_1+A_2+A_3)$, we can present the analytic
expression for the different components of $\widetilde{\sigma}^{(3);dabc}$
appearing in Eq.~(\ref{eq:sigmaall}) at
zero temperature, using the approximation of a linear
dispersion relation around the Dirac points, as
\begin{eqnarray}
  {\cal S}_1
  &=&\frac{1}{|\mu|}A_0\,,\label{eq:s1}\\
  {\cal S}_2
  (\vartheta_3) &=& {\cal
    G}_{\mu}\left(\vartheta_3\right)\frac{A_0}{\vartheta_3^2}-\frac{1}{|\mu|}\frac{A_0}{\vartheta_3}\,,\\
  {\cal S}_3
  (\vartheta_0)
  &=&{\cal H}_\mu(\vartheta_0)\frac{A_3}{\vartheta_0}-\frac{1}{|\mu|}\frac{A_3}{\vartheta_0}\,,\\
  {\cal S}_4
  (\vartheta_0,\vartheta_3) &=&-{\cal
    G}_\mu(\vartheta_3)\frac{\vartheta_3A_2+\vartheta_2A_3}{\vartheta_2^2\vartheta_3^2}\notag\\
&&+{\cal
  G}_{\mu}(\vartheta_0)\frac{(\vartheta_0+\vartheta_2)A_2+\vartheta_2A_3}{\vartheta_0^2\vartheta_2^2}
\notag\\
&&-{\cal H}_{\mu}(\vartheta_0)\frac{A_2}{\vartheta_0\vartheta_2}+\frac{1}{|\mu|}\frac{A_3}{\vartheta_0\vartheta_3}\,,
\end{eqnarray}    
\begin{eqnarray}
  {\cal S}_5
  (\vartheta) &=& {\cal H}_\mu(\vartheta)\frac{A_0}{\vartheta}+{\cal I}_\mu(\vartheta)A_1-\frac{1}{|\mu|}\frac{A_0}{\vartheta}\,,\\
  {\cal S}_6
  (\vartheta,\vartheta_3) &=&-{\cal
  G}_\mu(\vartheta_3)\frac{\vartheta
  A_0}{\vartheta_3^2(\vartheta^2-\vartheta_3^2)}\notag\\
&&+ {\cal
  G}_\mu(\vartheta)\frac{\vartheta_3A_0}{\vartheta^2(\vartheta^2-\vartheta_3^2)}+\frac{1}{|\mu|}\frac{A_0}{\vartheta\vartheta_3}\,,\label{eq:s6}\\
{\cal S}_7
(\vartheta,\vartheta_0) &=&{\cal
  H}_\mu(\vartheta_0)\left(\frac{A_2}{\vartheta_1^2}-\frac{A_3}{\vartheta_0\vartheta_1}\right)\notag\\
&+&{\cal
  H}_\mu(\vartheta)
\left(\frac{A_3}{\vartheta\vartheta_1}-\frac{A_2}{\vartheta_1^2}\right)\notag\\
&-&  {\cal I}_\mu(\vartheta)\frac{A_2}{\vartheta_1}+ \frac{1}{|\mu|}\frac{A_3}{\vartheta\vartheta_0}\,,
\end{eqnarray}
and 
\begin{eqnarray}
  && {\cal S}_8
(\vartheta,\vartheta_0,\vartheta_3) \notag\\
&=& {\cal G}_\mu(\vartheta_3)
  \left[\frac{A_2}{(\vartheta-\vartheta_3)\vartheta_2^2\vartheta_3}\right.\notag\\
&&\hspace{1.3cm}\left.+\frac{\vartheta^2\vartheta_2+\vartheta_3^3+\vartheta\vartheta_3(-3\vartheta_0+2\vartheta_3)}{(\vartheta-\vartheta_3)^3\vartheta_2^2\vartheta_3^2}A_3\right]\notag\\
  &+&{\cal G}_\mu(\vartheta_0) \left[-
\frac{\vartheta_0\vartheta_1+\vartheta_1\vartheta_2-\vartheta_0\vartheta_2}{\vartheta_0^2\vartheta_1^2\vartheta_2^2}A_2\right.\notag\\
&&\hspace{1.3cm}\left.-\frac{\vartheta_1\vartheta_2-\vartheta_0^2-\vartheta_0\vartheta_2}{\vartheta_1^2\vartheta_0^2\vartheta_2^2}A_3\right]\notag\\
&+&{\cal
  G}_\mu(\vartheta)\left[-\frac{1}{\vartheta\vartheta_1^2(\vartheta-\vartheta_3)}A_2\right.\notag\\
&&\hspace{1.3cm}\left.-\frac{5\vartheta^2+\vartheta_3(\vartheta_0+\vartheta_3)-\vartheta(3\vartheta_0+4\vartheta_3)}{\vartheta\vartheta_1^2(\vartheta-\vartheta_3)^3}A_3\right]\notag\\
&+&{\cal H}_\mu(\vartheta_0)\left(\frac{A_2}{\vartheta_0\vartheta_1\vartheta_2}-\frac{A_3}{\vartheta_1^2\vartheta_2}\right)\notag\\
&+&{\cal
  H}_\mu(\vartheta)\frac{4\vartheta^2-3\vartheta\vartheta_0-2\vartheta\vartheta_3+\vartheta_0\vartheta_3}{\vartheta\vartheta_1^2(\vartheta-\vartheta_3)^2}A_3\notag\\
&+&{\cal
  I}_\mu(\vartheta)\frac{A_3}{\vartheta_1(\vartheta-\vartheta_3)}\notag\\
&-&\frac{1}{|\mu|}\frac{A_3}{\vartheta\vartheta_0\vartheta_3}\,.\label{eq:s8}
\end{eqnarray}
where $\vartheta_2=\vartheta_0-\vartheta_3$, $\vartheta_1=\vartheta-\vartheta_0$, and
\begin{eqnarray}
  {\cal H}_\mu(\vartheta) &=& \frac{1}{2|\mu|-\vartheta}+\frac{1}{2|\mu|+\vartheta}\,,\\
  {\cal I}_\mu(\vartheta) &=&
  \frac{1}{(2|\mu|+\vartheta)^2}-\frac{1}{(2|\mu|-\vartheta)^2}\,.
\end{eqnarray}
For the details see Appendix
\ref{app:formula}. 

Using the nonzero independent components, the third order current in
Eq.~(\ref{eq:j3rd}) can be written as 
\begin{eqnarray}
  \bm J^{(3)}(t) =&& \int\frac{d\omega_1d\omega_2d\omega_3}{(2\pi)^3}
  e^{-i(\omega_1+\omega_2+\omega_3)t}  \notag\\
   \times \Big[&& \sigma^{(3);xxyy}(\omega_1,\omega_2,\omega_3)\bm
  E(\omega_1) \bm E(\omega_2)\cdot\bm E(\omega_3) \notag\\
   + &&\sigma^{(3);xyxy}(\omega_1,\omega_2,\omega_3)\bm
  E(\omega_2) \bm E(\omega_1)\cdot\bm E(\omega_3) \notag\\
   + &&\sigma^{(3);xyyx}(\omega_1,\omega_2,\omega_3)\bm
  E(\omega_3) \bm E(\omega_1)\cdot\bm E(\omega_2) \Big]\,.\quad\quad
\end{eqnarray}

\subsection{Divergences and limits\label{app:limit}}
The results for $\widetilde{\sigma}^{(3);dabc}(\omega _{1},\omega
_{2},\omega _{3})$ show a complicated dependence on the $\omega _{j}$,
on the $\Gamma _{i/e}^{(j)}$, and on $\mu$. The expressions in
Eqs.~(\ref{eq:s1}-\ref{eq:s8}) seem to exhibit a number of
divergences, 
but some of them are only apparent: For example, there seem to be divergences
when $\vartheta -\vartheta_{3}=0$, but a careful collection of terms shows that
even in the absence of relaxation $\lim_{\delta \rightarrow 0}\widetilde{\sigma}%
^{(3);dabc}(-\omega ,\omega +\delta ,\omega _{3})$ is finite. Some of the
divergences are of course real: There are divergences for $2\left\vert \mu
\right\vert \pm \vartheta =0$ in the functions $\mathcal{G}\left( \vartheta
\right) $, $\mathcal{H}\left( \vartheta \right) $, and $\mathcal{I}\left(
\vartheta \right) $, which lead to divergences in
$\widetilde{\sigma}^{(3);dabc}(\omega _{1},\omega _{2},\omega
_{3})$. These are associated
with interband optical transitions, and for nonvanishing relaxation they
occur at frequencies removed from the real axis; we will see how some of
them affect the structure of $\sigma^{(3);dabc}(\omega _{1},\omega
_{2},\omega _{3})$ in Sections \ref{sec:tn} and \ref{sec:sn}. There are also divergences
associated with $\vartheta +\vartheta_{3}=0$. In the absence of relaxation
these occur when $\omega _{1}+\omega _{2}+2\omega _{3}=0$, and lead to a
divergence in $\widetilde{\sigma}^{(3);dabc}(-\omega _{2}-2\omega _{3}+\delta
,\omega _{2},\omega _{3})$ as $\delta ^{-1}$. \ A special case of these is
when $\vartheta =0$ and $\vartheta _{j}=0$ for $j=0,1,2,$ or $3$. \ Some of the
associated conductivity terms, such as $\sigma ^{(3);dabc}(-\omega ,\omega
,\omega )$ and $\sigma ^{(3);dabc}(-\omega, -\omega, 2\omega )$  will
be considered in Section \ref{sec:tn}.

All of these
divergences only occur at complex frequencies in the presence of relaxation,
and have their analogs in gapped systems. Of a different nature are the
divergences that arise as $\left\vert \mu \right\vert \rightarrow 0$. 
While in a semiclassical calculation and in the absence of relaxation the
intraband third order nonlinear response coefficient that can be extracted from
the full nonlinear response is divergent  \cite{Europhys.Lett._79_27002_2007_Mikhailov} as $\left\vert \mu
\right\vert ^{-1}$, one might hope that in the presence of relaxation this
would be ameliorated. Yet in general it is not. To see this, we
reorganize the unsymmetrized conductivity to write 
\begin{eqnarray}
\widetilde{\sigma}^{(3);dabc}(\omega_1,\omega_2,\omega_3) &=&
\widetilde{\sigma}^{(3);dabc}_A(\omega_1,\omega_2,\omega_3) \notag\\ 
&+&
\widetilde{\sigma}^{(3);dabc}_B(\omega_1,\omega_2,\omega_3) \notag\\
&+&
\widetilde{\sigma}^{(3);dabc}_C(\omega_1,\omega_2,\omega_3)\label{eq:tildesigma}
\end{eqnarray}
where $\widetilde{\sigma}_A$ includes all terms involving ${\cal G}_\mu$,  $\widetilde{\sigma}_B$ includes
all terms involving ${\cal H}_\mu$ and ${\cal I}_\mu$, 
and the remainder, $\widetilde{\sigma}_C$, includes all terms proportional to
$|\mu|^{-1}$. {Similar separations are also used for the symmetrized
  conductivity $\sigma^{(3);dabc}$.} The term $\widetilde{\sigma}_C$ can be simplified to yield
\begin{eqnarray}
  \widetilde{\sigma}_C^{(3);dabc}(\omega_1,\omega_2,\omega_3)&=& \frac{\sigma_3}{|\mu|}\left(\frac{A_0}{\nu_0}-\frac{A_3}{\vartheta_0}\right)\notag\\
&\times&\frac{(\Gamma_{e}^{(3)}-\Gamma_{i}^{(3)})(\Gamma_{e}^{(1)}-\Gamma_{i}^{(1)})}{\nu\vartheta
    \nu_3\vartheta_3} \,.\label{eq:sigmac}
\end{eqnarray}
Note that even for finite relaxation we have $\tilde{\sigma}%
_{C}^{(3);dabc}(\omega _{1},\omega _{2},\omega _{3})$ diverging as $%
\left\vert \mu \right\vert \rightarrow 0$, for general frequencies $(\omega
_{1},\omega _{2},\omega _{3})$, when $\Gamma _{e}^{(j)}\neq \Gamma
_{i}^{(j)} $ for both $j=1$ and $j=3$. At least within the simple
description of relaxation we adopt here, the perturbation theory seems to
demand that either the first or third order relaxation rates (or both) must
not distinguish between intraband and interband relaxation to achieve a
finite result as $\left\vert \mu \right\vert \rightarrow 0$. This is at
least consistent with the physical intuition that the distinction between
intraband and interband motion is blurred as $\left\vert \mu \right\vert
\rightarrow 0$, in any case for electrons near the Fermi level, and any
reasonable theory should respect that; recall that in our phenomenological
description of relaxation all carriers share the same $\Gamma _{e}^{(j)}$
and $\Gamma _{i}^{(j)}$. But clearly a more sophisticated theory  is in
order to address the limit $\left\vert \mu \right\vert \rightarrow
0$. 

More evidence for the blurring of the distinction
between intraband and interband motion as $\left\vert \mu \right\vert
\rightarrow 0$ can be seen from how the contributions to $\widetilde{\sigma}_{C}^{(3);dabc}(\omega _{1},\omega _{2},\omega _{3})$ arise. \ The term in $\widetilde{\sigma}^{(3);dabc}(\omega _{1},\omega _{2},\omega _{3})$ that
contains only contributions from the formal intraband ($s_{1}=s_{2}$)
component of Eq.~(\ref{eq:her}) is the term proportional to $\mathcal{S}_{1}$; it varies
with $\left\vert \mu \right\vert $ as $\left\vert \mu \right\vert
^{-1}$, which is qualitatively different than the variation as $|\mu|$
of the corresponding Drude term in the linear
conductivity. Yet as $\left\vert \mu \right\vert \rightarrow 0$ the contribution to $\widetilde{\sigma}^{(3);dabc}(\omega _{1},\omega _{2},\omega _{3})$ involving only the formal
interband ($s_{1}\neq s_{2}$) component of Eq.~(\ref{eq:her}), that is proportional to $\mathcal{S}_{6}$, also becomes important;\ while it includes contributions
from $\mathcal{G}_{\mu }(\theta _{3})$ and $\mathcal{G}_{\mu }(\theta )$,
there is also a term proportional to $\left\vert \mu \right\vert
^{-1}$. The formally ``mixed'' terms, $\mathcal{S}_{j}$, with
$j=2,3,4,5,7,8,$ also provide terms proportional to $\left\vert \mu \right\vert ^{-1}$. The
summation of all these terms, all formally involving different proportions
of interband and intraband contributions, gives Eq.~(\ref{eq:sigmac}); the $\left\vert \mu
\right\vert ^{-1}$ behavior in $\widetilde{\sigma}_{C}(\omega _{1},\omega
_{2},\omega _{2})$ cannot be associated with motion that is just formally
intraband. 

Now note that $\tilde{\sigma}_{C}(\omega _{1},\omega _{2},\omega _{2})$ vanishes as all the $\Gamma
_{i/e}^{(n)}$ vanish. Yet here we physically \textit{would} expect to
recover the relaxation free, semiclassical
result  \cite{Europhys.Lett._79_27002_2007_Mikhailov} of a perturbative response
divergent as $\left\vert \mu \right\vert ^{-1}$, for $\hbar \omega _{i}\ll
\left\vert \mu \right\vert $, and the result associated in that calculation with purely
intraband motion. \ And we do recover it here, but in a nontrivial way:
Although $\widetilde{\sigma}_{C}^{(3)}(\omega _{1},\omega _{2},\omega _{3})$
vanishes, when the other contributions to $\sigma ^{(3);dabc}(\omega
_{1},\omega _{2},\omega _{3})$ are assembled and the limit $\hbar \omega
_{i}\ll \left\vert \mu \right\vert $ taken we find 
\begin{equation}
  \sigma^{(3);dabc}(\omega_1,\omega_2,\omega_3) = \frac{i\sigma_3 A_0}{6|\mu|\hbar^3\omega_1\omega_2\omega_3
    } + O(\omega_i^{-1})\,,
\end{equation}
in which the leading term is exactly the same as the contribution
proportional to the $\mathcal{S}_{1}$ term, and which agrees with the
relaxation free, semiclassical calculation  \cite{Europhys.Lett._79_27002_2007_Mikhailov} involving only intraband
motion. While the physically appropriate result of purely intraband,
semiclassical motion is recovered in this limit as it should be, the
connection to formally intraband, interband, and mixed responses in $\sigma
^{(3);dabc}(\omega _{1},\omega _{2},\omega _{3})$ is less than direct.

From Eqs.~(\ref{eq:tildesigma}, \ref{eq:sigmac}) we can also study more generally the
limits as the relaxation rates are allowed to vanish. Here we discuss the
simple case where the intraband and interband relaxation rates are the same
for all orders, but perhaps different than each other: $\ \Gamma
_{i}^{(j)}=\Gamma _{i}$ and $\Gamma _{e}^{(j)}=\Gamma _{e}$. We find that
as $\Gamma _{i/e}\rightarrow 0$ we recover from ${\sigma}
_{A}^{(3);dabc}(\omega _{1},\omega _{2},\omega _{3})$ the results derived
earlier  \cite{NewJ.Phys._16_53014_2014_Cheng} in the absence of relaxation. We find that in this limit
the contributions to ${\sigma}_{B}^{(3);dabc}(\omega _{1},\omega _{2},\omega
_{3})$ involving nonresonant transitions scale as $\Gamma_{i}$. 
For resonant transitions, there are two cases that require further attention: (i) Taking
$\omega_{\text{comb}}$ to be a possible frequency combination appearing 
in the expression in Eqs.~(\ref{eq:s1}-\ref{eq:s8}), resonant transitions (real or virtual) occur as
$|\hbar\omega_{\text{comb}}|=2|\mu|$. Then the function
${\cal H}_\mu$ or ${\cal I}_\mu$ becomes  ${\cal
  H}_{\mu}(\omega_{\text{comb}}+i\hbar^{-1}\Gamma_e)\propto \Gamma_e^{-1}$ or ${\cal
  I}_{\mu}(\omega_{\text{comb}}+i\hbar^{-1}\Gamma_e)\propto\Gamma_e^{-2}$
respectively, and then $\sigma^{(3);dabc}_B\propto\Gamma_i\Gamma_e^{-1}$; its
limit depends on the sequence of limits of $\Gamma_i\to0$ and
$\Gamma_e\to0$, and so there seems to be
no single well-defined relaxation free limit within this phenomenological
theory. (ii) For some $\omega_{\text{comb}}=0$, there can be divergences that occur at real
frequencies in the absence of relaxation; below we discuss the
behavior of $\sigma^{(3);dabc}$ near these divergences by considering
the frequencies in the neighborhood of some of them.

\subsection{Finite temperature\label{sec:tem}}
In calculating the response of a system to optical radiation, two
effects of the temperature are usually considered: its role in
establishing the initial electron distribution, and how it affects
relaxation rates. In this work, the latter is
implicit in our choice of relaxation rates. In our perturbative calculation, the
former can be taken into account in the following simple way: Explicitly displaying the
chemical potential and temperature dependence, we write $n_{s\bm
  k}(\mu, T)$ for the electron distribution at equilibrium, and
$\sigma^{(3)}(\mu,T)$ for the nonlinear conductivity. By using 
\begin{equation}
  n_{s\bm k}(\mu, T) = \int_{-\infty}^{\infty} dx F_\mu(x, T)\frac{\partial}{\partial x}
  n_{s\bm k}(x, 0) 
\end{equation}
with $F_\mu(x, T)=[1+e^{\beta(x-\mu)}]^{-1}$, the conductivity at
finite temperature can be related to the zero 
temperature conductivity via
\begin{eqnarray}
  {\sigma}^{(3)}(\mu, T) &=& \int_{-\infty}^{\infty} dx F_\mu(x,T) \frac{\partial }{\partial x}
  {\sigma}^{(3)}(x,
  0) \notag\\
&=& \beta \int_{-\infty}^{\infty} dx F_\mu(x,T)\left[1-F_\mu(x,T)\right] \sigma^{(3)}(x, 0)\,.\quad\quad\label{eq:tem}
\end{eqnarray}
Here the second line is obtained by using the
partial integration and the condition
${\sigma}^{(3)}(x\to\pm\infty,0)=0$. Because
$F_\mu(x,T)[1-F_\mu(x,T)]$ is a pulse 
function located at $x=\mu$ with a width of the order of the thermal
energy, the conductivity at finite temperature 
$T$ can be obtained by averaging the zero temperature values over the
chemical potential in an energy window with a width of the order of
magnitude of the thermal energy. In a case where the chemical potential $\mu$ and
the frequencies $\{\omega_i\}$ are chosen to be away from resonant transitions, the
conductivity is a smooth function around $\mu$. Considering that the
thermal energy $k_BT$ is only about $\sim 25.8$~meV at room
temperature, the conductivity at room temperature is close to the
value at zero temperature away from resonant transitions. However, around resonant transitions where 
the conductivity diverges, the effects of finite temperature can be
important. In Eqs.~(\ref{eq:s1}) to ~(\ref{eq:s8}), the chemical potential
appears in the functions ${\cal G}_\mu$, ${\cal H}_\mu$, and ${\cal I}_\mu$ in
$\widetilde{\sigma}_A$ and $\widetilde{\sigma}_B$, and as 
${|\mu|}^{-1}$ in $\widetilde{\sigma}_C$. Therefore, the conductivity
at finite temperature is determined by applying Eq.~(\ref{eq:tem}) to
these quantities.  The temperature effects on the contributions due to
the functions  ${\cal  G}_\mu$, ${\cal H}_\mu$, and ${\cal I}_\mu$ are discussed in Appendix~\ref{app:tem}.

Note that the treatment of the $\widetilde{\sigma}_{C}$ term requires particular care,
because at finite temperature there are always electrons initially near the
Dirac points, and they will lead to the same prediction for divergent
response that Eq.~(\ref{eq:sigmac}) indicates for electrons near the Dirac points at zero
temperature in an undoped sample. To show this explicitly, from
Eq.~(\ref{eq:tem}), ${|\mu|}^{-1}$ should be replaced by 
\begin{equation}
{|\mu|}^{-1} \longrightarrow  \beta \int_{-\infty}^{\infty} dx F_\mu(x,T)[1-F_\mu(x,T)]\frac{1}{|x|}\,.
\end{equation}
However, this diverges due to the singularity of the integrand at
$x=0$. Based on Eq.~(\ref{eq:sigmac}) where this term is nonzero only at
$\Gamma_i^{(j)}\neq\Gamma_e^{(j)}$, the divergence shows that either
the perturbation theory or the assumption of unequal intraband
and interband relaxation times in undoped graphene is not adequate,
and more realistic treatments of the scattering and temperature are
required. Nonetheless, from a full numerical solution of Eq.~(\ref{eq:kbe0}) and (\ref{eq:dampingrho})
  \cite{unpublishnumeric}, we find that contributions from the ${|\mu|^{-1}}$ term only give
a small contribution to the total conductivity at finite temperature.
Thus, at least at the level of the full SBE, whatever the final
description of relaxation yields for the $|\mu|^{-1}$ term 
it will not lead to significant contributions. So for our finite
temperature calculations we somewhat arbitrarily take
\begin{equation}
\frac{1}{|\mu|} \rightarrow \frac{1}{\sqrt{\mu^2+(k_BT)^2}}\,.
\end{equation}

\subsection{Pulse response}
Because most nonlinear experiments are
carried out using laser pulses, the optical response close to the 
divergences mentioned above is determined by the pulse shape. Except for the
$|\mu|^{-1}$ divergences just discussed, the inclusion of the
relaxation parameters $\Gamma_{i/e}^{(n)}$ moves the divergent frequencies off
the real axis.  Yet it is necessary to investigate the pulse 
effects when the energy broadening of the pulse is larger than the
broadening characterized by those relaxation parameters. For a field
associated with pulses of a fixed polarization 
\begin{equation}
\bm E(t) = \sum_i \bm E_{\omega_i}
p_{\omega_i}(t) e^{-i\omega_it}
\end{equation}
with the time domain envelope function
$p_{\omega_i}(t)$, the Fourier transform is
\begin{eqnarray}
  \bm E(\omega) &=& \sum_i \bm E_{\omega_i} P_{\omega_i}(\omega-\omega_i) 
\end{eqnarray}
with the frequency domain envelope function 
\begin{equation*}
P_{\omega_i}(\omega) =
\int dt p_{\omega_i}(t) e^{i\omega t}\,.
\end{equation*}
The third order current in both time and frequency domain can be written as
\begin{eqnarray*}
  J^d(t) &=& \sum_{lmn}e^{-i(\omega_l+\omega_m+\omega_n)t}
  C_{\omega_l,\omega_m,\omega_n}^{dabc}(t) E_{\omega_l}^a
  E_{\omega_m}^b E_{\omega_n}^c\,,\\
  J^d(\omega) &=& \sum_{lmn} {\cal
    C}_{\omega_l,\omega_m,\omega_n}^{dabc}(\omega-\omega_l-\omega_m-\omega_n) E_{\omega_l}^a
  E_{\omega_m}^b E_{\omega_n}^c\,,
\end{eqnarray*}
with 
\begin{eqnarray}
&&  C_{\omega_l,\omega_m,\omega_n}^{dabc}(t)\notag\\
 &=& \int\frac{d\delta_l d\delta_m
   d\delta_n}{(2\pi)^3} P_{\omega_l}(\delta_l)
 P_{\omega_m}(\delta_m)P_{\omega_n}(\delta_n)e^{-i(\delta_l+\delta_m+\delta_n)t}\notag\\
&&\times 
  \sigma^{(3);dabc}(\omega_l+\delta_l,\omega_m+\delta_m,\omega_n+\delta_n)\,,\label{eq:pulsect}
\end{eqnarray}
and
\begin{eqnarray}
&&{\cal C}_{\omega_l,\omega_m,\omega_n}^{dabc}(\delta)\notag\\
&=&  \int\frac{d\delta_l d\delta_m
   }{(2\pi)^2} P_{\omega_l}(\delta_l)
 P_{\omega_m}(\delta_m)P_{\omega_n}(\delta-\delta_l-\delta_m)\notag\\
&&\times 
  \sigma^{(3);dabc}(\omega_l+\delta_l,\omega_m+\delta_m,\omega_n+\delta-\delta_l-\delta_m)\,.~~
\end{eqnarray}

We will be particularly interested in two  special cases: \\
(i) For $\delta_i$ sufficiently small and $\sigma^{(3);dabc}$
sufficiently slowly varying in its frequency dependence so that
\begin{equation}
  \sigma^{(3);dabc}(\omega_l+\delta_l,\omega_m+\delta_m,\omega_n+\delta_n)
  \approx \sigma^{(3);dabc}(\omega_l,\omega_m,\omega_n)\,,
\end{equation}
over the frequency components of the envelope functions, the
 current response is given by
\begin{equation}
  C_{\omega_l,\omega_m,\omega_n}^{dabc}(t) =
  \sigma^{(3);dabc}(\omega_l,\omega_m,\omega_n) p_{\omega_l}(t)
  p_{\omega_m}(t)p_{\omega_n}(t)\,. 
\end{equation}
For a Gaussian pulse $p_{\omega_i}(t) = e^{-t^2/\Delta_i^2}$ which
gives $P_{\omega_i}(\omega)=\sqrt{\pi}\Delta_i e^{-\omega^2\Delta_i^2/4}$, we get
\begin{eqnarray}
  C^{dabc}_{\omega_l,\omega_m,\omega_n}(t) &\approx&
  \sigma^{(3);dabc}(\omega_l,\omega_m,\omega_n)
  e^{-t^2/\Delta^2}\,,\\
  {\cal
    C}^{dabc}_{\omega_l,\omega_m,\omega_n}(\delta)&\approx&\sigma^{(3);dabc}(\omega_l,\omega_m,\omega_n)\sqrt{\pi}
  \Delta e^{-(\delta \Delta/2)^2}\,.\quad
\end{eqnarray}
with $\Delta^{-2} = \Delta_l^{-2}+\Delta_m^{-2}+\Delta_n^{-2}$. In
this case, the generated currents are also Gaussian in their time and
frequency dependence.

\begin{figure}[t]
\centering
\includegraphics[width=6cm]{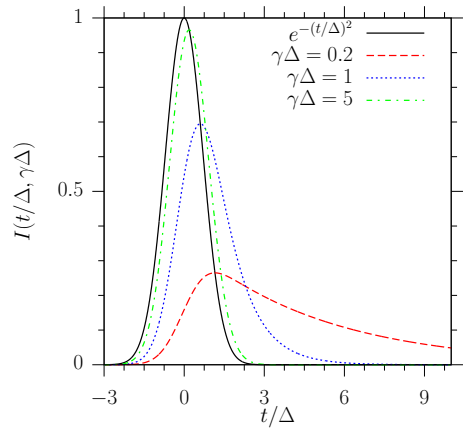}
\caption{(color online) Time evolution of $I(t/\Delta,\gamma\Delta)$ for $\gamma\Delta=0.2$ (red), 1 (blue), and 5 (green). The Gaussian pulse is
  plotted as black curve.}
\label{fig:pulse}
\end{figure}

(ii) For singular behavior
\begin{equation}
 \sigma^{(3);dabc}(\omega_l+\delta_l,\omega_m+\delta_m,\omega_n+\delta_n)
 \approx \frac{i\eta^{(3);dabc}(\omega_l,\omega_m,\omega_n)}
 {\delta_l+\delta_m+\delta_n + i\gamma}\,.\label{eq:pulse2}
\end{equation}
where $\gamma$ contains contributions from the relaxation parameters,
the optical coefficient $C^{dabc}_{\omega_l,\omega_m,\omega_n}(t)$
satisfies 
\begin{eqnarray}
  \Big(\frac{\partial }{\partial
    t} &&+ \gamma\Big) C_{\omega_l,\omega_m,\omega_n}^{dabc}(t) =\notag\\
&&  \eta^{(3);dabc}(\omega_l,\omega_m,\omega_n) p_{\omega_l}(t) p_{\omega_m}(t)p_{\omega_n}(t)\,.
\end{eqnarray}
The solution of this equation is
\begin{eqnarray}
 C_{\omega_l,\omega_m,\omega_n}^{dabc}(t) &=&  \eta^{(3);dabc}(\omega_l,\omega_m,\omega_n)
 \int_{-\infty}^0d\tau e^{\gamma \tau} \notag\\
&&\times p_{\omega_l}(t+\tau)
  p_{\omega_m}(t+\tau) p_{\omega_n}(t+\tau) \,,
\end{eqnarray}
and 
\begin{eqnarray}
  {\cal C}_{\omega_l,\omega_m,\omega_n}^{dabc}(\delta)
  &=&\frac{i\eta^{(3);dabc}(\omega_l,\omega_m,\omega_n)
  }{\delta+i\gamma} \notag\\
&&\times \int dt e^{i\omega t}  p_{\omega_l}(t)
  p_{\omega_m}(t) p_{\omega_n}(t)\,.
\end{eqnarray}
For a Gaussian pulse, we get 
\begin{eqnarray*}
  C_{\omega_l,\omega_m,\omega_n}^{dabc}(t) &=&
  \eta^{(3);dabc}(\omega_l,\omega_m,\omega_n) \frac{I(t/\Delta,\Delta\gamma)}{\gamma}\,,\\
  {\cal C}_{\omega_l,\omega_m,\omega_n}^{dabc}(\delta)
  &=&\frac{i\eta^{(3);dabc}(\omega_l,\omega_m,\omega_n)
  }{\delta+i\gamma} \sqrt{\pi}\Delta e^{-(\delta\Delta)^2/{4}}\,.
\end{eqnarray*}
where $I(x,y)=\frac{\sqrt{\pi}}{2} y e^{-xy}
e^{y^2/4}
\left[1+\text{Erf}\left(x-y/2\right)\right]$,  and $\text{Erf}(x)$ is
the error function.  In the absence of relaxation, we have $\lim\limits_{\gamma\to0}\frac{I(t/\Delta,\Delta\gamma)}{\gamma} =
{\sqrt{\pi}\Delta}/{2}[1+\text{Erf}(t/\Delta)]$, which is a constant
$\sqrt{\pi}\Delta$ as $t\to\infty$. This means that the current is
nonzero even after the optical pulses have passed, indicating
that current injection has occurred.  For finite $\gamma$, $C_{\omega_l,\omega_m,\omega_n}^{dabc}(t)$
at $t\to\infty$ is zero, but the injected current can still persist
for some time. Fig.~\ref{fig:pulse} shows the dependence of the
current response on the pulse width.  For a very long pulse,
$\gamma\Delta\gg1$, the current response  has a shape that is nearly
Gaussian; however, for $\gamma\Delta<1$, when the energy broadening of
the pulse is larger than the relaxation rate, the current response obviously deviates
from Gaussian shape, and can last long after the excitation pulses are passed.

\section{Third order optical nonlinearities\label{sec:tn}}
To illustrate how relaxation affects the third order optical
nonlinearities, in the sample calculations presented below we assume equal relaxation rates for all orders of
response, putting $\Gamma_i^{(n)}=\Gamma_i$ and
$\Gamma_e^{(n)}=\Gamma_e$, and consider four sets of parameters: (a) $\Gamma_i=\Gamma_e=0$, (b)
$\Gamma_i=\Gamma_e=33$~meV, (c) $\Gamma_i=65$~meV and 
$\Gamma_e=0.5$~meV, which are parameters used by
Gu {\it et al.}  \cite{Nat.Photon._6_554_2012_Gu}, (d) $\Gamma_i=0.5$~meV and
$\Gamma_e=65$~meV.  We define set (a) by the limit $\Gamma_i=\Gamma_e\to 0$, which 
recovers our relaxation free calculation  \cite{NewJ.Phys._16_53014_2014_Cheng}.

\subsection{Third harmonic generation}
For monochromatic incident light with frequency $\omega$, light is
nonlinearly generated to lowest order at the third harmonic frequency
$3\omega$ and at the fundamental frequency $\omega$. 
The first is described by the conductivity
$\sigma^{(3);dabc}(\omega,\omega,\omega)$; the second corresponds to
 Kerr effects and two photon absorption, both described by
$\sigma^{(3);dabc}(-\omega,\omega,\omega)$, and can be considered as a
nonlinear correction to the linear optical response. In this section, we
consider THG.

\begin{figure*}[t]
\centering
\includegraphics[width=11.8cm]{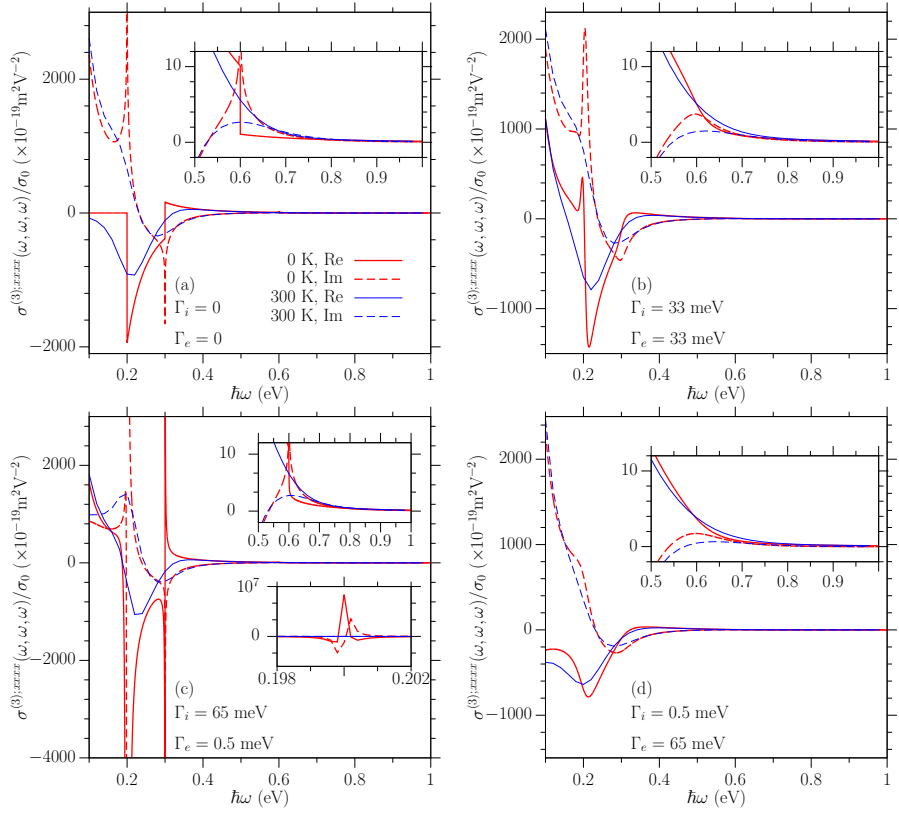}
\caption{(color online) Spectra of
  $\sigma^{(3);xxxx}(\omega,\omega,\omega)$ at zero (thick red 
  curves) and room (thin blue curves) temperatures for different relaxation parameters: (a) $\Gamma_i=\Gamma_e=0$, (b)
  $\Gamma_i=\Gamma_e=33$~meV, (c) $\Gamma_i=65$~meV and
  $\Gamma_e=0.5$~meV, (d) $\Gamma_i=0.5$~meV and
  $\Gamma_e=65$~meV. The real (imaginary) parts of the conductivity
  are given by the solid (dashed) curves; we have taken
  $|\mu|=0.3$~eV. The insets focus on  results in the region $[0.5,
  1]$~eV. In (c) the fine structure in region $[0.198,0.202]$~eV  is also displayed.}
\label{fig:thg00}
\end{figure*}

The conductivity tensor for THG only has one independent component 
\begin{eqnarray}
&&\sigma^{(3);xxyy}(\omega,\omega,\omega) =
\sigma^{(3);xyxy}(\omega,\omega,\omega)\notag\\
&=&\sigma^{(3);xyyx}(\omega,\omega,\omega)=\sigma^{(3);xxxx}(\omega,\omega,\omega)/3\,.
\end{eqnarray}
The induced current responsible for the THG is
\begin{equation}
  \bm J^{(3);d}_{\text{THG}}(t) = e^{-i3\omega t}
  \sigma^{(3);xxxx}(\omega,\omega,\omega) \bm E_{\omega} \bm
  E_{\omega}\cdot\bm E_{\omega} + c.c\,.
\end{equation}
In Fig.~(\ref{fig:thg00}) we give the result for
$\sigma^{(3);xxxx}(\omega,\omega,\omega)$ at $|\mu|=0.3$~eV for zero and room temperature.
We first look at the results for zero temperature. The relaxation-free
results are given as the thick (red) curves in Fig.~\ref{fig:thg00}
(a). This figure shows the step
function of the real parts and the logarithmic divergence of the
imaginary parts at three resonant photon energies 
$\hbar\omega=0.2$, $0.3$, and $0.6$~eV, which correspond to the
resonant transitions for which the chemical potential gap $2|\mu|$ matches the
energies of three photons, two photons, and one photon, respectively  \cite{NewJ.Phys._16_53014_2014_Cheng}. 
With relaxation included, the conductivity is a smooth
function of $\omega$, and plotted in Fig.~\ref{fig:thg00}
(b), (c), and (d). Some common effects induced by the relaxations are
shown: (i) the divergent peaks of the imaginary parts of the
conductivity in Fig.~\ref{fig:thg00} (a) become finite and
broadened, (ii) the step functions of the real parts become
continuous, (iii) the real parts become finite as $\hbar\omega <
{2|\mu|}/{3} = 0.2$~eV, and increase rapidly with decreasing
frequency. They receive contributions not only from intraband transitions,
describing Drude-like effects, but also from the interband transitions
due to the linear dispersion relation of graphene (for example, see the
prefactor $(\hbar\omega)^{-4}$ in Eq.~(\ref{eq:appthg})).

To illustrate the dominant features in these fine structures, we can
analytically expand the coefficients of the functions ${\cal G}_\mu$,
${\cal H}_\mu$, ${\cal I}_\mu$, and the $|\mu|^{-1}$ term in the
conductivity, for small relaxation parameters
$\Gamma_{i,e}/(\hbar\omega) \ll 1$, to write
\begin{eqnarray}
  \sigma^{(3);xxyy}(\omega,\omega,\omega) &=&
  \sigma^{(3);xxyy}_A(\omega)  +
  \sigma^{(3);xxyy}_B(\omega)  \notag\\
&+&
  \sigma^{(3);xxyy}_C(\omega)\,,
\end{eqnarray}
with
\begin{eqnarray}
\sigma^{(3);xxyy}_A&&(\omega) \approx \frac{i \sigma_3}{144(\hbar\omega)^4}\Big[17{\cal
    G}_\mu(\hbar\omega+i\Gamma_e)\notag\\
& &-64{\cal
      G}_\mu(2\hbar\omega+i\Gamma_e)+45{\cal
      G}_\mu(3\hbar\omega+i\Gamma_e)\Big]\,,\label{eq:appthg}\\
  \sigma^{(3);xxyy}_B&&(\omega) \approx \frac{\Gamma_i}{\hbar}
  \frac{\sigma_3}{36(\hbar\omega)^4}\Big[-8{\cal
      H}_\mu(2\hbar\omega+i\Gamma_e) \notag\\
&& +17{\cal
      H}_\mu(3\hbar\omega+i\Gamma_e)+ 3\omega{\cal
      I}_\mu(3\hbar\omega+i\Gamma_e)\Big]\,,\notag\\
\sigma^{(3);xxyy}_C&&(\omega)=-(\Gamma_{i}-\Gamma_{e})^2\dfrac{2i\sigma_3}{27(\hbar\omega)^5|\mu|}\,.\notag
\end{eqnarray}
In the relaxation-free limit as $\Gamma_{e,i}\to 0$,  $\sigma_B^{(3);xxyy}\to0$ 
and $\sigma_A^{(3);xxyy}$ recovers the results of our previous
work  \cite{NewJ.Phys._16_53014_2014_Cheng}. However, the relaxation-free limit of $\sigma_C^{(3);xxyy}$ strongly
depends on the details of the chemical potential and relaxation 
parameters; this is the contribution to $\sigma_C^{(3);xxyy}$ from the
general term discussed earlier in Eq.~(\ref{eq:sigmac}), which is
problematic unless $\Gamma_i=\Gamma_e$. For doped graphene where $\mu$
is finite, $\sigma^{(3);xxyy}_C$ goes to zero with decreasing
relaxation parameters; for graphene that is undoped or at low doping, a more
sophisticated treatment is in order, as discussed in Section \ref{app:limit}.
For the limit
$\Gamma_{i,e},|\mu|\ll \hbar\omega$,  the THG coefficient is approximated as
\begin{equation}
\sigma^{(3);xxyy}(\omega) \approx
\dfrac{-i\sigma_3}{72(\hbar\omega)^4}
  \left[\pi+\frac{16(\Gamma_{i}-\Gamma_{e})^2}{3\hbar\omega|\mu|}\right]\,,
\end{equation}
The term proportional to $\left\vert \mu \right\vert ^{-1}$ did not arise in
our previous calculation$^{5}$, where we assumed that
$\Gamma_{i,e}\rightarrow 0$ faster than $\mu \rightarrow 0$.  
Deferring the treatment of small doping to later studies, we focus
here on graphene with large enough chemical potential that $\sigma
_{C}^{(3);xxyy}(\omega )$ does not make a significant contribution to
the full third harmonic conductivity.

At room temperature, the conductivities for different relaxation parameters
look very similar to each other, and the fine structures caused by
the resonant transitions are smeared out. This can be understood by
the results in Appendix~\ref{app:tem}: temperature affects the
conductivity by smearing and lowering the peaks caused by functions ${\cal
  G}$, ${\cal H}$, and ${\cal I}$, which has an effect similar to increasing the
value of $\Gamma_e$. If we increase each $\Gamma_e$ by the thermal
energy of room temperature, the values of these new $\Gamma_e$ in
the four cases presented in Fig.~\ref{fig:thg00} are close, and it is not surprising that we get similar room temperature
results.

\subsection{Kerr effects and two photon absorption\label{sec:kerr}}
\begin{figure*}[t]
\centering
\includegraphics[width=11.8cm]{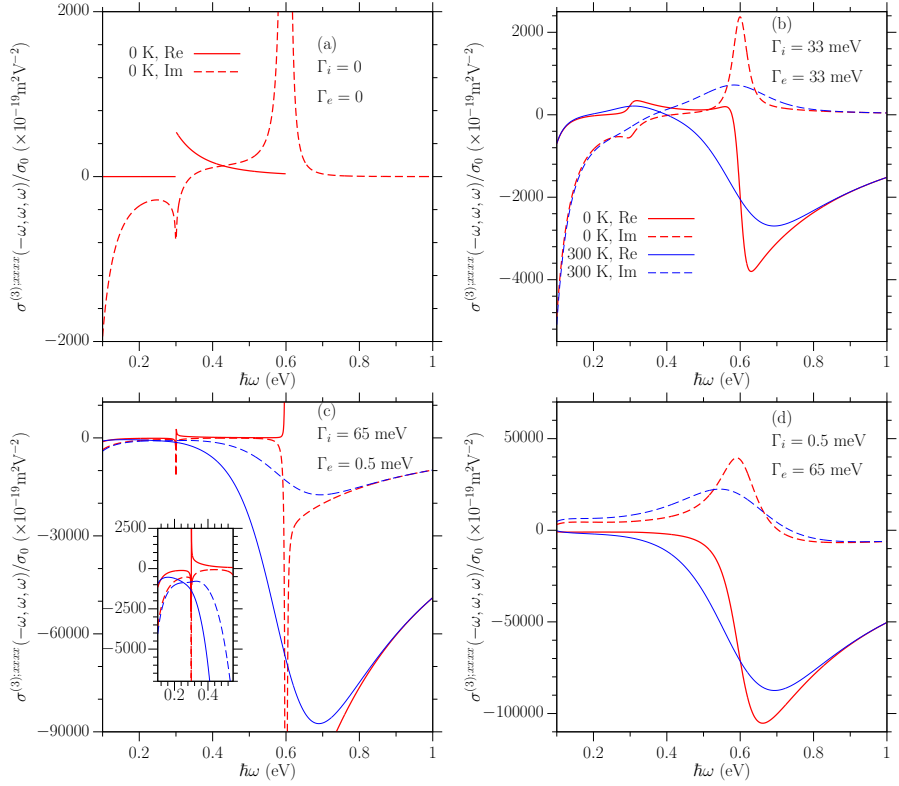}
\caption{(color online) Spectra of 
  $\sigma^{(3);xxxx}(-\omega,\omega,\omega)$ at zero (thick red
  curves) and room (thin blue curves) temperatures with different relaxation parameters: (a) $\Gamma_i=\Gamma_e=0$, (b)
  $\Gamma_i=\Gamma_e=33$~meV, (c) $\Gamma_i=65$~meV and
  $\Gamma_e=0.5$~meV, (d) $\Gamma_i=0.5$~meV and
  $\Gamma_e=65$~meV. The chemical potential is $|\mu|=0.3$~eV. The
  real (imaginary) parts of the conductivity are given by the solid (dashed) curves. The result in region $[0.1,
  0.55]$~eV of (c) is highlighted in the inset. In figure (a) the real part
  (solid curve) of $\sigma^{(3);xxxx}(-\omega ,\omega ,\omega )$
  diverges for all $\hbar \omega >2\left\vert \mu \right\vert $.}
\label{fig:tpa}
\end{figure*}
We now turn to the light nonlinearity generated at the same frequency
$\omega$ of the incident light. Taking $\bm E_{\omega}=\hat{\bm
  x}E_\omega^x + \hat{\bm y} E_{\omega}^y$, we write $\bm E_{\omega} =
\begin{pmatrix} E_{\omega}^x \\ E_{\omega}^y
\end{pmatrix}
$ and consider
\begin{equation}
  \bm E_{\omega} = E_\omega \begin{pmatrix} \cos\phi \\ \sin\phi e^{i\theta}
  \end{pmatrix}\,, \bm E_{-\omega} = E_{\omega}^{\ast} \begin{pmatrix} \cos\phi \\ \sin\phi e^{-i\theta}
  \end{pmatrix}\,.
\end{equation}
The nonlinear response at frequency $\omega$ is then given by 
\begin{eqnarray}
  \bm J^{(3)}(\omega) &=& 3E_{\omega}|E_{\omega}|^2 \bigg[ \sigma^{(3);xxxx}(\omega,\omega,-\omega)
  \begin{pmatrix} \cos\phi \\ \sin\phi e^{i\theta}
  \end{pmatrix} \notag\\
&& \hspace{-1cm}+ \sigma^{(3);xyyx}(\omega,\omega,-\omega)
i \sin(2\phi)\sin\theta \begin{pmatrix}
    \sin\phi e^{i\theta} \\ -\cos\phi
  \end{pmatrix}\bigg]\,.\quad\quad
\end{eqnarray}
For linearly polarized light ($\theta=0$), the second term vanishes;
the current from the first term has the same
polarization as the incident field, and gives an intensity dependent correction of the linear
conductivity $\sigma^{xx}_{\text{eff}}(\omega) = \sigma^{(1);xx}(\omega) +
\sigma_{nl}(\omega)$, with
\begin{equation}
  \sigma_{nl}(\omega) = 3\sigma^{(3);xxxx}(\omega,\omega,-\omega)
  |E_{\omega}|^2\,. \label{eq:nl}
\end{equation}
An effective nonlinear susceptibility can be introduced  \cite{Phys.Rev.Lett._105_097401_2010_Hendry,Phys.Rev.B_87_121406_2013_Kumar,ACSNano_7_8441_2013_Saeynaetjoki} $\chi_{nl}(\omega) =
\sigma_{nl}(\omega)/(-i\omega \epsilon_0 d_{\text{gr}})$, where the
effective thickness of graphene single layer $d_{\text{gr}}$ is taken
to be 3.3\AA  \cite{Phys.Rev.Lett._105_097401_2010_Hendry}; from this
an effective nonlinear refractive index $n_2$ and nonlinear loss 
$\beta_{\text{TPA}}$ can be extracted. In general, $\sigma^{(1);xx}$ has both real part and
imaginary parts, and the calculation of $n_2$ and $\beta_{\text{TPA}}$ should follow the results of del
Corso and Soles~[\onlinecite{J.Opt.Soc.Am.B_21_640_2004_Coso}]. 

In the limit of no relaxation, we showed
earlier  \cite{NewJ.Phys._16_53014_2014_Cheng} that 
$\sigma^{(3);dabc}(-\omega,\omega,\omega)$ has many divergences, and 
its behavior in the neighborhood of equal frequencies can be written as
\begin{eqnarray}
&&  \sigma^{(3);dabc}(-\omega,\omega+\delta_1,\omega+\delta_2)
\notag\\
&=&
  \frac{{\cal T}_1^{dabc}(\omega)}{\delta_1\delta_2} + \frac{{\cal
      T}_2^{dabc}(\omega;\delta_2)}{\delta_1} + \frac{{\cal
      T}_2^{dabc}(\omega;\delta_1)}{\delta_2}\notag\\
&& + {\cal
    T}_3^{dabc}(\omega;\delta_1,\delta_2)\,,\label{eq:tpawogamma}
\end{eqnarray}
where ${\cal T}_{1}^{dabc}$, ${\cal T}_{2}^{dabc}$, and 
${\cal T}_{3}^{dabc}$ are all smooth functions of $\delta_1$ and
$\delta_2$. The strength of the singularities is determined by ${\cal T}_{1}^{dabc}$ and ${\cal
  T}_2^{dabc}$, which are real functions and are only nonzero 
when the photon energy is greater than that for the onset of one-photon absorption 
($\hbar\omega>2|\mu|$). For fixed photon energy $\hbar\omega$,
the appearance of the divergence as $\mu$  decreases from $2|\mu|>\hbar\omega$ to $2|\mu|\le
\hbar\omega$ indicates that it is associated with the existence of
electrons (holes) at the $\bm k$ where one-photon absorption is possible. Physically, at these $\bm
k$ the third order correction to one-photon absorption would lead to
the perturbative description of the saturation, but in the absence of relaxation that
correction diverges, as it would for an inhomogeneously broadened
collection of two-level systems. At zero temperature,
the sharp Fermi surface can strictly exclude such electrons (holes)
for $2|\mu|>\hbar\omega$. However, at finite temperature thermal fluctuations will always place some
electrons (holes) at $\bm k$ where one-photon absorption can occur,
and so the divergence in the third order response will exist for any
photon energy.

Including relaxation parameters $\Gamma_i^{(j)}=\Gamma_i$ and $\Gamma_e^{(j)}=\Gamma_e$, 
$\sigma^{(3);dabc}(-\omega,\omega+\delta_1,\omega+\delta_2)$ includes
terms that are proportional to $t_1=(\delta_1+i \Gamma_i)^{-1}$, $t_2=(\delta_2+i
\Gamma_i)^{-1}$, $t_a=(\delta_1+\delta_2+2i\Gamma_e)^{-1}$, $t_1t_a$,
and $t_2t_a$. As any of these three quantities $\delta_1+i \Gamma_i$, $\delta_2+i
\Gamma_i$, or $\delta_1+\delta_2+2i\Gamma_e$ goes to zero,
$\sigma^{(3);dabc}(-\omega,\omega,\omega)$ diverges. But for nonzero
$\Gamma_i$ and $\Gamma_e$, $\sigma^{(3);dabc}(-\omega,\omega+\delta_1,\omega+\delta_2)$ is a
smooth function of real $\omega$, $\delta_1$, and $\delta_2$. For a pulse
response when the energy broadening of the pulse is less than the
relaxation energies, it is reasonable to set $\delta_1=\delta_2=0$, and
then the conductivity can be written as
\begin{eqnarray}
  \sigma^{(3);dabc}(-\omega,\omega, \omega) &=&
  \frac{{\cal L}_{1}^{dabc}(\omega)}{\Gamma_i\Gamma_e} +
  \frac{{\cal L}^{dabc}_{2}(\omega;\Gamma_e)}{\Gamma_i} \notag\\
&& +
  \frac{{\cal L}^{dabc}_3(\omega;\Gamma_i,\Gamma_e)}{\Gamma_e}\notag\\
&&+ {\cal L}^{dabc}_4(\omega;\Gamma_i,\Gamma_e)\,,
\end{eqnarray}
with 
\begin{eqnarray}
{\cal L}_1^{dabc}(\omega) &=& i\sigma_3 \frac{A_0}{12(\hbar\omega)^2}
\left[{\cal G}_{\mu}(\vartheta_+) + {\cal
    G}_{\mu}(\vartheta_-)\right]\,,\\
{\cal L}_2^{dabc}(\omega;\Gamma_e) &=& \frac{\sigma_3A_0}{12(\hbar\omega)^2}\left\{
\frac{\vartheta_+}{\vartheta_-^2}{\cal
  G}_{\mu}(\vartheta_-) \notag\right.\\
&&+
\frac{5\vartheta_+^2-3\theta_-^2}{2\vartheta_+^3}{\cal
    G}_{\mu}(\vartheta_+)\notag\\
&&\left.
  +\frac{2\hbar\omega}{\vartheta_+^2}\left[\hbar\omega {\cal
      H}_{\mu}(\vartheta_+) +
    \frac{4\Gamma_e^2}{\vartheta_-|\mu|}\right]\right\}\,,
\end{eqnarray}
\begin{eqnarray}
{\cal L}_3^{dabc}&&(\omega;\Gamma_i,\Gamma_e) = \frac{\sigma_3\Gamma_i}{6(\hbar\omega)^2\nu_+}\Big\{
  i(A_0-A_1){\cal
  G}_{\mu}(i\Gamma_e) \notag\\
&&- \frac{\Gamma_i\Gamma_e}{|\mu|(4\mu^2+\Gamma_e^2)\nu_+}\left(A_1+A_0+\frac{2\nu_+}{\nu_-}A_1\right)\Big\}\,,
\end{eqnarray}
The full expression of ${\cal L}_4^{dabc}(\omega;\Gamma_i,\Gamma_e)$
is complicated; we can achieve a good approximation by setting
$\Gamma_i=0$, for which
\begin{eqnarray}
&&{\cal L}_4^{dabc}(\omega;0,\Gamma_e) \notag\\
&=&
\frac{i\sigma_3}{12(\hbar\omega)^4}\Big\{2(A_1-A_0){\cal
    G}_\mu(i\Gamma_e) + 4(3A_1+A_0){\cal
    G}_\mu(\vartheta_+)\notag\\
&&-(4A_1+5A_0){\cal G}_\mu(\vartheta_-)  -
  8(A_1+A_0){\cal G}_\mu(2\hbar\omega+i\Gamma_e) \notag\\
&&- \frac{4\mu (\hbar\omega)^2}{(\vartheta_+^2-4\mu^2)^2}\left[(A_1+4A_0)\vartheta_++A_0\frac{\vartheta_+^2-4\mu^2}{\hbar\omega}\right]\Big\}\,.\quad\quad
\end{eqnarray}
In these expressions, we used $\vartheta_{\pm}=\pm\hbar\omega+i\Gamma_e$ and $\nu_{\pm} =
\pm\hbar\omega+i\Gamma_i$. 

In Fig.~\ref{fig:tpa}, the
photon energy dependence of $\sigma^{(3);dabc}(-\omega,\omega,\omega)$
is plotted for different relaxation parameters, with chemical potential
$|\mu|=0.3$~eV at zero and room temperatures. 
Fig.~\ref{fig:tpa} (a) gives the relaxation-free calculation, which is done as
$\lim\limits_{\Gamma_i=\Gamma_e\to0}\sigma^{(3);dabc}(-\omega,\omega,\omega)$. Three
regimes are apparent: (1) $\hbar\omega<|\mu|$, in which both
one- and two- photon absorption are absent, and the real part of the
conductivity is zero. The imaginary part  at low photon energy scales as $(\hbar\omega)^{-3}$. At $\hbar\omega=|\mu|$, the real part shows a step
function, while the imaginary part shows a logarithmic divergence. 
(2) $|\mu|<\hbar\omega<2|\mu|$,
in which two-photon absorption is present but  one-photon
absorption is still absent. The real part of the conductivity here
scales as $(\hbar\omega)^{-4}$. 
Around $\hbar\omega=2|\mu|$, the
imaginary part shows a divergence $(\hbar\omega-2|\mu|)^{-2}$. For
frequencies satisfying  $\hbar\omega<2|\mu|$, if the graphene is
subject to a Gaussian pulse sufficiently narrow in frequency, the
nonlinear current induced will still have a shape that is
approximately Gaussian, and characterizing the nonlinear response to a
pulse by Eq.~(\ref{eq:nl}) makes sense. (3) $\hbar\omega>2|\mu|$, where both two- and
one-photon absorption are present. The imaginary part of 
the conductivity diverges as $(\hbar\omega-2|\mu|)^{-2}$ around
$\hbar\omega=2|\mu|$, and the real part diverges for the entire region
$\hbar\omega>2|\mu|$. At finite temperature and in the absence of relaxation, the real part
diverges for {\it any} photon energy $\hbar\omega$. As we discussed
after Eq.~(\ref{eq:tpawogamma}), the divergence of the
real part of $\sigma^{(3);dabc}(-\omega,\omega,\omega)$ is induced
by the existence of electrons (holes) at the $\bm k$ where one-photon absorption
occurs; at zero temperature, these electrons (holes) only exist when the chemical potential
$|\mu|<\hbar\omega/2$, while at finite temperature, they exist at any
chemical potential due to thermal fluctuations.

In Fig.~\ref{fig:tpa} (b), (c), and (d), we present the results for
the same relaxation parameters as those adopted in the THG
calculation. Relaxation affects the conductivity in a complex way,
but there are some qualitative features that can be identified: (i) In
the neighborhood of the divergences that arise in the relaxation-free 
calculation, including the divergent regime $\hbar \omega >2\left\vert \mu
\right\vert $ and the special frequency $\hbar \omega =\left\vert \mu
\right\vert $, both the real
and imaginary parts of the conductivity are lowered and  are
everywhere finite. In Fig.~\ref{fig:tpa}
(b), (c), and (d), we find that a larger $\Gamma_e$ gives lower and
broader peaks at $\hbar\omega=|\mu|$ and $2|\mu|$. (ii) For the
relaxation parameters used here, the real part of the conductivity is 
negative for $\hbar\omega>2|\mu|$. Because of the presence in this
frequency range of one-photon absorption, which is always
positive, the two-photon absorption processes indicated by the real
part of $\sigma^{(3);dabc}(-\omega,\omega,\omega)$ can be understood
as a correction to the simple linear prediction of the absorption. In
fact, we can find a range of electric fields large enough so that 
$\sigma^{da}_{\text{eff}}(\omega)$ is negative; for a field anywhere
near or above this strength the perturbative result is naturally
suspect. (iii) Even for frequencies in the range $\left\vert \mu 
\right\vert <\hbar \omega <2\left\vert \mu \right\vert $, where only
two-photon absorption is present in the absence of relaxation, the real part
of the nonlinear conductivity $\sigma ^{(3);dabc}(-\omega ,\omega ,\omega )$
can be negative. Yet in the presence of relaxation the linear conductivity 
$\sigma^{(1);xx}(\omega )$ acquires a real part in this frequency range,
and the real part of $\sigma _{\text{eff}}^{xx}(\omega )$, for example, is
always positive for small enough electric field amplitudes, indicating
absorption. However, these results indicate the sensitivity to the
relaxation parameters of both the third order conductivity, and its
interplay with the first order conductivity, and a more sophisticated
description of the scattering is clearly in order.

At room temperature, the peaks or divergences are further
broadened. For a given frequency $\omega$, the regime $\hbar\omega>2|\mu|$ always
contributes to a finite temperature calculation due to the average
over the chemical potential. The absolute values of the real part of the
conductivity in the regime $\hbar\omega<2|\mu|$ also significantly
increase.

\subsection{Two-color coherent current injection\label{sec:coh}}
\begin{figure}[t]
\centering
\includegraphics[width=8cm]{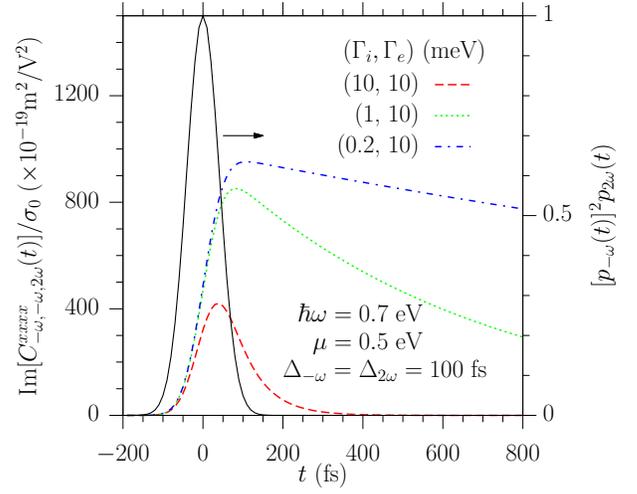}
\caption{(color online) Time evolution of
  $\text{Im}[C^{xxxx}_{-\omega,-\omega,2\omega}(t)]$ for Gaussian pulses with
  $\Delta_{-\omega}=\Delta_{2\omega}=100$~fs for different relaxation
  parameters $(\Gamma_i,\Gamma_e)=(0.2, 10)$~meV (blue chain
  curve), $(1.0, 10)$~meV (green dotted
  curve), and $(10, 10)$~meV (red dashed curve), $(10, 1)$~meV, and
  $(10,0.2)$~meV. The last three cases overlap with the red dashed curve. The pulse
  width corresponds to an energy broadening
  $\hbar\Delta_{-\omega}=6.6$~meV. Our calculations show
  $\text{Re}[C^{xxxx}_{-\omega,-\omega,2\omega}(t)]$ is negligible on
  this scale.}
\label{fig:pulsecoh}
\end{figure}
\begin{figure*}[t]
\centering
\includegraphics[width=11.8cm]{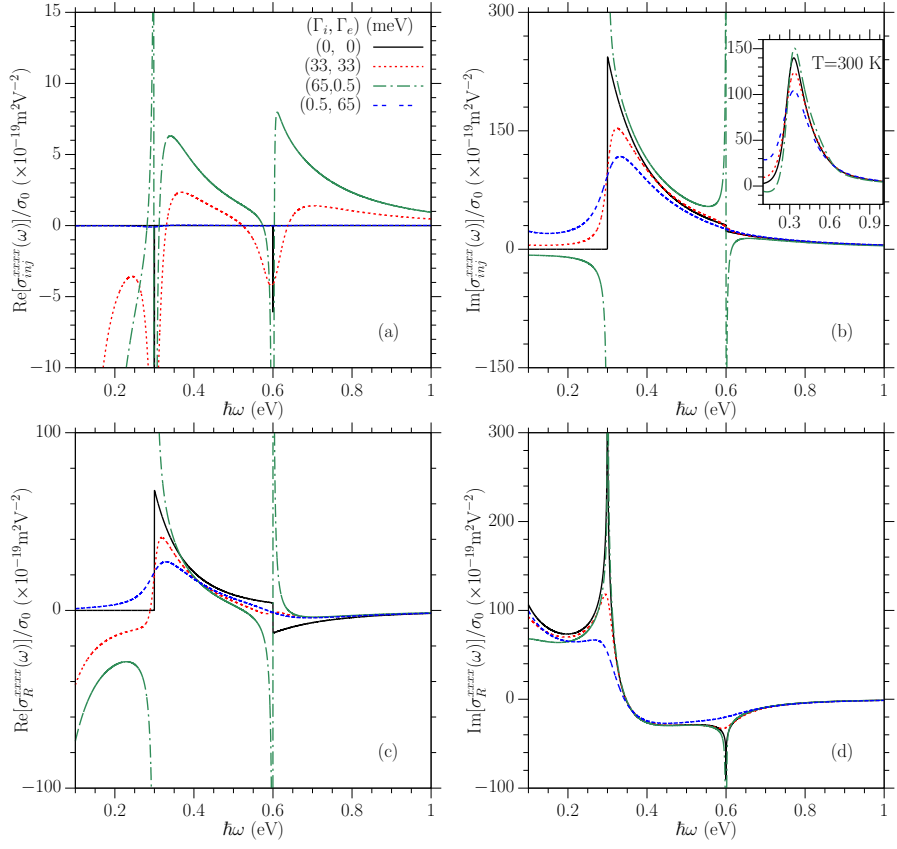}
\caption{(color online) Spectra of $\sigma_{inj}^{xxxx}(\omega)$ (a, b)
  and $\sigma_{R}^{xxxx}(\omega)$ (c, d) for different relaxation 
  parameters $\Gamma_i$ and $\Gamma_e$ at temperature $T=0$: Black
  solid curves:  $\Gamma_i=\Gamma_e=0$; Red dotted curves: $\Gamma_i=\Gamma_e=33$~meV;
  green chain curves: $\Gamma_i=65$~meV, $\Gamma_e=0.5$~meV; Blue dashed
  curves: $\Gamma_i=0.5$~meV, $\Gamma_e=65$~meV. The chemical
  potential is $|\mu|=0.3$~eV. The result of
  $\text{Im}[\sigma_{inj}^{xxxx}(\omega)]$ at 300~K is shown in the inset of (b).}
\label{fig:coh}
\end{figure*}
Now we turn to two-color coherent current injection, with frequencies
$\omega_1=\omega_2=-\omega$ and 
$\omega_3=2\omega$. In the relaxation-free calculation, the conductivity
$\sigma^{(3);dabc}(-\omega,-\omega,2\omega)$ diverges, and it is the
divergence that describes the current injection. In 
fact, in the neighborhood of these frequencies, the conductivity can
be written as
\begin{equation}
  \sigma^{(3);dabc}(-\omega,-\omega,2\omega+\delta\omega) =
  \frac{i\eta^{dabc}(\omega)}{3\delta\omega} + \sigma^{(3);dabc}_R(\omega)\,,
\end{equation}
where the injected current is determined by a well-behaved function
$\eta^{dabc}(\omega)$, and $\sigma^{(3);dabc}_R(\omega)$ is a smooth  
function of $\delta\omega$. With the inclusion of relaxation, the conductivity
$\sigma^{(3);dabc}(-\omega,-\omega,2\omega)$  itself is well
behaved. The divergence term in the relaxation-free limit becomes a
term similar to the right hand side of
Eq.~(\ref{eq:pulse2}). To check whether $\Gamma_i$ and $\Gamma_e$ have
the same importance for the injection, we give the pulse calculations
of $C^{xxxx}_{-\omega,-\omega,2\omega}(t)$ (see Eq.~\ref{eq:pulsect})
in Fig.~\ref{fig:pulsecoh} for different relaxation parameters. After the laser pulse, the
current response persists for times associated with $\Gamma_i$, showing that the contribution from $\Gamma _{i}$
dominates the relaxation of the injected current, as might be
expected. To highlight this, we write 
\begin{eqnarray}
&&  \sigma^{(3);dabc}(-\omega+\delta_1,-\omega+\delta_2,2\omega+\delta_3)\notag\\
&=& \frac{i
    \eta^{dabc}(\omega)}{3(\delta_1+\delta_2+\delta_3+i\hbar^{-1}\Gamma_i)}
  + \sigma^{(3);dabc}_{R}(\omega)\,,\label{eq:tcci}
\end{eqnarray}
From Eq.~(\ref{eq:sigmaall}), only terms ${\cal S}_{1-4}^{dabc}$ including
$(\delta_1+\delta_2+\delta_3+i\Gamma_i)^{-1}$ contribute to
$\eta^{dabc}(\omega)$. By writing
$\sigma_{inj}^{dabc}(\omega)\equiv{\hbar\eta^{dabc}(\omega)}/{\text{1eV}}$,
the first term in Eq.~(\ref{eq:tcci}) becomes
\begin{equation}
\frac{ 1\text{eV}
}{3(\hbar\delta_1+\hbar\delta_2+\hbar\delta_3+i\Gamma_i)}
i\sigma_{inj}^{dabc}(\omega)\label{eq:tcci1}
\end{equation}
with
\begin{eqnarray}
\sigma_{inj}^{dabc}(\omega) &\approx&
\frac{i\sigma_3}{\text{1eV} } \Bigg\{-\frac{2A_3+A_0}{2(\hbar\omega)^3}\text{Im}[{\cal
  G}_\mu(\hbar\omega+i\Gamma_e)]\notag\\
&&+\frac{A_3+A_0}{(\hbar\omega)^3}\text{Im}[{\cal
  G}_\mu(2\hbar\omega+i\Gamma_e)]\notag\\
&&+\frac{6\Gamma_e A_3 + 5 \Gamma_iA_0}{4(\hbar\omega)^4} \text{Re}[{\cal
  G}_\mu(\hbar\omega+i\Gamma_e)]\notag\\
&&-\frac{3\Gamma_e(A_0+A_3)+\Gamma_iA_0}{4(\hbar\omega)^4}\text{Re}[{\cal
  G}_\mu(2\hbar\omega+i\Gamma_e)]\notag\\
&&-\frac{\Gamma_i(A_3+A_0)}{4(\hbar\omega)^3}{\cal
  H}_\mu(-2\hbar\omega+i\Gamma_e) \notag\\
&&+
\frac{\Gamma_i(A_3+2A_0)}{4(\hbar\omega)^3}{\cal
  H}_\mu(\hbar\omega+i\Gamma_e)\Bigg\}\,,
\end{eqnarray}
where terms proportional to $|\mu|^{-1}$ are neglected. 
As $\Gamma_{i},\Gamma_e\to0$, only terms involving $\text{Im}[{\cal
  G}_\mu(\omega)]\propto\theta(|\omega|-2|\mu|)$ remain;
$\sigma^{dabc}_{inj}(\omega)$ is a pure imaginary quantity, and is
consistent with our previous work  \cite{NewJ.Phys._16_53014_2014_Cheng}. With
relaxation included,  terms involving $\text{Re}[{\cal G}_\mu]$ and ${\cal H}_\mu$ appear. 

In Fig.~\ref{fig:coh} (a) and (b) we plot the photon energy dependence of
$\sigma_{inj}^{xxxx}(\omega)$ for different relaxation parameters at zero
and room temperature. The real part is much smaller than the
imaginary part. At zero temperature, we see in Fig.~\ref{fig:coh} (b)
that for $\Gamma_e=0.5$~meV there are fine structures in the spectrum of
$\text{Im}[\sigma^{xxxx}_{inj}(\omega)]$ around $\hbar\omega=|\mu|$
and $\hbar\omega=2|\mu|$; it is due to the ${\cal H}_\mu$ terms. As
can be seen in the inset of Fig.~\ref{fig:coh} (b), finite temperature and
finite $\Gamma_e$ lead to similar broadening and lowering of the peaks. 

We also plot $\sigma^{xxxx}_{R}(\omega)$ in Fig.~\ref{fig:coh} (c)
and (d). The amplitude of $\sigma^{xxxx}_{R}$ is of the same order of
magnitude as that of $\sigma^{xxxx}_{inj}$. However, due to the prefactor $1\text{eV}/(3\Gamma
_{i})$, which relates $\sigma _{inj}^{xxxx}$ to its contribution to $\sigma
^{(3);dabc}(-\omega ,-\omega ,2\omega )$ in Eq.~(\ref{eq:tcci1}), 
$\sigma^{xxxx}_{inj}$ dominates for small $\Gamma_i$, usually taken
to be a few tens of meV. 

\subsection{Parametric frequency conversion\label{sec:pfc}}
\begin{figure*}[t]
\centering
\includegraphics[width=11.8cm]{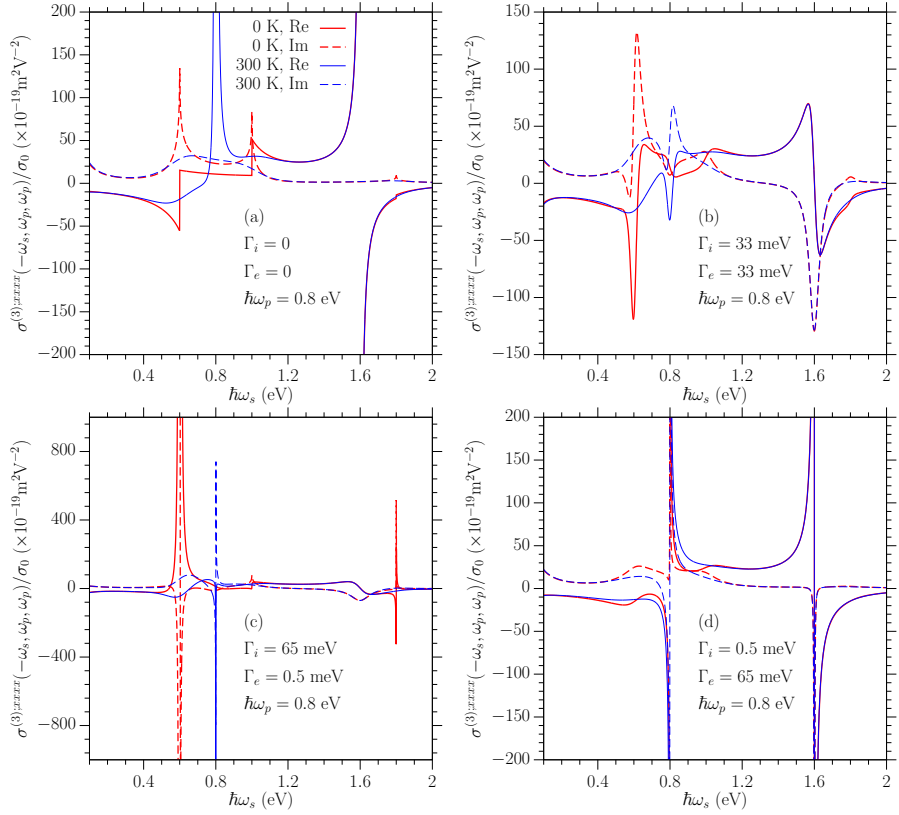}
\caption{(color online) Spectra of $\sigma^{(3);xxxx}(-\omega_s,\omega_p,\omega_p)$ for
different relaxation parameters at zero (thick red curves) and room
temperature (thin blue curves). Solid (dashed) curves give their real
(imaginary) parts separately. In the calculation, the chemical
potential is $|\mu|=0.5$~eV, and the pump photon energy is
$\hbar\omega_p=0.8$~eV. }
\label{fig:pfc}
\end{figure*}
Third order nonlinearities can lead to the appearance of new
frequencies via parametric frequency conversion, which is described by
$\sigma^{(3);dabc}(-\omega_s,\omega_p,\omega_p)$. Here $\omega_p$
is the frequency of a strong pump field, and $\omega_s$ is the signal
frequency converted by interaction with the pump to an idler frequency
$\omega_i=2\omega_p-\omega_s$. Possible resonant transitions occur as any of the frequencies $|\omega_p|$, $|\omega_s|$,
$|\omega_p-\omega_s|$, $|2\omega_p|$, or $|2\omega_p-\omega_s|$
equal $2|\mu|/\hbar$. In Fig.~\ref{fig:pfc} we plot the dependence
of $\sigma^{(3);xxxx}(-\omega_s,\omega_p,\omega_p)$ on  $\omega_s$
for different relaxation parameters at $\hbar\omega_p=0.8$~eV and $|\mu|=0.5$~eV. 

At zero temperature, the calculations show peaks/step functions for resonant transitions at $\hbar\omega_{s1} =
2\hbar\omega_p-2|\mu|=0.6$~eV, $\hbar\omega_{s2}=2|\mu|=1.0$~eV, or
$\hbar\omega_{s3}=2|\mu|+\hbar\omega_p=1.8$~eV, both with and without the
inclusion of relaxation. Around these resonant
transitions, the behavior of the conductivity can be analyzed as following:

(1) Around $\omega_s = \omega_{s1} + \delta\omega$, the idler photon energy $\hbar\omega_i=2|\mu|-\hbar\delta\omega$
  is close to the onset of the one-photon absorption. By taking
  $\vartheta=\hbar\omega_i+i\Gamma_e=2|\mu|-\hbar\delta\omega+i\Gamma_e$, the conductivity as
  $\delta\omega\to0$ is determined by  functions ${\cal
    G}_{\mu}(\vartheta)\sim \ln (\hbar\delta\omega-i\Gamma_e)$, ${\cal
    H}_{\mu}(\vartheta)\sim (\hbar\delta\omega-i\Gamma_e)^{-1}\Gamma_i $, and
  ${\cal
    I}_{\mu}(\vartheta)\sim(\hbar\delta\omega-i\Gamma_e)^{-2}\Gamma_i$. In the
  relaxation-free limit, only ${\cal
    G}_{\mu}(\vartheta)$ contributes a
  logarithmic divergence to the imaginary part, and a step change in the
  real part [in Fig.~\ref{fig:pfc} (a)] for nonzero
  $\delta\omega$. With the inclusion of relaxation, we can distinguish
  three different types of qualitative behavior, shown in  Fig.~\ref{fig:pfc} (b) - (d), based on the relative magnitude
  of $\Gamma_i$ and $\Gamma_e$: (b) $\Gamma_i=\Gamma_e$, all functions
  contribute; (c)  $\Gamma_e\ll \Gamma_i$, ${\cal
    I}_{\mu}(\vartheta)$ dominates; (d)  $\Gamma_e\gg \Gamma_i$, where for the values chosen the
  relaxation is large enough to smear out these resonances. 

  (2)  Around $\omega_s = \omega_{s2} + \delta\omega$, the signal
  frequency is close to the onset of the one-photon absorption. For
  non-resonant transitions in a usual semiconductor, $\omega_{s_2}$
  and $\omega_{s_1}$ are interchangeable frequencies to give the same
    conductivity of parametric
    frequency conversion  \cite{boyd_nonlinearoptics}; here in graphene they
  yield asymmetric peaks because the resonant transitions dominate. In
  the limit of no relaxation, the conductivity shows
  a logarithmic divergence that is easily smeared out by the inclusion
  of small relaxation parameters.

(3) Around $\omega_s=\omega_{s3}+\delta\omega$: By taking
  $\vartheta=-\hbar\omega_s+\hbar\omega_p+i\Gamma_e=-2|\mu|-\hbar\delta\omega+i\Gamma_e$, the conductivity as
  $\delta\omega\to0$ is determined by  functions ${\cal
    G}_{\mu}(\vartheta)\sim \ln (\hbar\delta\omega-i\Gamma_e)$ and ${\cal
    H}_{\mu}(\vartheta)\sim
  (\hbar\delta\omega-i\Gamma_e)^{-1}\Gamma_i$. In the limit of no
  relaxation, ${\cal G}_{\mu}$ gives a small peak. For $\Gamma_e\ll\Gamma_i$, the peak from ${\cal H}_{\mu}$ is stronger
  but very narrow. 

(4) At finite temperature there is a further smearing of the peaks
around the resonances, as described in Appendix~\ref{app:tem}. 

Besides these resonant transitions, two singularities are apparent:
(i) the singularity around $\omega_s=2\omega_p$, which corresponds to two-color
coherent current injection. The singularity is not determined by the
behavior of the functions ${\cal G}$, ${\cal H}$, ${\cal I}$, but by
the coefficients that premultiply them in Eqs.~(\ref{eq:s1}) to
(\ref{eq:s8}). Around $\omega_s=2\omega_p$ we put
$\omega_s=2\omega_p+\delta\omega$ and find
\begin{eqnarray*}
&&  \sigma^{(3);xxxx}(-2\omega_p+\delta\omega,\omega_p,\omega_p)
  \notag\\
&=&
  \frac{1\text{eV}}{\hbar\delta\omega+i\Gamma_i}i\sigma^{(3);xxxx}_{inj}(\omega_p) + \sigma^{(3);xxxx}_{R}(\omega_p)\,,
\end{eqnarray*}
an equation similar to Eq.~(\ref{eq:tcci}) discussed in Section \ref{sec:coh}. 
Two-color coherent current
injection requires both one-photon absorption (for $\omega_s$) and two photon
absorption (for $\omega_p$), {\it i.e.}, $\hbar\omega_p>|\mu|$. The
parameters we have adopted in Fig.~\ref{fig:pfc} fulfill this criterion, and thus the
singularity appears.  Finite
temperatures do not qualitatively affect this singularity because it is not related
to the chemical potential. 
(ii) The strong response around $\omega _{s}=\omega _{p}$ is related to the third order correction to
one-photon absorption, and only appears at finite temperature. Since $|\mu| <\hbar\omega_{p}<2|\mu|$,
at zero temperature only two-photon absorption is present and there is no
one-photon absorption. However, at finite temperature thermal fluctuations
will place electrons where one-photon absorption can occur, and the third
order correction to that will lead, in the absence of relaxation, to a
divergent result as discussed in Section \ref{sec:kerr}; in the presence of relaxation
the result will not be divergent but very large, describing the saturation
of the one-photon absorption at the level of the third order response.

\subsection{Comparison between calculations and experiments}
Experiments have already
extracted values of the effective third order susceptibilities of
THG  \cite{Phys.Rev.B_87_121406_2013_Kumar,ACSNano_7_8441_2013_Saeynaetjoki,Phys.Rev.X_3_021014_2013_Hong}, 
two-photon absorption  \cite{Opt.Lett._37_1856_2012_Zhang,Phys.Rev.X_3_021014_2013_Hong}, Kerr
effects  \cite{NanoLett._11_2622_2011_Yang,Nat.Photon._6_554_2012_Gu,NanoLett._11_5159_2011_Wu},
and parameter frequency conversion
  \cite{Phys.Rev.Lett._105_097401_2010_Hendry} at some photon
energies. The nonlinear conductivities we have calculated here are 
related to the effective
susceptibility by   \cite{Phys.Rev.Lett._105_097401_2010_Hendry,NewJ.Phys._16_53014_2014_Cheng}
\begin{equation}
  \chi^{(3);dabc}_{\text{eff}}(\omega_1,\omega_2,\omega_3) =
  \frac{ \sigma^{(3);dabc}(\omega_1,\omega_2,\omega_3)}{-i(\omega_1+\omega_2+\omega_3)\epsilon_0d_{\text{gr}}}\,.
\end{equation}
We first look at
the THG, for which the experimental technique 
is perhaps the most mature, and the extracted values can likely be considered more reliable
than those from other effects. For a reasonable chemical
potential estimated from the sample preparation, the calculations
without relaxation parameters  \cite{NewJ.Phys._16_53014_2014_Cheng} yield theoretical
results for the nonlinear conductivity about two orders of magnitude
smaller than the value extracted from experiments. Here we have found that calculations at
finite temperature for different sets of relaxation parameters (see the
insets of Fig.~\ref{fig:thg00}) are almost the same as calculations
at zero temperature and neglecting relaxation. 

For Kerr effects, because of the existence of divergent terms in the
expressions and the probably very low chemical potential in
experiments, it is not surprising that we could fit the nonlinear
susceptibility at one photon energy by tuning the relaxation
parameters. The complicated dependence is shown in
Fig.~\ref{fig:tpa}. As $\hbar\omega>2|\mu|$, the nonlinear
conductivity at both zero and room temperatures can vary many orders
of magnitude, depending on the relaxation parameters adopted.  

For parametric frequency conversion observed in the experiment by Hendry {\it et
  al.}  \cite{Phys.Rev.Lett._105_097401_2010_Hendry}, with parameters $\hbar\omega_p=1.31$~eV,
$\hbar\omega_s=1.05$~eV, and assuming a low chemical potential
$|\mu|=0.1$~eV, we checked the dependence of the conductivity on the
relaxation parameters $\Gamma_i$ and $\Gamma_e$ in the range of $[0, 60]$~meV. We find the
dependence is weak and the calculated values are still smaller than their
claimed values by two orders of magnitude   \cite{Phys.Rev.Lett._105_097401_2010_Hendry}.

Admittedly the measured effective susceptibilities for parametric
frequency conversion, Kerr effects and two photon absorption, and THG
show a strong dependence on 
the measurement method, light frequency, pulse duration, and perhaps sample
preparation. Yet even taking this into account, the
conclusion that the theoretical results are about two orders of magnitude smaller
than the measured results is inescapable. These discrepancies
could arise for a number of reasons, including: (1) The samples in many experiments are not 
suspended graphene, but graphene on a substrate or in solution. Thus
there may have been contributions to the optical nonlinearity from the
interaction between the graphene sheet and its environment, which may be
crucial considering that graphene is a one-atom thick material.  (2)
Thermal effects    \cite{boyd_nonlinearoptics}  caused by a high repetition rate of laser pulses, as used in
$Z$-scan experiments, may play an important role  \cite{Opt.Lett._36_2086_2011_Liu,Opt.Express_21_7511_2013_Zhang}. (3)
Because of the zero gap of graphene and the intense laser beams used
in experiments, saturation   \cite{Adv.Funct.Mater._19_3077_2009_Bao}
induced by one and/or two photon absorption can make necessary a
treatment more sophisticated than that of perturbation theory. Zhang
{\it et al.}  \cite{Opt.Lett._36_4569_2011_Zhang} used the density
matrix method to study four wave mixing in undoped
graphene in the saturation regime, and found an effective $\chi^{(3)}_{\text{eff}}$ about
$10^{-17}$~m$^2$/V$^2$, and decreasing with increasing light
intensity. Additional calculations for different third-order nonlinear
effects in graphene in the saturation regime are needed to assess the
impact of saturation on the theoretical nonlinearities. (4) The calculation at the independent particle level,
which works well as a starting point for most gapped semiconductors, may fail in
graphene, and it may be necessary to do a more realistic calculation,
including the full band structure, and the detailed effects of
scattering and the electron-electron interactions.

\section{DC current induced second order nonlinearity\label{sec:sn}}
\begin{table*}[t]
  \centering
  \begin{tabular}[t]{|c|c||c|c|}
    \hline
    unsymmetrized $\widetilde{\sigma}$& relaxation parameters  & unsymmetrized $\widetilde{\sigma}$ & relaxation parameters\\
    \hline
    &
    $\Gamma_{i}^{(3)}=\Gamma_{i}^{\text{dc}}$   &
    $\widetilde{\sigma}^{(3);dacb}(\omega_1,0,\omega_2)$& \\
    $\widetilde{\sigma}^{(3);dabc}(\omega_1,\omega_2,0)$ &
    $\Gamma_{e}^{(3)}=\Gamma_{e}^{\text{dc}}$ &     $\widetilde{\sigma}^{(3);dbca}(\omega_1,0,\omega_2)$ &$\Gamma_{i}^{(j)}=\Gamma_{i}^{\text{op}}$\\
    $\widetilde{\sigma}^{(3);dbac}(\omega_2,\omega_1,0)$&
    $\Gamma_{i}^{(1,2)}=\Gamma_{i}^{\text{op}}$ &     $\widetilde{\sigma}^{(3);dcab}(0,\omega_1,\omega_2)$ &
    $\Gamma_{e}^{(j)}=\Gamma_{e}^{\text{op}}$ \\
    &
    $\Gamma_{e}^{(1,2)}=\Gamma_{e}^{\text{op}}$ &     $\widetilde{\sigma}^{(3);dcba}(0,\omega_2,\omega_1)$ &\\
    \hline
  \end{tabular}
  \caption{Relaxation parameters used in the different processes
    associated with the dc current
    induced second order nonlinearity.}
  \label{tab:dampings}
\end{table*}
We now turn to the limiting case where one of the electric fields is a dc
field, taking $\omega_3=0$. The calculation of
$\widetilde{\sigma}^{(3);dabc}(\omega_1,\omega_2,0)$ from
Eq.~(\ref{eq:sigmaall}) includes a term proportional to
\begin{equation}
\dfrac{1}{\hbar\omega_3+i\Gamma_i^{(3)}} \rightarrow
\dfrac{1}{i\Gamma_i^{(3)}}\,.
\end{equation}
Therefore a nonzero relaxation $\Gamma_i^{(3)}$ for
the dc field is necessary to set up a steady state with a dc charge
current in graphene. For other transitions included in $\widetilde{\sigma}^{(3);dacb}(\omega_1,0, \omega_2)$ and
$\widetilde{\sigma}^{(3);dcab}(0,\omega_1,\omega_2)$, it is not
necessary to include relaxation associated with the dc field, because the dc field acts on the
optical excitation with frequency $\omega_2$ and $\omega_1+\omega_2$
respectively; these only survive during the optical pulse. We list the
relaxation parameters used in calculating the unsymmetrized 
conductivities in Table~\ref{tab:dampings}.
The third order conductivity of interest here, which we can refer to
as the dc-induced second
order optical conductivity, can be written as
\begin{equation}
  \sigma^{(3);dabc}(\omega_1,\omega_2,0) =
  \frac{1\text{eV}}{3\Gamma_i^{\text{dc}}}\sigma_{J}^{dabc}(\omega_1,\omega_2)
  + \sigma_{E}^{dabc}(\omega_1,\omega_2)\,.\label{eq:sigma2c}
\end{equation}
The first term includes all contributions that diverge as
$1/\Gamma_i^{\text{dc}}$, which are only involved in calculating
$\widetilde{\sigma}^{(3);dabc}(\omega_1,\omega_2,0)$ and
$\widetilde{\sigma}^{(3);dbac}(\omega_2,\omega_1,0)$; they both occur with the dc charge current. Thus we can associate it
with the dc current-induced second
order conductivity, in that it is second order in the optical fields
at $\omega_1$ and $\omega_2$. The second term includes all other
contributions, and we can associate it with a dc field-induced second
order conductivity, which exists even for a gapped
semiconductor without doping.  
Examining Eq.~(\ref{eq:sigmaall}), we see that $\sigma_{J}^{dabc}(\omega_1,\omega_2)$ is
independent of $\Gamma_i^{\text{dc}}$ and $\Gamma_e^{\text{dc}}$, and
it can be written
as 
\begin{equation*}
\sigma_{J}^{dabc}(\omega_1,\omega_2)=\frac{i\sigma_3}{\text{1eV}}S_{J}^{dabc}(\omega_1,\omega_2)
\end{equation*}
with
\begin{widetext}
\begin{eqnarray}
  S_{J}^{dabc}(\omega_1,\omega_2)  &=&
  \left[\frac{A_1}{(\hbar\omega_2)^2} +
    \frac{A_3}{\hbar\omega_1+i\Gamma_{e}^{\text{op}}}\left(\frac{1}{\hbar\omega_1+\hbar\omega_2+i\Gamma_{i}^{\text{op}}}-\frac{1}{\hbar\omega_2}\right)\right]{\cal
    H}_{\mu}(\hbar\omega_1+i\Gamma_{e}^{\text{op}})
  \notag\\ 
  &+&  \left[\frac{A_2}{(\hbar\omega_1)^2} +
    \frac{A_3}{\hbar\omega_2+i\Gamma_{e}^{\text{op}}}\left(\frac{1}{\hbar\omega_1+\hbar\omega_2+i\Gamma_{i}^{\text{op}}}-\frac{1}{\hbar\omega_1}\right)\right]{\cal
    H}_{\mu}(\hbar\omega_2+i\Gamma_{e}^{\text{op}})
  \notag\\
&+&\left[-\frac{A_1}{(\hbar\omega_2)^2}-\frac{A_2}{(\hbar\omega_1)^2}-\frac{A_1+A_2+A_3}{\hbar\omega_1+\hbar\omega_2+i\Gamma_{e}^{\text{op}}}\left(\frac{1}{\hbar\omega_1+i\Gamma_{i}^{\text{op}}}+\frac{1}{\hbar\omega_2+i\Gamma_{i}^{\text{op}}}\right)\notag\right.\\
&&\left.+\frac{A_3}{\hbar\omega_1+\hbar\omega_2+i\Gamma_{e}^{\text{op}}}\left(\frac{1}{\hbar\omega_1}+\frac{1}{\hbar\omega_2}\right)\right]{\cal
  H}_\mu(\hbar\omega_1+\hbar\omega_2+i\Gamma_{e}^{\text{op}})\notag\\
&+&\left[\left(\frac{1}{\hbar\omega_2+i\Gamma_{i}^{\text{op}}}-\frac{1}{\hbar\omega_2}\right)A_1+\left(\frac{1}{\hbar\omega_1+i\Gamma_{i}^{\text{op}}}-\frac{1}{\hbar\omega_1}\right)A_2\right]{\cal
  I}_\mu(\hbar\omega_1+\hbar\omega_2+i\Gamma_{e}^{\text{op}})\notag\\
&+&\left(\frac{1}{\hbar\omega_1+\hbar\omega_2+i\Gamma_{e}^{\text{op}}}-\frac{1}{\hbar\omega_1+\hbar\omega_2+i\Gamma_{i}^{\text{op}}}\right)\left[-A_0\left(\frac{1}{\hbar\omega_1+i\Gamma_{i}^{\text{op}}}+\frac{1}{\hbar\omega_2+i\Gamma_{i}^{\text{op}}}\right)\notag\right.\\
&&\left.+A_3\left(\frac{1}{\hbar\omega_1+i\Gamma_{e}^{\text{op}}}+\frac{1}{\hbar\omega_2+i\Gamma_{e}^{\text{op}}}\right)\right]\frac{1}{|\mu|}
\label{eq:cshn}
\end{eqnarray}
\end{widetext}

As we show below, the values of $\sigma _{J}$ and $\sigma _{E}$ are typically of the
same order of magnitude; hence it is the value of $\Gamma _{i}^{dc}$ that
determines whether the dc-current induced second order conductivity or the
dc-field induced second order conductivity makes the larger contribution to
the dc-induced second order conductivity $\sigma ^{(3)dabc}(\omega
_{1},\omega _{2},0)$. We can get a rough estimation of $\Gamma_i^{\text{dc}}$ from the graphene
mobility $\mu_m$.  The dc limit of the optical conductivity can be obtained from
Eq.~(\ref{eq:sigmaxx})  as 
\begin{equation}
  \sigma^{(1);xx}(0) \approx \frac{4\sigma_0|\mu|}{\pi\Gamma_i^{\text{dc}}}\,.
\end{equation}
The connection between the mobility and conductivity can be
written as $\sigma^{(1);xx}(0) = N_e |e| \mu_m$ with the carrier density $N_e = \frac{|\mu|^2}{\pi(\hbar
  v_F)^2}$ obtained from the linear dispersion. Hence, we get 
\begin{equation}
  \Gamma_i^{\text{dc}} = \frac{\hbar v_F^2 |e|}{|\mu|\mu_m}\,.
\end{equation}
For a sample with mobility $\mu_m=10^3$~cm$^2$/(V$\cdot$s) and
chemical potential $\mu=0.5$~eV, $\Gamma_i^{\text{dc}}$ is about
$10$~meV.  We will see below that for samples with such mobilities the
dc-current induced effects will typically dominate the dc induced second
order conductivity.

\begin{figure*}[t]
\centering
\includegraphics[width=10cm]{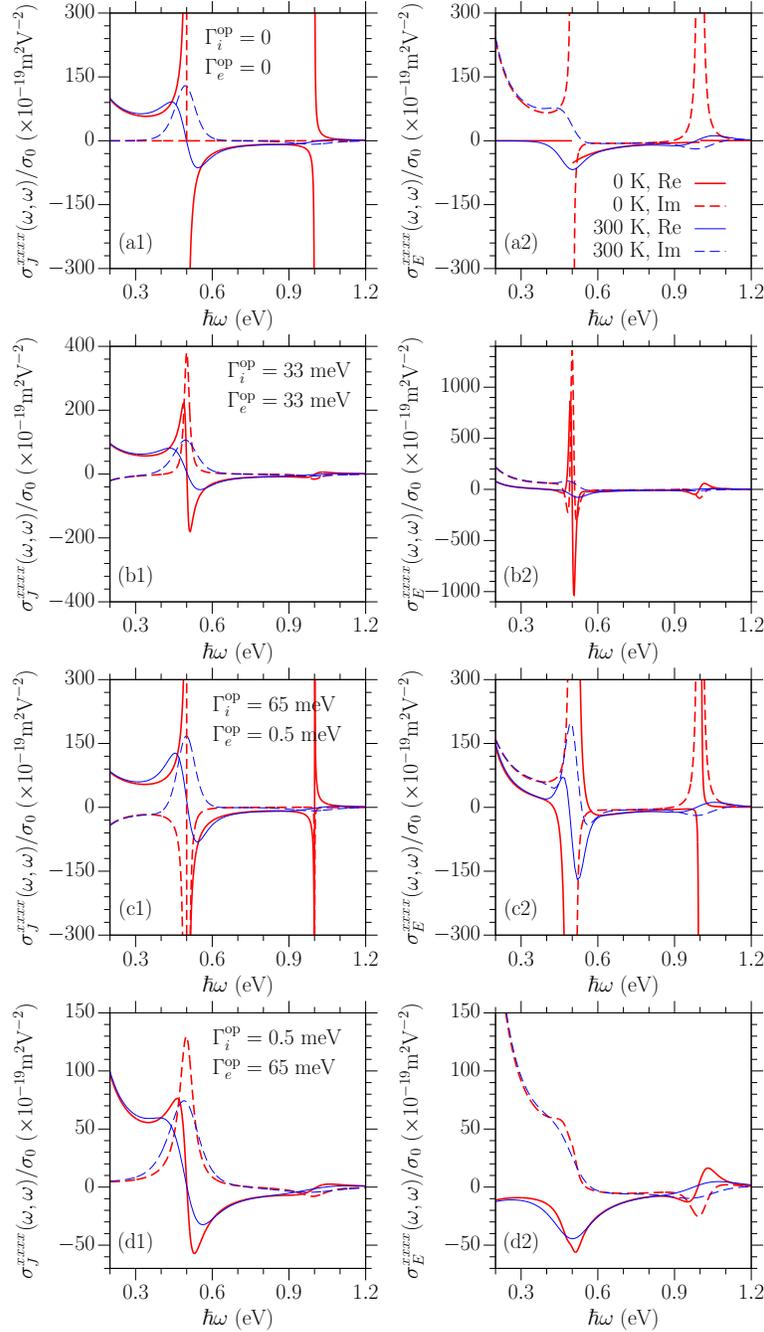}
\caption{(color online) Spectra of $\sigma_{J}^{xxxx}(\omega,\omega)$ (left column) and
$\sigma_{E}^{xxxx}(\omega,\omega)$ (right column) for
different optical relaxation parameters at zero temperature (thick red
curves) and at room temperatures (thin blue curves); $|\mu|=0.5$~eV. Solid (dashed) curves give their real (
imaginary) parts. In calculating $\sigma_E^{xxxx}(\omega,\omega)$,
$\Gamma_e^{\text{dc}}=\Gamma_e^{\text{op}}$ and
$\Gamma_i^{\text{dc}}=0$.}
\label{fig:cshg}
\end{figure*}
\begin{figure*}[t]
\centering
\includegraphics[width=11.8cm]{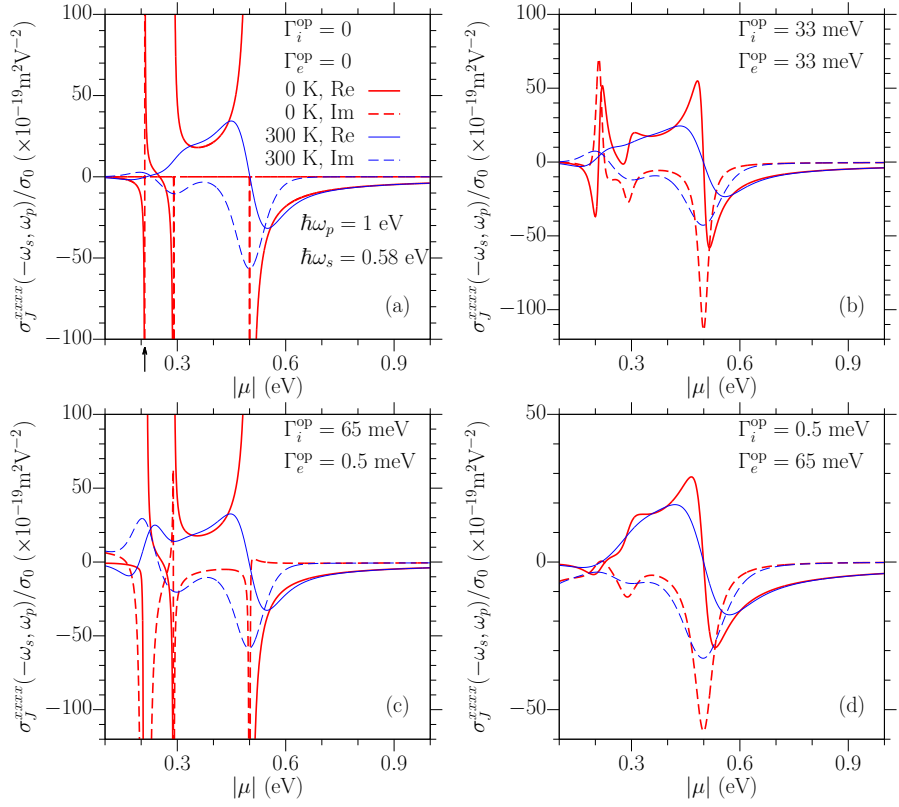}
\caption{(color online) Chemical potential $|\mu|$ dependence of $\sigma_{J}^{dabc}(-\omega_s,\omega_p)$ for
different optical relaxation parameters at zero temperature (thick red
curves) and at room temperatures (thin blue curves); $\hbar\omega_s=0.58$~eV and
$\hbar\omega_p=1$~eV. The real (imaginary) parts are given
by solid (dashed) curves. The arrow in (a) indicates the vertical line there
corresponding to the derivative of a $\delta$ function.}
\label{fig:cdf}
\end{figure*}
\subsection{DC current induced SHG\label{sec:cshg}}
We first consider dc-induced second harmonic generation,
governed by $\sigma ^{(3);dabc}(\omega ,\omega ,0)$. For monochromatic light at
frequency $\omega $, the second order optically induced current is
given by
\begin{equation}
  \bm J^{(3)}_{\text{SHG}}(\omega) = 2\sigma_{\text{SHG}}^A(\omega)
  \bm E_{\omega} \bm E_{\omega}\cdot\bm E_{\text{dc}} + \sigma_{\text{SHG}}^B(\omega)\bm
  E_{\text{dc}} \bm E_{\omega}\cdot \bm E_{\omega}\,.\label{eq:totalcshgj}
\end{equation}
There the two nonzero components are $\sigma_{\text{SHG}}^A(\omega) = 3
\sigma^{(3);xxyy}(\omega,\omega,0)$ and $\sigma_{\text{SHG}}^B(\omega) =
3\sigma^{(3);xyyx}(\omega,\omega,0)$. Correspondingly, each of them
includes two parts:  the dc-current induced second harmonic generation (CSHG)
$\sigma^{(3)}_J$ and the dc-field induced second harmonic
generation (EFISH) $\sigma^{(3)}_E$.

In Fig.~\ref{fig:cshg} we plot the photon energy
dependence of $\sigma_{J/E}^{dabc}(\omega,\omega)$ for $|\mu|=0.5$~eV
and different values of optical relaxation parameters
$\Gamma_e^{\text{op}}$ and $\Gamma_i^{\text{op}}$ at zero and room
temperature. Two resonant peaks appear for both $\sigma_{J}^{dabc}$ and
$\sigma_{E}^{dabc}$, one at $\hbar\omega=|\mu|$ and one at
$2|\mu|$. The first corresponds to the second harmonic resonant with
the onset of one-photon absorption,  and the second to the fundamental resonant with the onset of
one-photon absorption; the first peak leads to a higher response
coefficients than the second. In general, $\sigma_J^{xxxx}$ and $\sigma_E^{xxxx}$ are of the same order of
magnitude. Therefore, in a high mobility graphene sample with a small
$\Gamma_i^{\text{dc}}$, the contribution of
$\sigma_{J}^{dabc}(\omega,\omega)$ dominates $\sigma^{(3);dabc}(\omega,\omega, 0)$ because
of the prefactor ${\text{1eV}}/{\Gamma_i^{\text{dc}}}$ (see Eq.~(\ref{eq:sigma2c})). 

From Eq.~(\ref{eq:cshn}), we see that the first resonance in $\sigma_J^{dabc}(\omega)$ is determined  by ${\cal
  H}_\mu(2\hbar\omega+i\Gamma_e^{\text{op}})$  and
${\cal I}_{\mu}(2\hbar\omega+i\Gamma_e^{\text{op}})$; the other
resonance is determined only by ${\cal H}_\mu(\hbar\omega+i\Gamma_e^{\text{op}})$. Obviously, for both transitions
smaller values of $\Gamma_e^{\text{op}}$ result in a larger value and a sharper
peak. At room temperature, these peaks are broadened and lowered. The
vertical line at $\hbar\omega=|\mu|$ in Fig.~\ref{fig:cshg} (a1) comes
from  $\text{Im}[{\cal H}_\mu(2\hbar\omega+i\Gamma_e^{\text{op}})]$ as
$\Gamma_e^{\text{op}}\to0$, which is proportional to
$\delta(\hbar\omega-|\mu|)$; the peak in Fig.~\ref{fig:cshg} (c1) shows the fine structure of $\text{Im}[{\cal
  I}_\mu(2\hbar\omega+i\Gamma_e^{\text{op}})]$ for
$\Gamma_e^{\text{op}}=0.5$~meV (see
Appendix~\ref{app:tem}). However, it is interesting to note that
these two fine structures undergo important changes at room temperature:
In Fig.~\ref{fig:cshg} (a1), we see that the first fine structure
leads to a peak with broadened width; in Fig.~\ref{fig:cshg} (c1), we
see that the second fine structure leads to a sign change around
$\hbar\omega=|\mu|$ when the temperature increases from zero to room
temperature. 

\subsection{DC current induced difference frequency}
A counterpart of the third order parametric frequency conversion
discussed in Section \ref{sec:pfc} is difference frequency generation which
is, like second harmonic generation, a second order nonlinear effect
that can be induced in graphene when applying a dc field. With a
strong pump at frequency $\omega_p$, difference frequency generation
converts a signal frequency $\omega_s$ to a new frequency $\omega_p -
\omega_s$; the response is determined by
$\sigma^{(3);dabc}(-\omega_s,\omega_p,0)$.  
Similar to dc-induced second harmonic generation, there are current and
electric field contributions to dc-induced difference frequency
generation. As we found in Section \ref{sec:cshg} for dc-induced
second harmonic generation, the current contribution should dominate
the dc-induced difference frequency generation in a high mobility
sample.  As an example, we plot the chemical potential dependence of
$\sigma^{(3);dabc}_J(-\omega_s,\omega_p)$ for different optical
relaxation parameters in Fig.~\ref{fig:cdf} for $\hbar\omega_p=1$~eV (with
a wavelength of about $1.24~\mu$m) and $\hbar\omega_s=0.58$~eV (with a 
wavelength of about $2.1~\mu$m). 
For vanishing optical relaxation parameters
($\Gamma_i^{\text{op}}=\Gamma_e^{\text{op}}=0$), it is clear from
Fig.~\ref{fig:cdf} (a) that there are three resonant transitions in the plotted chemical potential
range: $|\mu_1|=(\hbar\omega_p-\hbar\omega_s)/2=0.21$~eV, $|\mu_2|=\hbar\omega_s/2=0.29$~eV,
and $|\mu_3|=\hbar\omega_p/2=0.5$~eV. Without relaxation, the imaginary
part of the conductivity is always zero except at these three resonant
transitions (shown as vertical lines): the first is given by
$\lim\limits_{\delta\to0}\text{Im}[{\cal
  I}_\mu(\hbar\omega_p-\hbar\omega_s+i\delta)]\propto\frac{d
  }{d |\mu|}\delta(\hbar\omega_p-\hbar\omega_s-|\mu|)$, the other two are
given by $\lim\limits_{\delta\to0}\text{Im}[{\cal
  H}_\mu(-\hbar\omega_s+i\delta)]\propto\delta(\hbar\omega_s-|\mu|)$ and $\lim\limits_{\delta\to0}\text{Im}[{\cal
  H}_\mu(\hbar\omega_p+i\delta)]\propto\delta(\hbar\omega_p-|\mu|)$. With
finite relaxation rates or at finite temperature, the vertical lines are
broadened to structures of finite strength and width.

\subsection{Comparison between calculations and experiments}
Bykov {\it et al.}  \cite{Phys.Rev.B_85_121413_2012_Bykov} observed that SHG radiation from a
graphene/SiO$_2$/Si(001) substrate strongly depends on the applied
current density in the graphene layer, which is attributed to the CSHG
effect of graphene. A similar structure was also studied by An {\it et
  al.}  \cite{NanoLett._13_2104_2013_An,Phys.Rev.B_89_115310_2014_An},
who could measure the
radiation from different locations on the graphene sheet; they interpreted
the result as EFISH, where the electric field is induced by
current-associated trapped charge at the graphene/SiO$_2$
interface. Because of the interface contribution to the SHG
radiation  \cite{Appl.Phys.Lett._95_261910_2009_Dean,Phys.Rev.B_82_125411_2010_Dean,Phys.Rev.B_85_121413_2012_Bykov,NanoLett._13_2104_2013_An,Phys.Rev.B_89_115310_2014_An},
the contribution of the current related SHG from the graphene is hard
to extract.

The best way of measuring the dc-induced second order nonlinearity of
graphene, without any background contribution from interface effects,
would be to mount graphene in a symmetric structure; this can be difficult. However,
within the framework of the experiments of the type that have already
been done, we can suggest a strategy that might help identify the in-plane 
 graphene CSHG(EFISH) by the azimuthal angle dependence of the
 generated signal. For
linearly polarized light with $\bm E_{\omega} =
E_{\omega}\begin{pmatrix}\cos\phi\\\sin\phi \end{pmatrix}
$ and $\bm E_{\text{dc}} =E_{\text{dc}} \begin{pmatrix}\cos\phi_J\\\sin\phi_J
\end{pmatrix}
$, Eq.~(\ref{eq:totalcshgj}) becomes
\begin{eqnarray}
  \bm J^{(3)}_{\text{SHG}}(\omega) &=&  E_{\omega}^2E_{\text{dc}}\Big\{
  \sigma_1(\omega) 
  \begin{pmatrix} \cos(2\phi-\phi_J) \\ \sin(2\phi-\phi_J)
  \end{pmatrix} \notag\\
  &&+ [\sigma_1(\omega) + \sigma_2(\omega)]\begin{pmatrix} \cos\phi_J \\ \sin\phi_J
  \end{pmatrix} \Big\}\,.
\end{eqnarray}
We see that the Cartesian components of the induced current vary as
cosinusoidal functions of $0\phi$ (that is, independent of $\phi )$ and $
2\phi $. In a short-hand notation, we will characterize these as $0\phi$
and $2\phi $ dependences. In most experiments
  \cite{Appl.Phys.Lett._95_261910_2009_Dean,Phys.Rev.B_82_125411_2010_Dean,Phys.Rev.B_85_121413_2012_Bykov,NanoLett._13_2104_2013_An,Phys.Rev.B_89_115310_2014_An},
the graphene sample is mounted on a SiO$_2$/Si substrate, where the interface between SiO$_2$ and
Si gives an interface-induced SHG and the bulk Si gives an electric
quadrupole/magnetic dipole induced SHG. But for different crystal
orientations of the Si substrate, the dependence of the combined
interface and bulk contributions on azimuthal angle will be different   \cite{Phys.Rev.B_35_1129_1987_Sipe}: For the (111) face,
the second harmonic radiation depends on the angle as $0\phi$ and
$3\phi$; for the (001) face, the dependence is $0\phi$ and $4\phi$; while
for the (110) face, the dependence becomes $0\phi$, $2\phi$, and
$4\phi$. Therefore, from the
azimuthal dependence of the SHG signal it might be possible to
distinguish the graphene CSHG (EFISH) from the interface
contributions, for example by putting graphene on top of different
SiO$_2$/Si structures, one with the (111) face of Si normal to the
interface and one with the (001) face normal. Because of the same origin of CSHG and
EFISH in graphene, they would have the same angle dependence, so such
experiments would not help to distinguish between these different
contributions from graphene; but for a heavily doped and high mobility
graphene sample our calculations show that
the CSHG should dominate.

\section{Conclusion and Discussion\label{sec:con}}
Perturbative analyses play a central role in nonlinear optics. Even when
the electrons in a material are treated as independent, and relaxation is
only described phenomenologically, the calculated response tensors that
relate the induced polarization or current to powers of the applied fields
indicate the nonlinear optical effects that are allowed, and point to where
resonances can lead to an interesting dependence on time and
frequency. Often more sophisticated models of the electron dynamics are required, and
sometimes the perturbative framework itself is insufficient to address the
physics of interest. But even then these kinds of perturbative treatments
provide a starting point for more realistic calculations.

In this paper we have provided such a treatment of the nonlinear third order
optical response of doped graphene, with the main goal of investigating the
effects of phenomenological relaxation parameters, finite temperature, and
laser pulse width on the induced currents. We focused on the contributions
of optical transitions around the Dirac points, where the widely used linear
dispersion relation is a good approximation. By solving semiconductor
Bloch equations perturbatively, an analytic expression for general third
order conductivities was obtained at zero temperature, taking different
relaxation parameters for interband and intraband optical transitions. The
nonlinear conductivities at finite temperature were obtained by an
appropriate integration over the chemical potential. The conductivities
show a complicated dependence on photon energy, chemical potential, and the
relaxation parameters.

Even with the inclusion of relaxation we found that the perturbative
approach itself is problematic at vanishing chemical potential, as might be
expected from a similar result in the semiclassical limit   \cite{Europhys.Lett._79_27002_2007_Mikhailov}, except in
the special case that either first- or third-order interband and intraband
relaxation rates are set equal. The perturbative approach adopted is unproblematic for doped graphene
at zero temperature, but is a concern at finite temperature, since thermal
fluctuations always place some electrons or holes near the Dirac points. Yet
numerical calculations of the full semiconductor Bloch equations indicate
that the contribution of such electrons to the full optical response is
small, so this effect does not afflict our results.

We discussed in detail different nonlinear effects, including
third harmonic generation, Kerr effects and two-photon absorption, two-color
coherent current injection, parametric frequency conversion, and dc-current
and -field induced second harmonic generation and difference
frequency generation. The interband relaxation generally broadens and lowers the
resonant peaks, while the intraband relaxation plays an important role in
some of the effects, including two-color coherent current injection and the
dc-current induced second order nonlinearities. At room temperature most
of the resonant structures are smeared out.

We also considered the response of graphene to laser pulses. The optical
response depends in detail on the frequency width of the incident pulse and
the frequency structure of the response tensors.  The two natural limits are
(1) when the frequency structure of the response coefficients is rather flat
on the scale of the frequency width of the incident pulse, and the induced
current follows the injecting pulses, and (2) when there are divergences in
the response coefficients at frequencies close to the real axis, as for
two-color coherent control, the 
a dynamics associated directly with the relaxation processes.

Comparison of our results with experiments is difficult, since in many of the
reported experiments the graphene samples have not been characterized in the
linear regime, and neither the relaxation parameters nor even the chemical
potential have been identified. Results for some nonlinear response
coefficients, such as that describing the Kerr effect, are predicted by our
calculations to be so sensitive to these parameters that we cannot hazard a
comparison of theory to experiment. Yet our results for third harmonic
generation and parametric frequency conversion are insensitive enough to
these parameters that we can conclude our results are about two orders of
magnitude smaller than those extracted from experiments  \cite{NewJ.Phys._16_53014_2014_Cheng}, even with the
adoption of reasonable relaxation parameters. We speculated on the causes
of this disagreement; it is of course early days for both detailed
experimental and theoretical studies of such nonlinear effects in
graphene. But these disagreements may persist, and a long and difficult journey may
be necessary to understand the details of the full nonlinear optical
response of graphene. Even so we can expect that, as in the study of the
nonlinear optical response of other materials, the kind of perturbative
calculation we have presented here will provide a useful port of
embarkation, paving the way to a better physical insight in the
complex nonlinear optical response of graphene.

\acknowledgments
This work has been supported by the EU-FET grant GRAPHENICS (618086),
by the ERC-FP7/2007-2013 grant 336940, by the FWO-Vlaanderen project
G.A002.13N, by the Natural Sciences and Engineering Research Council of
Canada, by VUB-Methusalem, VUB-OZR, and IAP-BELSPO under grant IAP
P7-35.

\appendix
\section{Perturbation solution of Semiconductor Bloch equation\label{app:formula}}
We describe the electronic states in graphene by the tight binding
model employing carbon $2p_z$ orbitals with only nearest neighbor
coupling. Neglecting the overlap between different $p_z$
orbitals, the band structure that results is electron-hole symmetric
with energies $\varepsilon_{+\bm
  k} = -\varepsilon_{-\bm k}$ where $+(-)$ is the band index for
$\pi^{\ast}(\pi)$ bands, and the Berry connections satisfy
$\bm\xi_{s\bar{s}\bm k} = \bm\xi_{\bar{s}s\bm k} \equiv \bm r_{\bm
  k}$ and $\bm \xi_{ss\bm k} = \bm \xi_{\bar{s}\bar{s}\bm k}$ where
$s=\pm$ is the band index and $\bar{s}$ is the index of the band that
is not the $s$ band   \cite{NewJ.Phys._16_53014_2014_Cheng}. Up to
the dipole approximation of light-matter interaction, the SBE in
Eq.~(\ref{eq:kbe1}) can be expanded as 
\begin{eqnarray}
  \hbar\frac{\partial \rho_{ss\bm k}^{(n)}(t)}{\partial t} &=& i e\bm E(t)\cdot\bm
  r_{\bm k}(\rho_{\bar{s}s\bm k}^{(n-1)}-\rho_{s\bar{s}\bm
    k}^{(n-1)})\notag\\
&-&  e\bm
  E(t)\cdot\bm\nabla_{\bm k}\rho_{ss\bm k}^{(n-1)} -\Gamma_{i}^{(n)} \rho_{ss\bm
    k}^{(n)}\,,\notag\\
  \hbar\frac{\partial \rho_{s\bar{s}\bm k}^{(n)}(t)}{\partial t} &=& -i
  s\epsilon_{\bm k}\rho_{s\bar{s}\bm k}^{(n)}(t) +i e\bm E(t)\cdot\bm
  r_{\bm k}(\rho_{\bar{s}\bar{s}\bm k}^{(n-1)}-\rho_{ss\bm
    k}^{(n-1)})\notag\\
&-&  e\bm
  E(t)\cdot\bm\nabla_{\bm k}\rho_{s\bar{s}\bm k}^{(n-1)} -\Gamma_{e}^{(n)}\rho_{s\bar{s}\bm
    k}^{(n)}\,,\label{eq:kb}
\end{eqnarray}
with $\epsilon_{\bm k}=\varepsilon_{+\bm k}-\varepsilon_{-\bm
  k}$. In Eq.~(\ref{eq:kb}) the terms involving $\bm r_{\bm k}$ give the
  interband contribution, and the terms involving $\bm\nabla_{\bm k}$ give
the intraband contribution. Treating the electric field term
perturbatively, the first three terms  are expanded as
\begin{eqnarray}
  \rho_{s_1s_2\bm k}^{(1)}(t) &=&\int
  \frac{d\omega_3}{2\pi} (-e) E_{\omega_3}^c e^{-i\omega_3t}  {\cal P}^{(1);c}_{s_1s_2\bm
    k}(\omega_3) \notag\\
  \rho_{s_1s_2\bm k}^{(2)}(t)&=& \int\frac{d\omega_2d\omega_3}{(2\pi)^2} (-e)^2 E_{\omega_2}^bE_{\omega_3}^ce^{-i\omega_0t}{\cal P}^{(2);bc}_{s_1s_2\bm
    k}(\omega_2,\omega_3) \notag\\
  \rho_{s_1s_2\bm k}^{(3)}(t)&=&
  \int\frac{d\omega_1d\omega_2d\omega_3}{(2\pi)^3} (-e)^3
  E_{\omega_1}^aE_{\omega_2}^bE_{\omega_3}^ce^{-i\omega t}\notag\\
&&\times {\cal P}^{(3);abc}_{s_1s_2\bm
    k}(\omega_1,\omega_2,\omega_3)\,,
\end{eqnarray}
with $\omega_0=\omega_2+\omega_3$ and
$\omega=\omega_1+\omega_0$. By substituting the above expansion into
Eq.~(\ref{eq:kb}), we get the following equations for ${\cal P}^{(i)}$. 

\noindent (1) The linear order terms are determined by 
\begin{eqnarray}
  \nu_3 {\cal P}^{(1);c}_{ss\bm
    k}(\omega_3)  &=& i\frac{\partial n_{s\bm k}}{\partial k_c}\,,\notag\\
  (\vartheta_3-s\epsilon_{\bm k}){\cal
    P}^{(1);c}_{s\bar{s}\bm k}(\omega_3) &=& -s r^c_{\bm k}\Delta
  n_{\bm k}\,.\label{eq:pert1}
\end{eqnarray}
Here we have put $\Delta n_{\bm
  k}=n_{+\bm k}-n_{-\bm k}$. The solutions are 
\begin{equation}
  {\cal P}^{(1);c}_{ss\bm k}(\omega_3) =
  \frac{i}{\nu_3}\frac{\partial n_{s\bm
      k}}{\partial k_c}\,,\quad
  {\cal P}^{(1);c}_{s\bar{s}\bm k}(\omega_3)
  =\frac{-sr_{\bm
      k}^c\Delta n_{\bm
      k}}{\vartheta_3-s\epsilon_{\bm k}}\,,
\end{equation}
which leads to the linear conductivity:
\begin{equation}
  \sigma^{(1);da}(\omega) =
  -e^2\sum_{s_1s_2}\int\frac{d\bm k}{4\pi^2}v_{s_2s_1\bm k}^d {\cal
    P}^{(1);a}_{s_1s_2\bm k}(\omega)\,.
\end{equation}
Here $\bm v_{s_1s_2\bm k}$ are the matrix elements of the velocity
operator, which satisfy $\bm v_{++\bm k}=-\bm v_{--\bm k}$ and $\bm
v_{+-\bm k}=-\bm v_{-+\bm k}$ in the tight binding model we have adopted. 

\noindent (2) The second order terms are determined by
\begin{eqnarray}
  \nu_0{\cal P}^{(2);bc}_{ss\bm k}(\omega_2,\omega_3) &=& r_{\bm
    k}^b[{\cal P}^{(1);c}_{\bar{s}s\bm k}(\omega_3) - {\cal
    P}^{(1);c}_{s\bar{s}\bm k}(\omega_3)] \notag\\
&&+ i\frac{\partial}{\partial
    k_b} {\cal P}^{(1);c}_{ss\bm k}(\omega_3)\,,\notag\\
  (\vartheta_3-s\epsilon_{\bm k}){\cal
    P}^{(2);bc}_{s\bar{s}\bm k}(\omega_2,\omega_3) &=& r^b_{\bm
    k}\left[{\cal P}^{(1);c}_{\bar{s}\bar{s}\bm k}(\omega_3) - {\cal
      P}^{(1);c}_{ss\bm k}(\omega_3)\right] \notag\\
&&+ i \frac{\partial
  }{\partial k_b} {\cal P}^{(1);c}_{s\bar{s}\bm k}(\omega_3)\,,\label{eq:pert2}
\end{eqnarray}
The solutions are
\begin{eqnarray}
{\cal P}^{(2);bc}_{ss\bm
  k}(\omega_2,\omega_3) &=&
\frac{i}{\nu_0}\left[\frac{i}{\nu_3}\frac{\partial^2 n_{s\bm k}}{\partial
    k_b\partial k_c} \right.\notag\\
&&\left.+ s r_{\bm k}^br_{\bm k}^c\Delta n_{\bm k}
  \left(\frac{1}{\vartheta_3+\epsilon_{\bm k}}+\frac{1}{\vartheta_3-\epsilon_{\bm k}}\right)\right]\,,\nonumber\\
{\cal P}^{(2);bc}_{s\bar{s}\bm
  k}(\omega_2,\omega_3)&=&\frac{-is}{\vartheta_0-s\epsilon_{\bm
    k}}\left[\frac{\partial}{\partial
  k_b}\left(\frac{r_{\bm k}^c\Delta n_{\bm
      k}}{\vartheta_3-s\epsilon_{\bm k}}\right)+\frac{r_{\bm
    k}^b}{\nu_3}\frac{\partial \Delta n_{\bm k}}{\partial
  k_c}\right]\,,\notag
\end{eqnarray}
Because the graphene crystal structure is centrosymmetric, its
second order conductivity is zero. 

\noindent (3) The third order terms are determined by 
\begin{eqnarray}
  \nu{\cal P}^{(3);abc}_{ss\bm k}
  &=& r_{\bm
    k}^b[{\cal P}^{(2);bc}_{\bar{s}s\bm k}(\omega_2,\omega_3) - {\cal
    P}^{(2);bc}_{s\bar{s}\bm k}(\omega_2,\omega_3)]\notag\\
  & +& i\frac{\partial}{\partial
    k_b} {\cal P}^{(2);bc}_{ss\bm k}(\omega_2,\omega_3)\,,\notag\\
  (\vartheta-s\epsilon_{\bm k}){\cal
    P}^{(3);abc}_{s\bar{s}\bm k}
  &=& r^b_{\bm
    k}\left[{\cal P}^{(2);bc}_{\bar{s}\bar{s}\bm k}(\omega_2,\omega_3) - {\cal
      P}^{(2);bc}_{ss\bm k}(\omega_2,\omega_3)\right] \notag\\
&+& i \frac{\partial
  }{\partial k_b} {\cal P}^{(2);bc}_{s\bar{s}\bm k}(\omega_2,\omega_3)\,,\label{eq:pert3}
\end{eqnarray}
The frequency dependence of ${\cal P}^{(3);dabc}_{s_1s_2\bm
  k}(\omega_1,\omega_2,\omega_3)$ is implicit. The solutions can be written as
\begin{eqnarray}
{\cal P}^{(3);abc}_{ss\bm
  k}
&=&\frac{1}{\nu\nu_0\nu_3}P^{abc}_{1;s\bm k} + \frac{1}{\nu\nu_0}
P^{abc}_{2;s\bm k}(\vartheta_3) \notag\\
&+& \frac{1}{\nu\nu_3}
P^{abc}_{3;s\bm k}(\vartheta_0) +
 \frac{1}{\nu}
 P^{abc}_{4;s\bm k}(\vartheta_0,\vartheta_3)\,, \notag\\
{\cal P}^{(3);abc}_{s\bar{s}\bm k}
&=& \frac{1}{\nu_0\nu_3} P^{abc}_{5;s\bm k}(\vartheta) +
\frac{1}{\nu_0} P^{abc}_{6;s\bm k}(\vartheta,\vartheta_3) \notag\\
&+& \frac{1}{\nu_3}P^{abc}_{7;s\bm k}(\vartheta,\vartheta_0) + P^{abc}_{8;s\bm k}(\vartheta,\vartheta_0,\vartheta_3)\,,
\end{eqnarray}
Here terms $P^{abc}_{i;s\bm k}$ with $i=1,\cdots,4$ are
related to the populations at band  $s$ and are given by
\begin{eqnarray*}
P^{abc}_{1;s\bm k} &=& -i\frac{\partial^3 n_{s\bm k}}{\partial
  k_a\partial k_b\partial k_c}\,,\\
P^{abc}_{2;s\bm k}(\vartheta_3) &=& - s\frac{\partial}{\partial
  k_a}\left[r_{\bm k}^br_{\bm k}^c\Delta n_{\bm
    k}\left(\frac{1}{\vartheta_3+\epsilon_{\bm k}}+\frac{1}{\vartheta_3-\epsilon_{\bm
        k}}\right)\right] \,,\\
P^{abc}_{3;s\bm k}(\vartheta_0) &=& -sr_{\bm k}^ar_{\bm
  k}^b\left(\frac{1}{\vartheta_0+\epsilon_{\bm
      k}}+\frac{1}{\vartheta_0-\epsilon_{\bm k}}\right)\frac{\partial
  \Delta n_{\bm k}}{\partial k_c} \,,\\
P^{abc}_{4;s\bm k}(\vartheta_0,\vartheta_3) &=& -s r_{\bm k}^a
\left[\frac{1}{\vartheta_0+\epsilon_{\bm k}}\frac{\partial}{\partial
    k_b}\left(\frac{r_{\bm k}^c\Delta n_{\bm
        k}}{\vartheta_3+\epsilon_{\bm k}}\right)\notag\right.\notag\\
&&\left.+\frac{1}{\vartheta_0-\epsilon_{\bm k}}\frac{\partial}{\partial
    k_b}\left(\frac{r_{\bm k}^c\Delta n_{\bm k}}{\vartheta_3-\epsilon_{\bm k}}\right)\right]\,,
\end{eqnarray*}
while the terms $P^{abc}_{i;s\bm k}$ with $i=5,\cdots,8$ are related
to the interband polarization and are given by
\begin{eqnarray*}
P^{abc}_{5;s\bm k}(\vartheta)&=& s\frac{r_{\bm k}^a}{\vartheta-s\epsilon_{\bm
    k}}\frac{\partial^2 \Delta n_{\bm k}}{\partial k_b\partial
  k_c}\,,\\
P^{abc}_{6;s\bm k}(\vartheta,\vartheta_3) &=& -2is\frac{r_{\bm k}^ar_{\bm k}^b r_{\bm k}^c \Delta n_{\bm k}}{\vartheta-s\epsilon_{\bm
    k}}\left(
\frac{1}{\vartheta_3+\epsilon_{\bm k}}+\frac{1}{\vartheta_3-\epsilon_{\bm
    k}}\right)\,,\\
P^{abc}_{7;s\bm k}(\vartheta,\vartheta_0)&=& \frac{i}{\vartheta-s\epsilon_{\bm
    k}}\frac{\partial}{\partial k_a}\left(r_{\bm k}^b \frac{\partial
    \Delta n_{\bm k}}{\partial k_c}\right)\,,\\
P^{abc}_{8;s\bm k}(\vartheta,\vartheta_0,\vartheta_3)&=& \frac{1}{\vartheta-s\epsilon_{\bm
    k}}\frac{\partial }{\partial k_a}\left\{ \frac{1}{\vartheta_0-s\epsilon_{\bm
      k}}\notag\right.\\
&&\left.\times\left[\frac{\partial }{\partial k_b}\left(\frac{r_{\bm k}^c \Delta n_{\bm
    k}}{\vartheta_3-s\epsilon_{\bm k}}\right)\right]\right\}\,.
\end{eqnarray*}
The unsymmetrized third order conductivity which follows from these
terms is
\begin{eqnarray}
  \widetilde{\sigma}^{(3);dabc}(\omega_1,\omega_2,\omega_3) &=&
  -e^4\sum_{s_1s_2}\int\frac{d\bm k}{4\pi^2}v_{s_2s_1\bm k}^d \notag\\
&&\times {\cal
    P}^{(3);abc}_{s_1s_2\bm k}(\omega_1,\omega_2,\omega_3)\,.
\end{eqnarray}
Then we find terms ${\cal S}_i$ in Eq.~(\ref{eq:sigmaall}) are given by
\begin{equation*}
{\cal S}_i^{dabc} = -(i\sigma_3 )^{-1}e^4\int\!\!\frac{d\bm k}{4\pi^2}v_{++\bm k}^d \left(P^{abc}_{i;+\bm
      k}-P^{abc}_{i;-\bm k}\right)\,,  \label{eq:sigma01}
\end{equation*}
 for $i=1,2,3,4$ and 
\begin{equation*}
{\cal S}_i^{dabc} = (i\sigma_3 )^{-1} e^4\int\!\!\frac{d\bm k}{4\pi^2}v_{+-\bm k}^d\left(P^{abc}_{i;+\bm k}-P^{abc}_{i;-\bm k}\right)\,,
  \label{eq:sigma0}
\end{equation*}
for $i=5,6,7,8$.

In this work, we only consider optical transitions around the Dirac points $\bm K=(\bm b_1+2\bm
b_2)/3$ or $\bm K^{\prime}=(\bm b_2+2\bm b_1)/3$ with $\bm b_1$ and
$\bm b_2$ the primitive reciprocal lattice vectors. These two Dirac
cones are connected by the inversion symmetry, and they lead to the same 
contribution to the conductivities we consider, whether linear or
third order. In the following, we
calculate the conductivity around $\bm K$ explicitly and get the total
results by considering both valley degeneracy $g_v=2$ and spin
degeneracy $g_s=2$.  Around the Dirac point $\bm K$, we approximate each
quantity up to its lowest order of $\bm k-\bm K$: The electronic dispersion is $\varepsilon_{s\bm K+\bm
k}=s\hbar v_Fk$, the velocity matrix elements are  $\bm v_{ss(\bm
K+\bm k)}\approx sv_{F}{\bm k}/{k}$ and $\bm v_{s\bar{s}(\bm K+\bm k)}\approx
isv_{F} \bm k\times {\hat{\bm z}}/{k}$, and the interband Berry
connection is $\bm r _{\bm K+\bm k}\approx \bm k \times {\hat{\bm
    z}}/{2k^{2}}$. The linear conductivity that results is given by Eq.~(\ref{eq:sigmaxx}). 
\begin{figure*}[t]
\centering
\includegraphics[width=\linewidth]{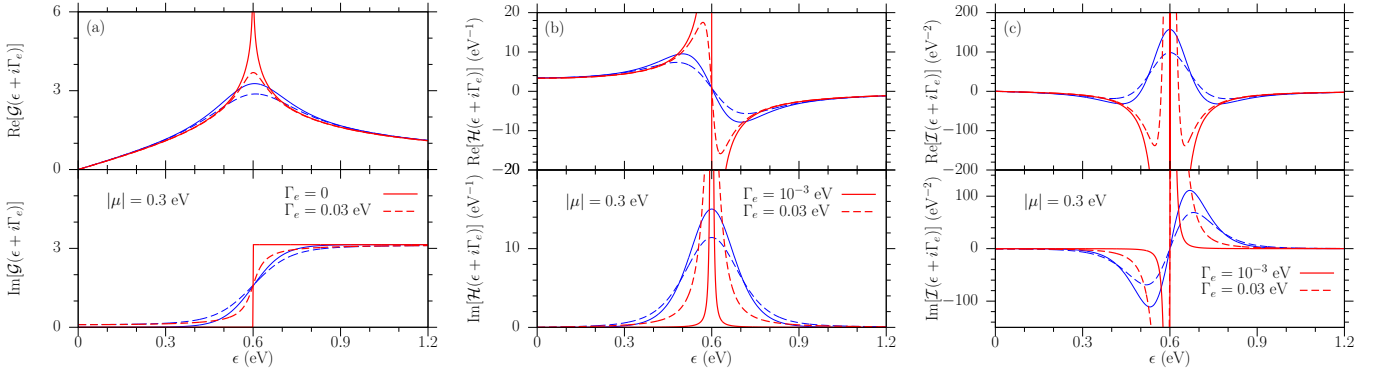}
\caption{(color online) $\epsilon$ dependence of (a) ${\cal G}_\mu(\epsilon+i\Gamma)$
  and $\overline{\cal G}_{\mu;T}(\epsilon+i\Gamma)$, (b)  ${\cal H}_\mu(\epsilon+i\Gamma)$
  and $\overline{\cal H}_{\mu;T}(\epsilon+i\Gamma)$, (c) ${\cal I}_\mu(\epsilon+i\Gamma)$
  and $\overline{\cal I}_{\mu;T}(\epsilon+i\Gamma)$. Solid (dashed) curves
  are for $\Gamma=0$ ($\Gamma=0.03$~eV), and thick red (thin blue) curves are
  for the function without (with) a temperature average. Other parameters
  are $|\mu|=0.3$~eV and $T=300$~K.}
\label{fig:tempghi}
\end{figure*}

In calculating the integrals over $\bm k$ necessary to evaluate the
third order conductivities, we use the relation 
\begin{eqnarray}
  \dfrac{\partial \epsilon_{\bm k}}{\partial k_a} &=& 2v_{++\bm
    k}^a\,, \\
  \dfrac{\partial r_{\bm k}^a}{\partial k_b} &=& \dfrac{2i}{\epsilon_{\bm
      k}^2}(v_{++\bm k}^av_{+-\bm k}^b + v_{+-\bm k}^b v_{++\bm k}^a)\,,\\
  \dfrac{\partial v_{++\bm k}^a}{\partial k_b} &=& -2 \dfrac{v_{+-\bm k}^a v_{+-\bm
      k}^b}{\epsilon_{\bm k}}\,,\\
 \dfrac{\partial v_{+-\bm k}^a}{\partial k_b} &=& -2\dfrac{ v_{++\bm k}^a v_{+-\bm
      k}^b}{\epsilon_{\bm k}}\,,
\end{eqnarray}
to expand the derivatives in ${\cal S}_i$, and find that all the
required integrations over
$\bm k$ can be related to
\begin{eqnarray}
  g_sg_v\frac{e^4}{\Omega} && \int\!\!\frac{d\bm k}{(2\pi)^2} v_{+-\bm k}^d v_{+-\bm k}^a v_{++\bm k}^b v_{++\bm
    k}^c\delta(\epsilon_{\bm k} -\Omega)=
  \frac{1}{4}\sigma_3 A_1
  \,,\notag\\
g_sg_v\frac{e^4}{\Omega}&&\int\!\!\frac{d\bm k}{(2\pi)^2}v_{+-\bm k}^d v_{+-\bm k}^a v_{+-\bm k}^b v_{+-\bm k}^c
  \delta(\epsilon_{\bm k}-\Omega) = 
\frac{1}{4}\sigma_3 A_0  \,,\notag
\end{eqnarray}
Using partial fractions and taking the integral over $\Omega$, we obtain all the
${\cal S}_i$ given in Eq.~(\ref{eq:s1}-\ref{eq:s8}) 

\section{Temperature Effects\label{app:tem}}
In this appendix we discuss how the temperature affects the
contributions to the conductivities from ${\cal G}_\mu(\epsilon+i\Gamma_e)$, ${\cal
  H}_\mu(\epsilon+i\Gamma_e)$, and ${\cal I}_\mu(\epsilon+i\Gamma_e)$.

(1) ${\cal G}_\mu(\epsilon+i\Gamma_e)$: At finite temperature, this is
replaced by $\overline{\cal G}_{\mu;T}(\epsilon+i\Gamma_e) $ with  
\begin{equation}
  \overline{\cal G}_{\mu;T}(\epsilon+i\Gamma_e) = \beta
  \int_{-\infty}^{\infty} dx F_\mu(x,T)[1-F_\mu(x,T)]{\cal G}_x(\epsilon+i\Gamma_e)\,.
\end{equation}
As $\Gamma_e\to 0$, ${\cal G}_{\mu}(\epsilon+i\Gamma_e)$ diverges
logarithmically at $\epsilon=\pm2|\mu|$, while $\overline{\cal
  G}_{\mu;T}(\epsilon+i\Gamma_e)$ is smooth. Both functions are smooth
as $\epsilon+i\Gamma_e\to0$. In
Fig.~\ref{fig:tempghi}~(a) both ${\cal G}_{\mu}(\epsilon+i\Gamma_e)$ and
$\overline{\cal G}_{\mu;T}(\epsilon+i\Gamma_e)$ are plotted for
$\Gamma_e=0$ and $\Gamma_e=0.03$~eV at $|\mu|=0.3$~eV and
$T=300$~K. Moving from zero to finite temperature, a finite $T$ has an effect
similar to the inclusion of relaxation: Both remove the singularity and broaden the
peak and step function. 

(2) ${\cal H}_\mu(\epsilon+i\Gamma_e)$: At finite temperature, this is
replaced by $\overline{\cal H}_{\mu;T}(\epsilon+i\Gamma_e) $ with  
\begin{equation}
  \overline{\cal H}_{\mu;T}(\epsilon+i\Gamma_e) = \beta
  \int_{-\infty}^{\infty} dx F_\mu(x,T)[1-F_\mu(x,T)]{\cal H}_x(\epsilon+i\Gamma_e)\,.
\end{equation}
At $|\mu|\to z_0$ with $2z_0=\epsilon+i\Gamma_e$ for $\epsilon>0$, ${\cal
  H}_{\mu}(\epsilon+i\Gamma_e)$ diverges as
$(|\mu|-z_0)^{-1}$. In
the relaxation free limit for nonzero $\epsilon$ we can write
\begin{equation}
\frac{1}{|\mu|-\epsilon-i\Gamma_e} \stackrel{\Gamma_e\to0}{\longrightarrow}
P.\frac{1}{|\mu|-\epsilon} + i\pi \delta(|\mu|-\epsilon)\,,
\end{equation}
where $P.$ means the integration takes the principal part; thus the imaginary
part of ${\cal H}_{\mu}$ tends to a $\delta$ function. However, both
the real and imaginary parts of $\overline{\cal H}_{\mu;T}$ are smooth
for $\epsilon>0$ or $\epsilon=0$ and $\Gamma_e\neq 0$. For small
$\Gamma_e$, $\overline{\cal H}_{\mu;T}(i\Gamma_e)\propto\ln\Gamma_e$. 

In Fig.~\ref{fig:tempghi}~(b) both ${\cal H}_{\mu}(\epsilon+i\Gamma_e)$ and
$\overline{\cal H}_{\mu;T}(\epsilon+i\Gamma_e)$ are plotted for
$\Gamma_e=10^{-3}$~eV and $\Gamma_e=0.03$~eV at $\mu=0.3$~eV and
$T=300$~K. The inclusion of finite temperature leads to a broadening
of the $\delta$-function-like imaginary part.

(3) ${\cal I}_\mu(\epsilon+i\Gamma_e)$: At finite temperature, this is
replaced by $\overline{\cal I}_{\mu;T}(\epsilon+i\Gamma_e) $ with  
\begin{equation}
  \overline{\cal I}_{\mu;T}(\epsilon+i\Gamma_e) = \beta
  \int_{-\infty}^{\infty} dx F_\mu(x,T)[1-F_\mu(x,T)]{\cal I}_x(\epsilon+i\Gamma_e)\,.
\end{equation}
At $|\mu|\to z_0$ with $2z_0=\epsilon+i\Gamma_e$ for $\epsilon>0$, ${\cal
  I}_{\mu}(\epsilon+i\Gamma_e)$ diverges as
$(|\mu|-z_0)^{-2}$. Around $\epsilon=2|\mu|$, $\text{Re}[{\cal
  I}_{\mu}(\epsilon+i\Gamma_e)]$ has two minima $\sim - \Gamma_e^{-2}/2$ around
$\epsilon\approx 2|\mu|\pm \sqrt{3}\Gamma_e$ and a maximum  $\sim
\Gamma_e^{-2}$ at $\epsilon=2|\mu|$, while $\text{Im}[{\cal
  I}_\mu(\epsilon+i\Gamma_e)]$ has two extrema at $\epsilon\sim2|\mu|\pm
\Gamma_e/\sqrt{3}$ with values $\sim\pm3\sqrt{3}/(8\Gamma_e^2)$. These
indicate that this function varies very fast around $\epsilon=2|\mu|$ for very small $\Gamma_e$.  In a manner
similar to the ${\cal H}$ function, at room temperature $\overline{\cal I}_{\mu;T}$ is a
smooth function with respect to $\epsilon>0$ for any $\Gamma_e\ge0$, and $\overline{\cal
  I}_{\mu;T}(i\Gamma_e)\propto\Gamma_e^{-1}$.

In Fig.~\ref{fig:tempghi}~(c) both ${\cal I}_{\mu}(\epsilon+i\Gamma_e)$ and
$\overline{\cal I}_{\mu;T}(\epsilon+i\Gamma_e)$ are plotted for
$\Gamma_e=10^{-3}$~eV and $\Gamma_e=0.03$~eV at $|\mu|=0.3$~eV and
$T=300$~K. 

\input{graphene_damping.bbl}
\end{document}

%% file: graphene_damping.bbl
%

%% file: graphene_damping.bbl
\begin{thebibliography}{63}%
\makeatletter
\providecommand \@ifxundefined [1]{%
 \@ifx{#1\undefined}
}%
\providecommand \@ifnum [1]{%
 \ifnum #1\expandafter \@firstoftwo
 \else \expandafter \@secondoftwo
 \fi
}%
\providecommand \@ifx [1]{%
 \ifx #1\expandafter \@firstoftwo
 \else \expandafter \@secondoftwo
 \fi
}%
\providecommand \natexlab [1]{#1}%
\providecommand \enquote  [1]{``#1''}%
\providecommand \bibnamefont  [1]{#1}%
\providecommand \bibfnamefont [1]{#1}%
\providecommand \citenamefont [1]{#1}%
\providecommand \href@noop [0]{\@secondoftwo}%
\providecommand \href [0]{\begingroup \@sanitize@url \@href}%
\providecommand \@href[1]{\@@startlink{#1}\@@href}%
\providecommand \@@href[1]{\endgroup#1\@@endlink}%
\providecommand \@sanitize@url [0]{\catcode `\\12\catcode `\$12\catcode
  `\&12\catcode `\#12\catcode `\^12\catcode `\_12\catcode `\%12\relax}%
\providecommand \@@startlink[1]{}%
\providecommand \@@endlink[0]{}%
\providecommand \url  [0]{\begingroup\@sanitize@url \@url }%
\providecommand \@url [1]{\endgroup\@href {#1}{\urlprefix }}%
\providecommand \urlprefix  [0]{URL }%
\providecommand \Eprint [0]{\href }%
\providecommand \doibase [0]{http://dx.doi.org/}%
\providecommand \selectlanguage [0]{\@gobble}%
\providecommand \bibinfo  [0]{\@secondoftwo}%
\providecommand \bibfield  [0]{\@secondoftwo}%
\providecommand \translation [1]{[#1]}%
\providecommand \BibitemOpen [0]{}%
\providecommand \bibitemStop [0]{}%
\providecommand \bibitemNoStop [0]{.\EOS\space}%
\providecommand \EOS [0]{\spacefactor3000\relax}%
\providecommand \BibitemShut  [1]{\csname bibitem#1\endcsname}%
\let\auto@bib@innerbib\@empty
\bibitem [{\citenamefont {Castro~Neto}\ \emph {et~al.}(2009)\citenamefont
  {Castro~Neto}, \citenamefont {Guinea}, \citenamefont {Peres}, \citenamefont
  {Novoselov},\ and\ \citenamefont
  {Geim}}]{Rev.Mod.Phys._81_109_2009_CastroNeto}%
  \BibitemOpen
  \bibfield  {author} {\bibinfo {author} {\bibfnamefont {A.~H.}\ \bibnamefont
  {Castro~Neto}}, \bibinfo {author} {\bibfnamefont {F.}~\bibnamefont {Guinea}},
  \bibinfo {author} {\bibfnamefont {N.~M.~R.}\ \bibnamefont {Peres}}, \bibinfo
  {author} {\bibfnamefont {K.~S.}\ \bibnamefont {Novoselov}}, \ and\ \bibinfo
  {author} {\bibfnamefont {A.~K.}\ \bibnamefont {Geim}},\ }\href {\doibase
  10.1103/RevModPhys.81.109} {\bibfield  {journal} {\bibinfo  {journal} {Rev.
  Mod. Phys.}\ }\textbf {\bibinfo {volume} {81}},\ \bibinfo {pages} {109}
  (\bibinfo {year} {2009})}\BibitemShut {NoStop}%
\bibitem [{\citenamefont {Bonaccorso}\ \emph {et~al.}(2010)\citenamefont
  {Bonaccorso}, \citenamefont {Sun}, \citenamefont {Hasan},\ and\ \citenamefont
  {Ferrari}}]{Nat.Photon._4_611_2010_Bonaccorso}%
  \BibitemOpen
  \bibfield  {author} {\bibinfo {author} {\bibfnamefont {F.}~\bibnamefont
  {Bonaccorso}}, \bibinfo {author} {\bibfnamefont {Z.}~\bibnamefont {Sun}},
  \bibinfo {author} {\bibfnamefont {T.}~\bibnamefont {Hasan}}, \ and\ \bibinfo
  {author} {\bibfnamefont {A.~C.}\ \bibnamefont {Ferrari}},\ }\href
  {http://dx.doi.org/10.1038/nphoton.2010.186} {\bibfield  {journal} {\bibinfo
  {journal} {Nat. Photon.}\ }\textbf {\bibinfo {volume} {4}},\ \bibinfo {pages}
  {611} (\bibinfo {year} {2010})}\BibitemShut {NoStop}%
\bibitem [{\citenamefont {Gu}\ \emph {et~al.}(2012)\citenamefont {Gu},
  \citenamefont {Petrone}, \citenamefont {McMillan}, \citenamefont {van~der
  Zande}, \citenamefont {Yu}, \citenamefont {Lo}, \citenamefont {Kwong},
  \citenamefont {Hone},\ and\ \citenamefont
  {Wong}}]{Nat.Photon._6_554_2012_Gu}%
  \BibitemOpen
  \bibfield  {author} {\bibinfo {author} {\bibfnamefont {T.}~\bibnamefont
  {Gu}}, \bibinfo {author} {\bibfnamefont {N.}~\bibnamefont {Petrone}},
  \bibinfo {author} {\bibfnamefont {J.~F.}\ \bibnamefont {McMillan}}, \bibinfo
  {author} {\bibfnamefont {A.}~\bibnamefont {van~der Zande}}, \bibinfo {author}
  {\bibfnamefont {M.}~\bibnamefont {Yu}}, \bibinfo {author} {\bibfnamefont
  {G.~Q.}\ \bibnamefont {Lo}}, \bibinfo {author} {\bibfnamefont {D.~L.}\
  \bibnamefont {Kwong}}, \bibinfo {author} {\bibfnamefont {J.}~\bibnamefont
  {Hone}}, \ and\ \bibinfo {author} {\bibfnamefont {C.~W.}\ \bibnamefont
  {Wong}},\ }\href {http://dx.doi.org/10.1038/nphoton.2012.147} {\bibfield
  {journal} {\bibinfo  {journal} {Nat. Photon.}\ }\textbf {\bibinfo {volume}
  {6}},\ \bibinfo {pages} {554} (\bibinfo {year} {2012})}\BibitemShut {NoStop}%
\bibitem [{\citenamefont {Glazov}\ and\ \citenamefont
  {Ganichev}(2014)}]{Phys.Rep._535_101_2014_Glazov}%
  \BibitemOpen
  \bibfield  {author} {\bibinfo {author} {\bibfnamefont {M.}~\bibnamefont
  {Glazov}}\ and\ \bibinfo {author} {\bibfnamefont {S.}~\bibnamefont
  {Ganichev}},\ }\href {\doibase 10.1016/j.physrep.2013.10.003} {\bibfield
  {journal} {\bibinfo  {journal} {Phys. Rep.}\ }\textbf {\bibinfo {volume}
  {535}},\ \bibinfo {pages} {101} (\bibinfo {year} {2014})}\BibitemShut
  {NoStop}%
\bibitem [{\citenamefont {Cheng}\ \emph
  {et~al.}(2014{\natexlab{a}})\citenamefont {Cheng}, \citenamefont
  {Vermeulen},\ and\ \citenamefont {Sipe}}]{NewJ.Phys._16_53014_2014_Cheng}%
  \BibitemOpen
  \bibfield  {author} {\bibinfo {author} {\bibfnamefont {J.~L.}\ \bibnamefont
  {Cheng}}, \bibinfo {author} {\bibfnamefont {N.}~\bibnamefont {Vermeulen}}, \
  and\ \bibinfo {author} {\bibfnamefont {J.~E.}\ \bibnamefont {Sipe}},\ }\href
  {\doibase 10.1088/1367-2630/16/5/053014} {\bibfield  {journal} {\bibinfo
  {journal} {New J. Phys.}\ }\textbf {\bibinfo {volume} {16}},\ \bibinfo
  {pages} {053014} (\bibinfo {year} {2014}{\natexlab{a}})}\BibitemShut
  {NoStop}%
\bibitem [{\citenamefont {Cheng}\ \emph
  {et~al.}(2014{\natexlab{b}})\citenamefont {Cheng}, \citenamefont
  {Vermeulen},\ and\ \citenamefont {Sipe}}]{Opt.Express_22_15868_2014_Cheng}%
  \BibitemOpen
  \bibfield  {author} {\bibinfo {author} {\bibfnamefont {J.~L.}\ \bibnamefont
  {Cheng}}, \bibinfo {author} {\bibfnamefont {N.}~\bibnamefont {Vermeulen}}, \
  and\ \bibinfo {author} {\bibfnamefont {J.~E.}\ \bibnamefont {Sipe}},\ }\href
  {\doibase 10.1364/OE.22.015868} {\bibfield  {journal} {\bibinfo  {journal}
  {Opt. Express}\ }\textbf {\bibinfo {volume} {22}},\ \bibinfo {pages} {15868}
  (\bibinfo {year} {2014}{\natexlab{b}})}\BibitemShut {NoStop}%
\bibitem [{\citenamefont {Novoselov}\ \emph {et~al.}(2004)\citenamefont
  {Novoselov}, \citenamefont {Geim}, \citenamefont {Morozov}, \citenamefont
  {Jiang}, \citenamefont {Zhang}, \citenamefont {Dubonos}, \citenamefont
  {Grigorieva},\ and\ \citenamefont {Firsov}}]{Science_306_666_2004_Novoselov}%
  \BibitemOpen
  \bibfield  {author} {\bibinfo {author} {\bibfnamefont {K.~S.}\ \bibnamefont
  {Novoselov}}, \bibinfo {author} {\bibfnamefont {A.~K.}\ \bibnamefont {Geim}},
  \bibinfo {author} {\bibfnamefont {S.~V.}\ \bibnamefont {Morozov}}, \bibinfo
  {author} {\bibfnamefont {D.}~\bibnamefont {Jiang}}, \bibinfo {author}
  {\bibfnamefont {Y.}~\bibnamefont {Zhang}}, \bibinfo {author} {\bibfnamefont
  {S.~V.}\ \bibnamefont {Dubonos}}, \bibinfo {author} {\bibfnamefont {I.~V.}\
  \bibnamefont {Grigorieva}}, \ and\ \bibinfo {author} {\bibfnamefont {A.~A.}\
  \bibnamefont {Firsov}},\ }\href {\doibase 10.1126/science.1102896} {\bibfield
   {journal} {\bibinfo  {journal} {Science}\ }\textbf {\bibinfo {volume}
  {306}},\ \bibinfo {pages} {666} (\bibinfo {year} {2004})}\BibitemShut
  {NoStop}%
\bibitem [{\citenamefont {Wang}\ \emph {et~al.}(2008)\citenamefont {Wang},
  \citenamefont {Zhang}, \citenamefont {Tian}, \citenamefont {Girit},
  \citenamefont {Zettl}, \citenamefont {Crommie},\ and\ \citenamefont
  {Shen}}]{Science_320_206_2008_Wang}%
  \BibitemOpen
  \bibfield  {author} {\bibinfo {author} {\bibfnamefont {F.}~\bibnamefont
  {Wang}}, \bibinfo {author} {\bibfnamefont {Y.}~\bibnamefont {Zhang}},
  \bibinfo {author} {\bibfnamefont {C.}~\bibnamefont {Tian}}, \bibinfo {author}
  {\bibfnamefont {C.}~\bibnamefont {Girit}}, \bibinfo {author} {\bibfnamefont
  {A.}~\bibnamefont {Zettl}}, \bibinfo {author} {\bibfnamefont
  {M.}~\bibnamefont {Crommie}}, \ and\ \bibinfo {author} {\bibfnamefont
  {Y.~R.}\ \bibnamefont {Shen}},\ }\href {\doibase 10.1126/science.1152793}
  {\bibfield  {journal} {\bibinfo  {journal} {Science}\ }\textbf {\bibinfo
  {volume} {320}},\ \bibinfo {pages} {206} (\bibinfo {year}
  {2008})}\BibitemShut {NoStop}%
\bibitem [{\citenamefont {Liu}\ \emph {et~al.}(2011{\natexlab{a}})\citenamefont
  {Liu}, \citenamefont {Liu},\ and\ \citenamefont
  {Zhu}}]{J.Mater.Chem._21_3335_2011_Liu}%
  \BibitemOpen
  \bibfield  {author} {\bibinfo {author} {\bibfnamefont {H.}~\bibnamefont
  {Liu}}, \bibinfo {author} {\bibfnamefont {Y.}~\bibnamefont {Liu}}, \ and\
  \bibinfo {author} {\bibfnamefont {D.}~\bibnamefont {Zhu}},\ }\href {\doibase
  10.1039/c0jm02922j} {\bibfield  {journal} {\bibinfo  {journal} {J. Mater.
  Chem.}\ }\textbf {\bibinfo {volume} {21}},\ \bibinfo {pages} {3335} (\bibinfo
  {year} {2011}{\natexlab{a}})}\BibitemShut {NoStop}%
\bibitem [{\citenamefont {W\"ulbern}\ \emph {et~al.}(2010)\citenamefont
  {W\"ulbern}, \citenamefont {Prorok}, \citenamefont {Hampe}, \citenamefont
  {Petrov}, \citenamefont {Eich}, \citenamefont {Luo}, \citenamefont {Jen},
  \citenamefont {Jenett},\ and\ \citenamefont
  {Jacob}}]{Opt.Lett._35_2753_2010_Wuelbern}%
  \BibitemOpen
  \bibfield  {author} {\bibinfo {author} {\bibfnamefont {J.~H.}\ \bibnamefont
  {W\"ulbern}}, \bibinfo {author} {\bibfnamefont {S.}~\bibnamefont {Prorok}},
  \bibinfo {author} {\bibfnamefont {J.}~\bibnamefont {Hampe}}, \bibinfo
  {author} {\bibfnamefont {A.}~\bibnamefont {Petrov}}, \bibinfo {author}
  {\bibfnamefont {M.}~\bibnamefont {Eich}}, \bibinfo {author} {\bibfnamefont
  {J.}~\bibnamefont {Luo}}, \bibinfo {author} {\bibfnamefont {A.~K.-Y.}\
  \bibnamefont {Jen}}, \bibinfo {author} {\bibfnamefont {M.}~\bibnamefont
  {Jenett}}, \ and\ \bibinfo {author} {\bibfnamefont {A.}~\bibnamefont
  {Jacob}},\ }\href {\doibase 10.1364/ol.35.002753} {\bibfield  {journal}
  {\bibinfo  {journal} {Opt. Lett.}\ }\textbf {\bibinfo {volume} {35}},\
  \bibinfo {pages} {2753} (\bibinfo {year} {2010})}\BibitemShut {NoStop}%
\bibitem [{\citenamefont {Matheisen}\ \emph {et~al.}(2014)\citenamefont
  {Matheisen}, \citenamefont {Waldow}, \citenamefont {Chmielak}, \citenamefont
  {Sawallich}, \citenamefont {Wahlbrink}, \citenamefont {Bolten}, \citenamefont
  {Nagel},\ and\ \citenamefont {Kurz}}]{Opt.Express_22_5252_2014_Matheisen}%
  \BibitemOpen
  \bibfield  {author} {\bibinfo {author} {\bibfnamefont {C.}~\bibnamefont
  {Matheisen}}, \bibinfo {author} {\bibfnamefont {M.}~\bibnamefont {Waldow}},
  \bibinfo {author} {\bibfnamefont {B.}~\bibnamefont {Chmielak}}, \bibinfo
  {author} {\bibfnamefont {S.}~\bibnamefont {Sawallich}}, \bibinfo {author}
  {\bibfnamefont {T.}~\bibnamefont {Wahlbrink}}, \bibinfo {author}
  {\bibfnamefont {J.}~\bibnamefont {Bolten}}, \bibinfo {author} {\bibfnamefont
  {M.}~\bibnamefont {Nagel}}, \ and\ \bibinfo {author} {\bibfnamefont
  {H.}~\bibnamefont {Kurz}},\ }\href {\doibase 10.1364/oe.22.005252} {\bibfield
   {journal} {\bibinfo  {journal} {Opt. Express}\ }\textbf {\bibinfo {volume}
  {22}},\ \bibinfo {pages} {5252} (\bibinfo {year} {2014})}\BibitemShut
  {NoStop}%
\bibitem [{\citenamefont {Ironside}\ \emph {et~al.}(1993)\citenamefont
  {Ironside}, \citenamefont {Aitchison},\ and\ \citenamefont
  {Arnold}}]{IEEEJ.QuantumElectron._29_2650_1993_Ironside}%
  \BibitemOpen
  \bibfield  {author} {\bibinfo {author} {\bibfnamefont {C.}~\bibnamefont
  {Ironside}}, \bibinfo {author} {\bibfnamefont {J.}~\bibnamefont {Aitchison}},
  \ and\ \bibinfo {author} {\bibfnamefont {J.}~\bibnamefont {Arnold}},\ }\href
  {\doibase 10.1109/3.250387} {\bibfield  {journal} {\bibinfo  {journal} {IEEE
  J. Quantum Electron.}\ }\textbf {\bibinfo {volume} {29}},\ \bibinfo {pages}
  {2650} (\bibinfo {year} {1993})}\BibitemShut {NoStop}%
\bibitem [{\citenamefont {Liu}\ \emph {et~al.}(2014)\citenamefont {Liu},
  \citenamefont {Wang},\ and\ \citenamefont
  {Fang}}]{Appl.Phys.Lett._104_111114_2014_Liu}%
  \BibitemOpen
  \bibfield  {author} {\bibinfo {author} {\bibfnamefont {W.}~\bibnamefont
  {Liu}}, \bibinfo {author} {\bibfnamefont {L.}~\bibnamefont {Wang}}, \ and\
  \bibinfo {author} {\bibfnamefont {C.}~\bibnamefont {Fang}},\ }\href {\doibase
  10.1063/1.4869466} {\bibfield  {journal} {\bibinfo  {journal} {Appl. Phys.
  Lett.}\ }\textbf {\bibinfo {volume} {104}},\ \bibinfo {pages} {111114}
  (\bibinfo {year} {2014})}\BibitemShut {NoStop}%
\bibitem [{\citenamefont {Gandomkar}\ and\ \citenamefont
  {Ahmadi}(2011)}]{Opt.Lett._36_3825_2011_Gandomkar}%
  \BibitemOpen
  \bibfield  {author} {\bibinfo {author} {\bibfnamefont {M.}~\bibnamefont
  {Gandomkar}}\ and\ \bibinfo {author} {\bibfnamefont {V.}~\bibnamefont
  {Ahmadi}},\ }\href {\doibase 10.1364/ol.36.003825} {\bibfield  {journal}
  {\bibinfo  {journal} {Opt. Lett.}\ }\textbf {\bibinfo {volume} {36}},\
  \bibinfo {pages} {3825} (\bibinfo {year} {2011})}\BibitemShut {NoStop}%
\bibitem [{\citenamefont {Hagan}\ \emph {et~al.}(1994)\citenamefont {Hagan},
  \citenamefont {Wang}, \citenamefont {Stegeman}, \citenamefont {Van~Stryland},
  \citenamefont {Sheik-Bahae},\ and\ \citenamefont
  {Assanto}}]{Opt.Lett._19_1305_1994_Hagan}%
  \BibitemOpen
  \bibfield  {author} {\bibinfo {author} {\bibfnamefont {D.~J.}\ \bibnamefont
  {Hagan}}, \bibinfo {author} {\bibfnamefont {Z.}~\bibnamefont {Wang}},
  \bibinfo {author} {\bibfnamefont {G.}~\bibnamefont {Stegeman}}, \bibinfo
  {author} {\bibfnamefont {E.~W.}\ \bibnamefont {Van~Stryland}}, \bibinfo
  {author} {\bibfnamefont {M.}~\bibnamefont {Sheik-Bahae}}, \ and\ \bibinfo
  {author} {\bibfnamefont {G.}~\bibnamefont {Assanto}},\ }\href {\doibase
  10.1364/ol.19.001305} {\bibfield  {journal} {\bibinfo  {journal} {Opt.
  Lett.}\ }\textbf {\bibinfo {volume} {19}},\ \bibinfo {pages} {1305} (\bibinfo
  {year} {1994})}\BibitemShut {NoStop}%
\bibitem [{\citenamefont {Ren}\ \emph {et~al.}(2013)\citenamefont {Ren},
  \citenamefont {Zhong}, \citenamefont {Chen},\ and\ \citenamefont
  {Li}}]{ChinesePhys.Lett._30_97301_2013_Ren}%
  \BibitemOpen
  \bibfield  {author} {\bibinfo {author} {\bibfnamefont {M.-L.}\ \bibnamefont
  {Ren}}, \bibinfo {author} {\bibfnamefont {X.-L.}\ \bibnamefont {Zhong}},
  \bibinfo {author} {\bibfnamefont {B.-Q.}\ \bibnamefont {Chen}}, \ and\
  \bibinfo {author} {\bibfnamefont {Z.-Y.}\ \bibnamefont {Li}},\ }\href
  {\doibase 10.1088/0256-307x/30/9/097301} {\bibfield  {journal} {\bibinfo
  {journal} {Chinese Phys. Lett.}\ }\textbf {\bibinfo {volume} {30}},\ \bibinfo
  {pages} {097301} (\bibinfo {year} {2013})}\BibitemShut {NoStop}%
\bibitem [{\citenamefont {Dean}\ and\ \citenamefont {van
  Driel}(2009)}]{Appl.Phys.Lett._95_261910_2009_Dean}%
  \BibitemOpen
  \bibfield  {author} {\bibinfo {author} {\bibfnamefont {J.~J.}\ \bibnamefont
  {Dean}}\ and\ \bibinfo {author} {\bibfnamefont {H.~M.}\ \bibnamefont {van
  Driel}},\ }\href {\doibase 10.1063/1.3275740} {\bibfield  {journal} {\bibinfo
   {journal} {Appl. Phys. Lett.}\ }\textbf {\bibinfo {volume} {95}},\ \bibinfo
  {eid} {261910} (\bibinfo {year} {2009})}\BibitemShut {NoStop}%
\bibitem [{\citenamefont {Dean}\ and\ \citenamefont {van
  Driel}(2010)}]{Phys.Rev.B_82_125411_2010_Dean}%
  \BibitemOpen
  \bibfield  {author} {\bibinfo {author} {\bibfnamefont {J.~J.}\ \bibnamefont
  {Dean}}\ and\ \bibinfo {author} {\bibfnamefont {H.~M.}\ \bibnamefont {van
  Driel}},\ }\href {\doibase 10.1103/physrevb.82.125411} {\bibfield  {journal}
  {\bibinfo  {journal} {Phys. Rev. B}\ }\textbf {\bibinfo {volume} {82}},\
  \bibinfo {pages} {125411} (\bibinfo {year} {2010})}\BibitemShut {NoStop}%
\bibitem [{\citenamefont {Bykov}\ \emph {et~al.}(2012)\citenamefont {Bykov},
  \citenamefont {Murzina}, \citenamefont {Rybin},\ and\ \citenamefont
  {Obraztsova}}]{Phys.Rev.B_85_121413_2012_Bykov}%
  \BibitemOpen
  \bibfield  {author} {\bibinfo {author} {\bibfnamefont {A.~Y.}\ \bibnamefont
  {Bykov}}, \bibinfo {author} {\bibfnamefont {T.~V.}\ \bibnamefont {Murzina}},
  \bibinfo {author} {\bibfnamefont {M.~G.}\ \bibnamefont {Rybin}}, \ and\
  \bibinfo {author} {\bibfnamefont {E.~D.}\ \bibnamefont {Obraztsova}},\ }\href
  {\doibase 10.1103/PhysRevB.85.121413} {\bibfield  {journal} {\bibinfo
  {journal} {Phys. Rev. B}\ }\textbf {\bibinfo {volume} {85}},\ \bibinfo
  {pages} {121413} (\bibinfo {year} {2012})}\BibitemShut {NoStop}%
\bibitem [{\citenamefont {An}\ \emph {et~al.}(2013)\citenamefont {An},
  \citenamefont {Nelson}, \citenamefont {Lee},\ and\ \citenamefont
  {Diebold}}]{NanoLett._13_2104_2013_An}%
  \BibitemOpen
  \bibfield  {author} {\bibinfo {author} {\bibfnamefont {Y.~Q.}\ \bibnamefont
  {An}}, \bibinfo {author} {\bibfnamefont {F.}~\bibnamefont {Nelson}}, \bibinfo
  {author} {\bibfnamefont {J.~U.}\ \bibnamefont {Lee}}, \ and\ \bibinfo
  {author} {\bibfnamefont {A.~C.}\ \bibnamefont {Diebold}},\ }\href {\doibase
  10.1021/nl4004514} {\bibfield  {journal} {\bibinfo  {journal} {Nano Lett.}\
  }\textbf {\bibinfo {volume} {13}},\ \bibinfo {pages} {2104} (\bibinfo {year}
  {2013})}\BibitemShut {NoStop}%
\bibitem [{\citenamefont {An}\ \emph {et~al.}(2014)\citenamefont {An},
  \citenamefont {Rowe}, \citenamefont {Dougherty}, \citenamefont {Lee},\ and\
  \citenamefont {Diebold}}]{Phys.Rev.B_89_115310_2014_An}%
  \BibitemOpen
  \bibfield  {author} {\bibinfo {author} {\bibfnamefont {Y.~Q.}\ \bibnamefont
  {An}}, \bibinfo {author} {\bibfnamefont {J.~E.}\ \bibnamefont {Rowe}},
  \bibinfo {author} {\bibfnamefont {D.~B.}\ \bibnamefont {Dougherty}}, \bibinfo
  {author} {\bibfnamefont {J.~U.}\ \bibnamefont {Lee}}, \ and\ \bibinfo
  {author} {\bibfnamefont {A.~C.}\ \bibnamefont {Diebold}},\ }\href {\doibase
  10.1103/physrevb.89.115310} {\bibfield  {journal} {\bibinfo  {journal} {Phys.
  Rev. B}\ }\textbf {\bibinfo {volume} {89}},\ \bibinfo {pages} {115310}
  (\bibinfo {year} {2014})}\BibitemShut {NoStop}%
\bibitem [{\citenamefont {Sipe}\ \emph {et~al.}(1987)\citenamefont {Sipe},
  \citenamefont {Moss},\ and\ \citenamefont {van
  Driel}}]{Phys.Rev.B_35_1129_1987_Sipe}%
  \BibitemOpen
  \bibfield  {author} {\bibinfo {author} {\bibfnamefont {J.}~\bibnamefont
  {Sipe}}, \bibinfo {author} {\bibfnamefont {D.}~\bibnamefont {Moss}}, \ and\
  \bibinfo {author} {\bibfnamefont {H.}~\bibnamefont {van Driel}},\ }\href
  {\doibase 10.1103/physrevb.35.1129} {\bibfield  {journal} {\bibinfo
  {journal} {Phys. Rev. B}\ }\textbf {\bibinfo {volume} {35}},\ \bibinfo
  {pages} {1129–1141} (\bibinfo {year} {1987})}\BibitemShut {NoStop}%
\bibitem [{\citenamefont
  {Mikhailov}(2007)}]{Europhys.Lett._79_27002_2007_Mikhailov}%
  \BibitemOpen
  \bibfield  {author} {\bibinfo {author} {\bibfnamefont {S.~A.}\ \bibnamefont
  {Mikhailov}},\ }\href {http://stacks.iop.org/0295-5075/79/i=2/a=27002}
  {\bibfield  {journal} {\bibinfo  {journal} {Europhys. Lett.}\ }\textbf
  {\bibinfo {volume} {79}},\ \bibinfo {pages} {27002} (\bibinfo {year}
  {2007})}\BibitemShut {NoStop}%
\bibitem [{\citenamefont {Mikhailov}\ and\ \citenamefont
  {Ziegler}(2008)}]{J.Phys.Condens.Matter_20_384204_Mikhailov}%
  \BibitemOpen
  \bibfield  {author} {\bibinfo {author} {\bibfnamefont {S.~A.}\ \bibnamefont
  {Mikhailov}}\ and\ \bibinfo {author} {\bibfnamefont {K.}~\bibnamefont
  {Ziegler}},\ }\href {http://stacks.iop.org/0953-8984/20/i=38/a=384204}
  {\bibfield  {journal} {\bibinfo  {journal} {J. Phys. Condens. Matter}\
  }\textbf {\bibinfo {volume} {20}},\ \bibinfo {pages} {384204} (\bibinfo
  {year} {2008})}\BibitemShut {NoStop}%
\bibitem [{\citenamefont {Glazov}(2011)}]{JETPLett._93_366_2011_Glazov}%
  \BibitemOpen
  \bibfield  {author} {\bibinfo {author} {\bibfnamefont {M.}~\bibnamefont
  {Glazov}},\ }\href {\doibase 10.1134/S0021364011070046} {\bibfield  {journal}
  {\bibinfo  {journal} {JETP Lett.}\ }\textbf {\bibinfo {volume} {93}},\
  \bibinfo {pages} {366} (\bibinfo {year} {2011})}\BibitemShut {NoStop}%
\bibitem [{\citenamefont
  {Mikhailov}(2011)}]{Phys.Rev.B_84_45432_2011_Mikhailov}%
  \BibitemOpen
  \bibfield  {author} {\bibinfo {author} {\bibfnamefont {S.~A.}\ \bibnamefont
  {Mikhailov}},\ }\href {\doibase 10.1103/physrevb.84.045432} {\bibfield
  {journal} {\bibinfo  {journal} {Phys. Rev. B}\ }\textbf {\bibinfo {volume}
  {84}},\ \bibinfo {pages} {045432} (\bibinfo {year} {2011})}\BibitemShut
  {NoStop}%
\bibitem [{\citenamefont {Lin}\ \emph {et~al.}(2014)\citenamefont {Lin},
  \citenamefont {Weng}, \citenamefont {Lyu}, \citenamefont {Tsai},\ and\
  \citenamefont {Su}}]{Appl.Phys.Lett._105_151605_2014_Lin}%
  \BibitemOpen
  \bibfield  {author} {\bibinfo {author} {\bibfnamefont {K.-H.}\ \bibnamefont
  {Lin}}, \bibinfo {author} {\bibfnamefont {S.-W.}\ \bibnamefont {Weng}},
  \bibinfo {author} {\bibfnamefont {P.-W.}\ \bibnamefont {Lyu}}, \bibinfo
  {author} {\bibfnamefont {T.-R.}\ \bibnamefont {Tsai}}, \ and\ \bibinfo
  {author} {\bibfnamefont {W.-B.}\ \bibnamefont {Su}},\ }\href {\doibase
  10.1063/1.4898065} {\bibfield  {journal} {\bibinfo  {journal} {Appl. Phys.
  Lett.}\ }\textbf {\bibinfo {volume} {105}},\ \bibinfo {pages} {151605}
  (\bibinfo {year} {2014})}\BibitemShut {NoStop}%
\bibitem [{\citenamefont {Wu}\ \emph {et~al.}(2012)\citenamefont {Wu},
  \citenamefont {Mao}, \citenamefont {Jones}, \citenamefont {Yao},
  \citenamefont {Zhang},\ and\ \citenamefont {Xu}}]{NanoLett._12_2032_2012_Wu}%
  \BibitemOpen
  \bibfield  {author} {\bibinfo {author} {\bibfnamefont {S.}~\bibnamefont
  {Wu}}, \bibinfo {author} {\bibfnamefont {L.}~\bibnamefont {Mao}}, \bibinfo
  {author} {\bibfnamefont {A.~M.}\ \bibnamefont {Jones}}, \bibinfo {author}
  {\bibfnamefont {W.}~\bibnamefont {Yao}}, \bibinfo {author} {\bibfnamefont
  {C.}~\bibnamefont {Zhang}}, \ and\ \bibinfo {author} {\bibfnamefont
  {X.}~\bibnamefont {Xu}},\ }\href {\doibase 10.1021/nl300084j} {\bibfield
  {journal} {\bibinfo  {journal} {Nano Lett.}\ }\textbf {\bibinfo {volume}
  {12}},\ \bibinfo {pages} {2032} (\bibinfo {year} {2012})}\BibitemShut
  {NoStop}%
\bibitem [{\citenamefont {Avetissian}\ \emph
  {et~al.}(2012{\natexlab{a}})\citenamefont {Avetissian}, \citenamefont
  {Avetissian}, \citenamefont {Mkrtchian},\ and\ \citenamefont
  {Sedrakian}}]{J.Nanophoton._6_61702_2012_Avetissian}%
  \BibitemOpen
  \bibfield  {author} {\bibinfo {author} {\bibfnamefont {H.~K.}\ \bibnamefont
  {Avetissian}}, \bibinfo {author} {\bibfnamefont {A.~K.}\ \bibnamefont
  {Avetissian}}, \bibinfo {author} {\bibfnamefont {G.~F.}\ \bibnamefont
  {Mkrtchian}}, \ and\ \bibinfo {author} {\bibfnamefont {K.~V.}\ \bibnamefont
  {Sedrakian}},\ }\href {\doibase 10.1117/1.jnp.6.061702} {\bibfield  {journal}
  {\bibinfo  {journal} {J. Nanophoton.}\ }\textbf {\bibinfo {volume} {6}},\
  \bibinfo {pages} {061702} (\bibinfo {year} {2012}{\natexlab{a}})}\BibitemShut
  {NoStop}%
\bibitem [{\citenamefont
  {Khurgin}(1995)}]{Appl.Phys.Lett._67_1113_1995_Khurgin}%
  \BibitemOpen
  \bibfield  {author} {\bibinfo {author} {\bibfnamefont {J.~B.}\ \bibnamefont
  {Khurgin}},\ }\href {\doibase 10.1063/1.114978} {\bibfield  {journal}
  {\bibinfo  {journal} {Appl. Phys. Lett.}\ }\textbf {\bibinfo {volume} {67}},\
  \bibinfo {pages} {1113} (\bibinfo {year} {1995})}\BibitemShut {NoStop}%
\bibitem [{\citenamefont {Hendry}\ \emph {et~al.}(2010)\citenamefont {Hendry},
  \citenamefont {Hale}, \citenamefont {Moger}, \citenamefont {Savchenko},\ and\
  \citenamefont {Mikhailov}}]{Phys.Rev.Lett._105_097401_2010_Hendry}%
  \BibitemOpen
  \bibfield  {author} {\bibinfo {author} {\bibfnamefont {E.}~\bibnamefont
  {Hendry}}, \bibinfo {author} {\bibfnamefont {P.~J.}\ \bibnamefont {Hale}},
  \bibinfo {author} {\bibfnamefont {J.}~\bibnamefont {Moger}}, \bibinfo
  {author} {\bibfnamefont {A.~K.}\ \bibnamefont {Savchenko}}, \ and\ \bibinfo
  {author} {\bibfnamefont {S.~A.}\ \bibnamefont {Mikhailov}},\ }\href {\doibase
  10.1103/PhysRevLett.105.097401} {\bibfield  {journal} {\bibinfo  {journal}
  {Phys. Rev. Lett.}\ }\textbf {\bibinfo {volume} {105}},\ \bibinfo {pages}
  {097401} (\bibinfo {year} {2010})}\BibitemShut {NoStop}%
\bibitem [{\citenamefont {S\"ayn\"atjoki}\ \emph {et~al.}(2013)\citenamefont
  {S\"ayn\"atjoki}, \citenamefont {Karvonen}, \citenamefont {Riikonen},
  \citenamefont {Kim}, \citenamefont {Mehravar}, \citenamefont {Norwood},
  \citenamefont {Peyghambarian}, \citenamefont {Lipsanen},\ and\ \citenamefont
  {Kieu}}]{ACSNano_7_8441_2013_Saeynaetjoki}%
  \BibitemOpen
  \bibfield  {author} {\bibinfo {author} {\bibfnamefont {A.}~\bibnamefont
  {S\"ayn\"atjoki}}, \bibinfo {author} {\bibfnamefont {L.}~\bibnamefont
  {Karvonen}}, \bibinfo {author} {\bibfnamefont {J.}~\bibnamefont {Riikonen}},
  \bibinfo {author} {\bibfnamefont {W.}~\bibnamefont {Kim}}, \bibinfo {author}
  {\bibfnamefont {S.}~\bibnamefont {Mehravar}}, \bibinfo {author}
  {\bibfnamefont {R.~A.}\ \bibnamefont {Norwood}}, \bibinfo {author}
  {\bibfnamefont {N.}~\bibnamefont {Peyghambarian}}, \bibinfo {author}
  {\bibfnamefont {H.}~\bibnamefont {Lipsanen}}, \ and\ \bibinfo {author}
  {\bibfnamefont {K.}~\bibnamefont {Kieu}},\ }\href {\doibase
  10.1021/nn4042909} {\bibfield  {journal} {\bibinfo  {journal} {ACS Nano}\
  }\textbf {\bibinfo {volume} {7}},\ \bibinfo {pages} {8441} (\bibinfo {year}
  {2013})}\BibitemShut {NoStop}%
\bibitem [{\citenamefont {Kumar}\ \emph {et~al.}(2013)\citenamefont {Kumar},
  \citenamefont {Kumar}, \citenamefont {Gerstenkorn}, \citenamefont {Wang},
  \citenamefont {Chiu}, \citenamefont {Smirl},\ and\ \citenamefont
  {Zhao}}]{Phys.Rev.B_87_121406_2013_Kumar}%
  \BibitemOpen
  \bibfield  {author} {\bibinfo {author} {\bibfnamefont {N.}~\bibnamefont
  {Kumar}}, \bibinfo {author} {\bibfnamefont {J.}~\bibnamefont {Kumar}},
  \bibinfo {author} {\bibfnamefont {C.}~\bibnamefont {Gerstenkorn}}, \bibinfo
  {author} {\bibfnamefont {R.}~\bibnamefont {Wang}}, \bibinfo {author}
  {\bibfnamefont {H.-Y.}\ \bibnamefont {Chiu}}, \bibinfo {author}
  {\bibfnamefont {A.~L.}\ \bibnamefont {Smirl}}, \ and\ \bibinfo {author}
  {\bibfnamefont {H.}~\bibnamefont {Zhao}},\ }\href {\doibase
  10.1103/PhysRevB.87.121406} {\bibfield  {journal} {\bibinfo  {journal} {Phys.
  Rev. B}\ }\textbf {\bibinfo {volume} {87}},\ \bibinfo {pages} {121406}
  (\bibinfo {year} {2013})}\BibitemShut {NoStop}%
\bibitem [{\citenamefont {Boyd}(2008)}]{boyd_nonlinearoptics}%
  \BibitemOpen
  \bibfield  {author} {\bibinfo {author} {\bibfnamefont {R.~W.}\ \bibnamefont
  {Boyd}},\ }\href@noop {} {\emph {\bibinfo {title} {Nonlinear Optics}}},\
  \bibinfo {edition} {3rd}\ ed.\ (\bibinfo  {publisher} {Academic},\ \bibinfo
  {year} {2008})\BibitemShut {NoStop}%
\bibitem [{\citenamefont {Hong}\ \emph {et~al.}(2013)\citenamefont {Hong},
  \citenamefont {Dadap}, \citenamefont {Petrone}, \citenamefont {Yeh},
  \citenamefont {Hone},\ and\ \citenamefont
  {Osgood}}]{Phys.Rev.X_3_021014_2013_Hong}%
  \BibitemOpen
  \bibfield  {author} {\bibinfo {author} {\bibfnamefont {S.-Y.}\ \bibnamefont
  {Hong}}, \bibinfo {author} {\bibfnamefont {J.~I.}\ \bibnamefont {Dadap}},
  \bibinfo {author} {\bibfnamefont {N.}~\bibnamefont {Petrone}}, \bibinfo
  {author} {\bibfnamefont {P.-C.}\ \bibnamefont {Yeh}}, \bibinfo {author}
  {\bibfnamefont {J.}~\bibnamefont {Hone}}, \ and\ \bibinfo {author}
  {\bibfnamefont {R.~M.}\ \bibnamefont {Osgood}},\ }\href {\doibase
  10.1103/PhysRevX.3.021014} {\bibfield  {journal} {\bibinfo  {journal} {Phys.
  Rev. X}\ }\textbf {\bibinfo {volume} {3}},\ \bibinfo {pages} {021014}
  (\bibinfo {year} {2013})}\BibitemShut {NoStop}%
\bibitem [{\citenamefont {Yang}\ \emph {et~al.}(2011)\citenamefont {Yang},
  \citenamefont {Feng}, \citenamefont {Wang}, \citenamefont {Huang},
  \citenamefont {Chen}, \citenamefont {Wee},\ and\ \citenamefont
  {Ji}}]{NanoLett._11_2622_2011_Yang}%
  \BibitemOpen
  \bibfield  {author} {\bibinfo {author} {\bibfnamefont {H.}~\bibnamefont
  {Yang}}, \bibinfo {author} {\bibfnamefont {X.}~\bibnamefont {Feng}}, \bibinfo
  {author} {\bibfnamefont {Q.}~\bibnamefont {Wang}}, \bibinfo {author}
  {\bibfnamefont {H.}~\bibnamefont {Huang}}, \bibinfo {author} {\bibfnamefont
  {W.}~\bibnamefont {Chen}}, \bibinfo {author} {\bibfnamefont {A.~T.~S.}\
  \bibnamefont {Wee}}, \ and\ \bibinfo {author} {\bibfnamefont
  {W.}~\bibnamefont {Ji}},\ }\href {\doibase 10.1021/nl200587h} {\bibfield
  {journal} {\bibinfo  {journal} {Nano Lett.}\ }\textbf {\bibinfo {volume}
  {11}},\ \bibinfo {pages} {2622} (\bibinfo {year} {2011})}\BibitemShut
  {NoStop}%
\bibitem [{\citenamefont {Zhang}\ \emph {et~al.}(2012)\citenamefont {Zhang},
  \citenamefont {Virally}, \citenamefont {Bao}, \citenamefont {Ping},
  \citenamefont {Massar}, \citenamefont {Godbout},\ and\ \citenamefont
  {Kockaert}}]{Opt.Lett._37_1856_2012_Zhang}%
  \BibitemOpen
  \bibfield  {author} {\bibinfo {author} {\bibfnamefont {H.}~\bibnamefont
  {Zhang}}, \bibinfo {author} {\bibfnamefont {S.}~\bibnamefont {Virally}},
  \bibinfo {author} {\bibfnamefont {Q.}~\bibnamefont {Bao}}, \bibinfo {author}
  {\bibfnamefont {L.~K.}\ \bibnamefont {Ping}}, \bibinfo {author}
  {\bibfnamefont {S.}~\bibnamefont {Massar}}, \bibinfo {author} {\bibfnamefont
  {N.}~\bibnamefont {Godbout}}, \ and\ \bibinfo {author} {\bibfnamefont
  {P.}~\bibnamefont {Kockaert}},\ }\href {\doibase 10.1364/OL.37.001856}
  {\bibfield  {journal} {\bibinfo  {journal} {Opt. Lett.}\ }\textbf {\bibinfo
  {volume} {37}},\ \bibinfo {pages} {1856} (\bibinfo {year}
  {2012})}\BibitemShut {NoStop}%
\bibitem [{\citenamefont {Wu}\ \emph {et~al.}(2011)\citenamefont {Wu},
  \citenamefont {Zhang}, \citenamefont {Yan}, \citenamefont {Bian},
  \citenamefont {Wang}, \citenamefont {Bai}, \citenamefont {Lu}, \citenamefont
  {Zhao},\ and\ \citenamefont {Wang}}]{NanoLett._11_5159_2011_Wu}%
  \BibitemOpen
  \bibfield  {author} {\bibinfo {author} {\bibfnamefont {R.}~\bibnamefont
  {Wu}}, \bibinfo {author} {\bibfnamefont {Y.}~\bibnamefont {Zhang}}, \bibinfo
  {author} {\bibfnamefont {S.}~\bibnamefont {Yan}}, \bibinfo {author}
  {\bibfnamefont {F.}~\bibnamefont {Bian}}, \bibinfo {author} {\bibfnamefont
  {W.}~\bibnamefont {Wang}}, \bibinfo {author} {\bibfnamefont {X.}~\bibnamefont
  {Bai}}, \bibinfo {author} {\bibfnamefont {X.}~\bibnamefont {Lu}}, \bibinfo
  {author} {\bibfnamefont {J.}~\bibnamefont {Zhao}}, \ and\ \bibinfo {author}
  {\bibfnamefont {E.}~\bibnamefont {Wang}},\ }\href {\doibase
  10.1021/nl2023405} {\bibfield  {journal} {\bibinfo  {journal} {Nano Lett.}\
  }\textbf {\bibinfo {volume} {11}},\ \bibinfo {pages} {5159} (\bibinfo {year}
  {2011})}\BibitemShut {NoStop}%
\bibitem [{\citenamefont {Sun}\ \emph {et~al.}(2010)\citenamefont {Sun},
  \citenamefont {Divin}, \citenamefont {Rioux}, \citenamefont {Sipe},
  \citenamefont {Berger}, \citenamefont {de~Heer}, \citenamefont {First},\ and\
  \citenamefont {Norris}}]{NanoLett._10_1293_2010_Sun}%
  \BibitemOpen
  \bibfield  {author} {\bibinfo {author} {\bibfnamefont {D.}~\bibnamefont
  {Sun}}, \bibinfo {author} {\bibfnamefont {C.}~\bibnamefont {Divin}}, \bibinfo
  {author} {\bibfnamefont {J.}~\bibnamefont {Rioux}}, \bibinfo {author}
  {\bibfnamefont {J.~E.}\ \bibnamefont {Sipe}}, \bibinfo {author}
  {\bibfnamefont {C.}~\bibnamefont {Berger}}, \bibinfo {author} {\bibfnamefont
  {W.~A.}\ \bibnamefont {de~Heer}}, \bibinfo {author} {\bibfnamefont {P.~N.}\
  \bibnamefont {First}}, \ and\ \bibinfo {author} {\bibfnamefont {T.~B.}\
  \bibnamefont {Norris}},\ }\href {\doibase 10.1021/nl9040737} {\bibfield
  {journal} {\bibinfo  {journal} {Nano Lett.}\ }\textbf {\bibinfo {volume}
  {10}},\ \bibinfo {pages} {1293} (\bibinfo {year} {2010})}\BibitemShut
  {NoStop}%
\bibitem [{\citenamefont {Sun}\ \emph {et~al.}(2012{\natexlab{a}})\citenamefont
  {Sun}, \citenamefont {Divin}, \citenamefont {Mihnev}, \citenamefont {Winzer},
  \citenamefont {Malic}, \citenamefont {Knorr}, \citenamefont {Sipe},
  \citenamefont {Berger}, \citenamefont {de~Heer}, \citenamefont {First},\ and\
  \citenamefont {Norris}}]{NewJ.Phys._14_105012_2012_Sun}%
  \BibitemOpen
  \bibfield  {author} {\bibinfo {author} {\bibfnamefont {D.}~\bibnamefont
  {Sun}}, \bibinfo {author} {\bibfnamefont {C.}~\bibnamefont {Divin}}, \bibinfo
  {author} {\bibfnamefont {M.}~\bibnamefont {Mihnev}}, \bibinfo {author}
  {\bibfnamefont {T.}~\bibnamefont {Winzer}}, \bibinfo {author} {\bibfnamefont
  {E.}~\bibnamefont {Malic}}, \bibinfo {author} {\bibfnamefont
  {A.}~\bibnamefont {Knorr}}, \bibinfo {author} {\bibfnamefont {J.~E.}\
  \bibnamefont {Sipe}}, \bibinfo {author} {\bibfnamefont {C.}~\bibnamefont
  {Berger}}, \bibinfo {author} {\bibfnamefont {W.~A.}\ \bibnamefont {de~Heer}},
  \bibinfo {author} {\bibfnamefont {P.~N.}\ \bibnamefont {First}}, \ and\
  \bibinfo {author} {\bibfnamefont {T.~B.}\ \bibnamefont {Norris}},\ }\href
  {http://stacks.iop.org/1367-2630/14/i=10/a=105012} {\bibfield  {journal}
  {\bibinfo  {journal} {New J. Phys.}\ }\textbf {\bibinfo {volume} {14}},\
  \bibinfo {pages} {105012} (\bibinfo {year} {2012}{\natexlab{a}})}\BibitemShut
  {NoStop}%
\bibitem [{\citenamefont {Sun}\ \emph {et~al.}(2012{\natexlab{b}})\citenamefont
  {Sun}, \citenamefont {Rioux}, \citenamefont {Sipe}, \citenamefont {Zou},
  \citenamefont {Mihnev}, \citenamefont {Berger}, \citenamefont {de~Heer},
  \citenamefont {First},\ and\ \citenamefont
  {Norris}}]{Phys.Rev.B_85_165427_2012_Sun}%
  \BibitemOpen
  \bibfield  {author} {\bibinfo {author} {\bibfnamefont {D.}~\bibnamefont
  {Sun}}, \bibinfo {author} {\bibfnamefont {J.}~\bibnamefont {Rioux}}, \bibinfo
  {author} {\bibfnamefont {J.~E.}\ \bibnamefont {Sipe}}, \bibinfo {author}
  {\bibfnamefont {Y.}~\bibnamefont {Zou}}, \bibinfo {author} {\bibfnamefont
  {M.~T.}\ \bibnamefont {Mihnev}}, \bibinfo {author} {\bibfnamefont
  {C.}~\bibnamefont {Berger}}, \bibinfo {author} {\bibfnamefont {W.~A.}\
  \bibnamefont {de~Heer}}, \bibinfo {author} {\bibfnamefont {P.~N.}\
  \bibnamefont {First}}, \ and\ \bibinfo {author} {\bibfnamefont {T.~B.}\
  \bibnamefont {Norris}},\ }\href {\doibase 10.1103/physrevb.85.165427}
  {\bibfield  {journal} {\bibinfo  {journal} {Phys. Rev. B}\ }\textbf {\bibinfo
  {volume} {85}},\ \bibinfo {pages} {165427} (\bibinfo {year}
  {2012}{\natexlab{b}})}\BibitemShut {NoStop}%
\bibitem [{Note1()}]{Note1}%
  \BibitemOpen
  \bibinfo {note} {Note in particular the footnote on the second page of Cheng
  {\protect \it et al.} \cite {NewJ.Phys._16_53014_2014_Cheng}, which points
  out a source of confusion in comparing some of the experimental work with the
  theoretical study of Hendry {\protect \it et al.}\cite
  {Phys.Rev.Lett._105_097401_2010_Hendry}{}}\BibitemShut {NoStop}%
\bibitem [{\citenamefont
  {Mikhailov}(2014)}]{Phys.Rev.B_90_241301_2014_Mikhailov}%
  \BibitemOpen
  \bibfield  {author} {\bibinfo {author} {\bibfnamefont {S.~A.}\ \bibnamefont
  {Mikhailov}},\ }\href {\doibase 10.1103/physrevb.90.241301} {\bibfield
  {journal} {\bibinfo  {journal} {Phys. Rev. B}\ }\textbf {\bibinfo {volume}
  {90}},\ \bibinfo {pages} {241301(R)} (\bibinfo {year} {2014})}\BibitemShut
  {NoStop}%
\bibitem [{Note2()}]{Note2}%
  \BibitemOpen
  \bibinfo {note} {Despite the claim\cite {Phys.Rev.B_90_241301_2014_Mikhailov}
  that the scalar potential treatment of THG leads to disagreement with our
  earlier work\cite {NewJ.Phys._16_53014_2014_Cheng}, we find \cite
  {commentcomparison} agreement between the two approaches}\BibitemShut
  {NoStop}%
\bibitem [{\citenamefont {Cheng}\ \emph {et~al.}()\citenamefont {Cheng},
  \citenamefont {Vermeulen},\ and\ \citenamefont {Sipe}}]{commentcomparison}%
  \BibitemOpen
  \bibfield  {author} {\bibinfo {author} {\bibfnamefont {J.~L.}\ \bibnamefont
  {Cheng}}, \bibinfo {author} {\bibfnamefont {N.}~\bibnamefont {Vermeulen}}, \
  and\ \bibinfo {author} {\bibfnamefont {J.~E.}\ \bibnamefont {Sipe}},\
  }\href@noop {} {}\bibinfo {note} {{}unpublished}\BibitemShut {NoStop}%
\bibitem [{\citenamefont {Avetissian}\ \emph
  {et~al.}(2012{\natexlab{b}})\citenamefont {Avetissian}, \citenamefont
  {Avetissian}, \citenamefont {Mkrtchian},\ and\ \citenamefont
  {Sedrakian}}]{Phys.Rev.B_85_115443_2012_Avetissian}%
  \BibitemOpen
  \bibfield  {author} {\bibinfo {author} {\bibfnamefont {H.~K.}\ \bibnamefont
  {Avetissian}}, \bibinfo {author} {\bibfnamefont {A.~K.}\ \bibnamefont
  {Avetissian}}, \bibinfo {author} {\bibfnamefont {G.~F.}\ \bibnamefont
  {Mkrtchian}}, \ and\ \bibinfo {author} {\bibfnamefont {K.~V.}\ \bibnamefont
  {Sedrakian}},\ }\href {\doibase 10.1103/physrevb.85.115443} {\bibfield
  {journal} {\bibinfo  {journal} {Phys. Rev. B}\ }\textbf {\bibinfo {volume}
  {85}},\ \bibinfo {pages} {115443} (\bibinfo {year}
  {2012}{\natexlab{b}})}\BibitemShut {NoStop}%
\bibitem [{\citenamefont {Avetissian}\ \emph
  {et~al.}(2013{\natexlab{a}})\citenamefont {Avetissian}, \citenamefont
  {Mkrtchian}, \citenamefont {Batrakov}, \citenamefont {Maksimenko},\ and\
  \citenamefont {Hoffmann}}]{Phys.Rev.B_88_165411_2013_Avetissian}%
  \BibitemOpen
  \bibfield  {author} {\bibinfo {author} {\bibfnamefont {H.~K.}\ \bibnamefont
  {Avetissian}}, \bibinfo {author} {\bibfnamefont {G.~F.}\ \bibnamefont
  {Mkrtchian}}, \bibinfo {author} {\bibfnamefont {K.~G.}\ \bibnamefont
  {Batrakov}}, \bibinfo {author} {\bibfnamefont {S.~A.}\ \bibnamefont
  {Maksimenko}}, \ and\ \bibinfo {author} {\bibfnamefont {A.}~\bibnamefont
  {Hoffmann}},\ }\href {\doibase 10.1103/physrevb.88.165411} {\bibfield
  {journal} {\bibinfo  {journal} {Phys. Rev. B}\ }\textbf {\bibinfo {volume}
  {88}},\ \bibinfo {pages} {165411} (\bibinfo {year}
  {2013}{\natexlab{a}})}\BibitemShut {NoStop}%
\bibitem [{\citenamefont {Avetissian}\ \emph
  {et~al.}(2013{\natexlab{b}})\citenamefont {Avetissian}, \citenamefont
  {Mkrtchian}, \citenamefont {Batrakov}, \citenamefont {Maksimenko},\ and\
  \citenamefont {Hoffmann}}]{Phys.Rev.B_88_245411_2013_Avetissian}%
  \BibitemOpen
  \bibfield  {author} {\bibinfo {author} {\bibfnamefont {H.~K.}\ \bibnamefont
  {Avetissian}}, \bibinfo {author} {\bibfnamefont {G.~F.}\ \bibnamefont
  {Mkrtchian}}, \bibinfo {author} {\bibfnamefont {K.~G.}\ \bibnamefont
  {Batrakov}}, \bibinfo {author} {\bibfnamefont {S.~A.}\ \bibnamefont
  {Maksimenko}}, \ and\ \bibinfo {author} {\bibfnamefont {A.}~\bibnamefont
  {Hoffmann}},\ }\href {\doibase 10.1103/physrevb.88.245411} {\bibfield
  {journal} {\bibinfo  {journal} {Phys. Rev. B}\ }\textbf {\bibinfo {volume}
  {88}},\ \bibinfo {pages} {245411} (\bibinfo {year}
  {2013}{\natexlab{b}})}\BibitemShut {NoStop}%
\bibitem [{\citenamefont {Aversa}\ and\ \citenamefont
  {Sipe}(1995)}]{Phys.Rev.B_52_14636_1995_Aversa}%
  \BibitemOpen
  \bibfield  {author} {\bibinfo {author} {\bibfnamefont {C.}~\bibnamefont
  {Aversa}}\ and\ \bibinfo {author} {\bibfnamefont {J.~E.}\ \bibnamefont
  {Sipe}},\ }\href {\doibase 10.1103/PhysRevB.52.14636} {\bibfield  {journal}
  {\bibinfo  {journal} {Phys. Rev. B}\ }\textbf {\bibinfo {volume} {52}},\
  \bibinfo {pages} {14636} (\bibinfo {year} {1995})}\BibitemShut {NoStop}%
\bibitem [{\citenamefont {Haug}\ and\ \citenamefont
  {Koch}(2004)}]{book_HaugKoch}%
  \BibitemOpen
  \bibfield  {author} {\bibinfo {author} {\bibfnamefont {H.}~\bibnamefont
  {Haug}}\ and\ \bibinfo {author} {\bibfnamefont {S.~W.}\ \bibnamefont
  {Koch}},\ }\href@noop {} {\emph {\bibinfo {title} {Quantum Theory of the
  Optical and Electronic Properties of Semiconductor}}}\ (\bibinfo  {publisher}
  {World Scientific Publishing Co. Pte. Ltd.},\ \bibinfo {year}
  {2004})\BibitemShut {NoStop}%
\bibitem [{\citenamefont {Malic}\ \emph {et~al.}(2011)\citenamefont {Malic},
  \citenamefont {Winzer}, \citenamefont {Bobkin},\ and\ \citenamefont
  {Knorr}}]{Phys.Rev.B_84_205406_2011_Malic}%
  \BibitemOpen
  \bibfield  {author} {\bibinfo {author} {\bibfnamefont {E.}~\bibnamefont
  {Malic}}, \bibinfo {author} {\bibfnamefont {T.}~\bibnamefont {Winzer}},
  \bibinfo {author} {\bibfnamefont {E.}~\bibnamefont {Bobkin}}, \ and\ \bibinfo
  {author} {\bibfnamefont {A.}~\bibnamefont {Knorr}},\ }\href {\doibase
  10.1103/physrevb.84.205406} {\bibfield  {journal} {\bibinfo  {journal} {Phys.
  Rev. B}\ }\textbf {\bibinfo {volume} {84}},\ \bibinfo {pages} {205406}
  (\bibinfo {year} {2011})}\BibitemShut {NoStop}%
\bibitem [{\citenamefont {Haug}\ and\ \citenamefont
  {Jauho}(2007)}]{QuantumKineticsinTransportandOpticsofSemiconductors}%
  \BibitemOpen
  \bibfield  {author} {\bibinfo {author} {\bibfnamefont {H.}~\bibnamefont
  {Haug}}\ and\ \bibinfo {author} {\bibfnamefont {A.-P.}\ \bibnamefont
  {Jauho}},\ }\href
  {http://www.amazon.com/Kinetics-Transport-Semiconductors-Springer-Solid-State/dp/3540735615%3FSubscriptionId%3D0JYN1NVW651KCA56C102%26tag%3Dtechkie-20%26linkCode%3Dxm2%26camp%3D2025%26creative%3D165953%26creativeASIN%3D3540735615}
  {\emph {\bibinfo {title} {Quantum Kinetics in Transport and Optics of
  Semiconductors (Springer Series in Solid-State Sciences)}}}\ (\bibinfo
  {publisher} {Springer},\ \bibinfo {year} {2007})\BibitemShut {NoStop}%
\bibitem [{\citenamefont {Zhang}\ and\ \citenamefont
  {Wu}(2013)}]{Phys.Rev.B_87_085319_2013_Zhang}%
  \BibitemOpen
  \bibfield  {author} {\bibinfo {author} {\bibfnamefont {P.}~\bibnamefont
  {Zhang}}\ and\ \bibinfo {author} {\bibfnamefont {M.~W.}\ \bibnamefont {Wu}},\
  }\href {\doibase 10.1103/PhysRevB.87.085319} {\bibfield  {journal} {\bibinfo
  {journal} {Phys. Rev. B}\ }\textbf {\bibinfo {volume} {87}},\ \bibinfo
  {pages} {085319} (\bibinfo {year} {2013})}\BibitemShut {NoStop}%
\bibitem [{\citenamefont {Ando}\ \emph {et~al.}(2002)\citenamefont {Ando},
  \citenamefont {Zheng},\ and\ \citenamefont
  {Suzuura}}]{J.Phys.Soc.Jpn._71_1318_2002_Ando}%
  \BibitemOpen
  \bibfield  {author} {\bibinfo {author} {\bibfnamefont {T.}~\bibnamefont
  {Ando}}, \bibinfo {author} {\bibfnamefont {Y.}~\bibnamefont {Zheng}}, \ and\
  \bibinfo {author} {\bibfnamefont {H.}~\bibnamefont {Suzuura}},\ }\href
  {\doibase 10.1143/jpsj.71.1318} {\bibfield  {journal} {\bibinfo  {journal}
  {J. Phys. Soc. Jpn.}\ }\textbf {\bibinfo {volume} {71}},\ \bibinfo {pages}
  {1318–1324} (\bibinfo {year} {2002})}\BibitemShut {NoStop}%
\bibitem [{\citenamefont {Mermin}(1970)}]{Phys.Rev.B_1_2362_1970_Mermin}%
  \BibitemOpen
  \bibfield  {author} {\bibinfo {author} {\bibfnamefont {N.}~\bibnamefont
  {Mermin}},\ }\href {\doibase 10.1103/physrevb.1.2362} {\bibfield  {journal}
  {\bibinfo  {journal} {Phys. Rev. B}\ }\textbf {\bibinfo {volume} {1}},\
  \bibinfo {pages} {2362} (\bibinfo {year} {1970})}\BibitemShut {NoStop}%
\bibitem [{\citenamefont {Sutherland}(2003)}]{HandbookofNonlinearOptics}%
  \BibitemOpen
  \bibfield  {author} {\bibinfo {author} {\bibfnamefont {R.~L.}\ \bibnamefont
  {Sutherland}},\ }\href
  {http://www.amazon.com/Handbook-Nonlinear-Optical-Science-Engineering/dp/0824742435%3FSubscriptionId%3D0JYN1NVW651KCA56C102%26tag%3Dtechkie-20%26linkCode%3Dxm2%26camp%3D2025%26creative%3D165953%26creativeASIN%3D0824742435}
  {\emph {\bibinfo {title} {Handbook of Nonlinear Optics}}}\ (\bibinfo
  {publisher} {CRC Press},\ \bibinfo {year} {2003})\BibitemShut {NoStop}%
\bibitem [{\citenamefont {Mikhailov}\ and\ \citenamefont
  {Ziegler}(2007)}]{Phys.Rev.Lett._99_16803_2007_Mikhailov}%
  \BibitemOpen
  \bibfield  {author} {\bibinfo {author} {\bibfnamefont {S.}~\bibnamefont
  {Mikhailov}}\ and\ \bibinfo {author} {\bibfnamefont {K.}~\bibnamefont
  {Ziegler}},\ }\href {\doibase 10.1103/physrevlett.99.016803} {\bibfield
  {journal} {\bibinfo  {journal} {Phys. Rev. Lett.}\ }\textbf {\bibinfo
  {volume} {99}},\ \bibinfo {pages} {016803} (\bibinfo {year}
  {2007})}\BibitemShut {NoStop}%
\bibitem [{\citenamefont {{Cheng}}\ \emph {et~al.}()\citenamefont {{Cheng}},
  \citenamefont {Vermeulen},\ and\ \citenamefont {Sipe}}]{unpublishnumeric}%
  \BibitemOpen
  \bibfield  {author} {\bibinfo {author} {\bibfnamefont {J.~L.}\ \bibnamefont
  {{Cheng}}}, \bibinfo {author} {\bibfnamefont {N.}~\bibnamefont {Vermeulen}},
  \ and\ \bibinfo {author} {\bibfnamefont {J.~E.}\ \bibnamefont {Sipe}},\
  }\href@noop {} {}\bibinfo {note} {{}in preparation}\BibitemShut {NoStop}%
\bibitem [{\citenamefont {del Coso}\ and\ \citenamefont
  {Solis}(2004)}]{J.Opt.Soc.Am.B_21_640_2004_Coso}%
  \BibitemOpen
  \bibfield  {author} {\bibinfo {author} {\bibfnamefont {R.}~\bibnamefont {del
  Coso}}\ and\ \bibinfo {author} {\bibfnamefont {J.}~\bibnamefont {Solis}},\
  }\href {\doibase 10.1364/josab.21.000640} {\bibfield  {journal} {\bibinfo
  {journal} {J. Opt. Soc. Am. B}\ }\textbf {\bibinfo {volume} {21}},\ \bibinfo
  {pages} {640} (\bibinfo {year} {2004})}\BibitemShut {NoStop}%
\bibitem [{\citenamefont {Liu}\ \emph {et~al.}(2011{\natexlab{b}})\citenamefont
  {Liu}, \citenamefont {Shi}, \citenamefont {Yan}, \citenamefont {Zhou},\ and\
  \citenamefont {Tian}}]{Opt.Lett._36_2086_2011_Liu}%
  \BibitemOpen
  \bibfield  {author} {\bibinfo {author} {\bibfnamefont {Z.-B.}\ \bibnamefont
  {Liu}}, \bibinfo {author} {\bibfnamefont {S.}~\bibnamefont {Shi}}, \bibinfo
  {author} {\bibfnamefont {X.-Q.}\ \bibnamefont {Yan}}, \bibinfo {author}
  {\bibfnamefont {W.-Y.}\ \bibnamefont {Zhou}}, \ and\ \bibinfo {author}
  {\bibfnamefont {J.-G.}\ \bibnamefont {Tian}},\ }\href {\doibase
  10.1364/ol.36.002086} {\bibfield  {journal} {\bibinfo  {journal} {Opt.
  Lett.}\ }\textbf {\bibinfo {volume} {36}},\ \bibinfo {pages} {2086} (\bibinfo
  {year} {2011}{\natexlab{b}})}\BibitemShut {NoStop}%
\bibitem [{\citenamefont {Zhang}\ \emph {et~al.}(2013)\citenamefont {Zhang},
  \citenamefont {Liu}, \citenamefont {Li}, \citenamefont {Ma}, \citenamefont
  {Chen}, \citenamefont {Tian}, \citenamefont {Xu},\ and\ \citenamefont
  {Chen}}]{Opt.Express_21_7511_2013_Zhang}%
  \BibitemOpen
  \bibfield  {author} {\bibinfo {author} {\bibfnamefont {X.-L.}\ \bibnamefont
  {Zhang}}, \bibinfo {author} {\bibfnamefont {Z.-B.}\ \bibnamefont {Liu}},
  \bibinfo {author} {\bibfnamefont {X.-C.}\ \bibnamefont {Li}}, \bibinfo
  {author} {\bibfnamefont {Q.}~\bibnamefont {Ma}}, \bibinfo {author}
  {\bibfnamefont {X.-D.}\ \bibnamefont {Chen}}, \bibinfo {author}
  {\bibfnamefont {J.-G.}\ \bibnamefont {Tian}}, \bibinfo {author}
  {\bibfnamefont {Y.-F.}\ \bibnamefont {Xu}}, \ and\ \bibinfo {author}
  {\bibfnamefont {Y.-S.}\ \bibnamefont {Chen}},\ }\href {\doibase
  10.1364/oe.21.007511} {\bibfield  {journal} {\bibinfo  {journal} {Opt.
  Express}\ }\textbf {\bibinfo {volume} {21}},\ \bibinfo {pages} {7511}
  (\bibinfo {year} {2013})}\BibitemShut {NoStop}%
\bibitem [{\citenamefont {Bao}\ \emph {et~al.}(2009)\citenamefont {Bao},
  \citenamefont {Zhang}, \citenamefont {Wang}, \citenamefont {Ni},
  \citenamefont {Yan}, \citenamefont {Shen}, \citenamefont {Loh},\ and\
  \citenamefont {Tang}}]{Adv.Funct.Mater._19_3077_2009_Bao}%
  \BibitemOpen
  \bibfield  {author} {\bibinfo {author} {\bibfnamefont {Q.}~\bibnamefont
  {Bao}}, \bibinfo {author} {\bibfnamefont {H.}~\bibnamefont {Zhang}}, \bibinfo
  {author} {\bibfnamefont {Y.}~\bibnamefont {Wang}}, \bibinfo {author}
  {\bibfnamefont {Z.}~\bibnamefont {Ni}}, \bibinfo {author} {\bibfnamefont
  {Y.}~\bibnamefont {Yan}}, \bibinfo {author} {\bibfnamefont {Z.~X.}\
  \bibnamefont {Shen}}, \bibinfo {author} {\bibfnamefont {K.~P.}\ \bibnamefont
  {Loh}}, \ and\ \bibinfo {author} {\bibfnamefont {D.~Y.}\ \bibnamefont
  {Tang}},\ }\href {\doibase 10.1002/adfm.200901007} {\bibfield  {journal}
  {\bibinfo  {journal} {Adv. Funct. Mater.}\ }\textbf {\bibinfo {volume}
  {19}},\ \bibinfo {pages} {3077} (\bibinfo {year} {2009})}\BibitemShut
  {NoStop}%
\bibitem [{\citenamefont {Zhang}\ and\ \citenamefont
  {Voss}(2011)}]{Opt.Lett._36_4569_2011_Zhang}%
  \BibitemOpen
  \bibfield  {author} {\bibinfo {author} {\bibfnamefont {Z.}~\bibnamefont
  {Zhang}}\ and\ \bibinfo {author} {\bibfnamefont {P.~L.}\ \bibnamefont
  {Voss}},\ }\href {\doibase 10.1364/OL.36.004569} {\bibfield  {journal}
  {\bibinfo  {journal} {Opt. Lett.}\ }\textbf {\bibinfo {volume} {36}},\
  \bibinfo {pages} {4569} (\bibinfo {year} {2011})}\BibitemShut {NoStop}%
\end{thebibliography}
